\documentclass[prd,showkeys,floatfix,twocolumn,amsmath,amssymb,floatfix]{revtex4}
\usepackage{graphicx}
\usepackage{multirow}
\usepackage{subfigure}
\usepackage{color}
\usepackage[colorlinks,citecolor=blue,anchorcolor=red,menucolor=red,linkcolor=red,filecolor=red,runcolor=red,urlcolor=blue,frenchlinks=red]{hyperref}
\usepackage{enumitem}

\makeatletter
\renewcommand{\@thesubfigure}{\hskip\subfiglabelskip}
\makeatother
\allowdisplaybreaks[3]

\begin{document}
%=====================================================================================
%=====================================================================================
\title{Is the $X17$ composed of four bare quarks?}
%=====================================================================================
%=====================================================================================
%

\author{Hua-Xing Chen}

\affiliation{
School of Physics, Southeast University, Nanjing 210094, China \\
School of Physics, Beihang University, Beijing 100191, China
}

\begin{abstract}
When studying tetraquark states using the method of QCD sum rules, we find some tetraquark currents, such as $\bar u_L \gamma_\mu d_L~\bar d_L \gamma^\mu u_L$ and $\bar u_L \gamma_\mu d_L~\bar d_R \gamma^\mu u_R$, etc. We calculate their correlation functions using the method of operator product expansion, and find that non-perturbative QCD effects do not contribute much to them. This indicates the possible existence of tetraquark states composed of bare quarks with abnormally small masses. We study the possible assignment of the $X17$ as such a state. We argue that the Atomki experiments might observe a new type of nuclear decay process, where sea quarks take away the extra energy of the nucleus. A unique feature of this tetraquark assignment is that we predict two almost degenerate states with significantly different widths.
\end{abstract}
\maketitle
\pagenumbering{arabic}
%
%
%
%=====================================================================================
%=====================================================================================
\section{Introduction}\label{sec:intro}
%=====================================================================================
%=====================================================================================
%

In Ref.~\cite{Krasznahorkay:2015iga} Krasznahorkay et al. studied the internal (electron-positron) pair creation process, and observed an anomaly in the decay of the excited $^8Be^*$ at 18.15~MeV with $J^P = 1^+$ into the ground-state $^8Be$ with $J^P = 0^+$. In this experiment, a highly significant structure with the mass $M = 16.70 \pm 0.35 \pm 0.5$~MeV was observed. Later in Ref.~\cite{Krasznahorkay:2019lyl} the same group announced a similar anomaly in the decay of the excited $^4He^*$ at 21.01~MeV with $J^P = 0^-$ into the ground-state $^4He$ with $J^P = 0^+$. In this experiment the same/similar structure was observed, whose mass and width were measured to be $M = 16.84 \pm 0.16 \pm 0.20$~MeV and $\Gamma \approx 3.9 \times 10^{-5}$~eV. However, this structure was not observed  in the bremsstrahlung reaction $e^-Z \to e^-ZX$ performed by NA64~\cite{Banerjee:2018vgk,Banerjee:2019hmi}, so its existence still needs to be confirmed in future experiments.

Since the above two mass values are both about 17~MeV, this structure is called the ``$X17$''. Its observations quickly attracted lots of interests from theorists, and various explanations were proposed, such as a fifth force~\cite{Feng:2016jff,Feng:2016ysn,Gu:2016ege,Neves:2016nek,Kahn:2016vjr,Fayet:2016nyc,Dror:2017nsg,Das:2019ycx,Jentschura:2020zlr}, new and/or dark matter~\cite{Alexander:2016aln,Battaglieri:2017aum,Liang:2016ffe,Jia:2016uxs,Kitahara:2016zyb,Ellwanger:2016wfe,Chen:2016tdz,Kozaczuk:2016nma,Chiang:2016cyf,Dror:2017ehi,Kozaczuk:2017per,Jia:2018mkc,DelleRose:2018pgm,DelleRose:2018eic,Hati:2020fzp}, and axions~\cite{Bauer:2017ris,Alves:2017avw,Kirpichnikov:2020tcf,Dusaev:2020gxi}, etc. More discussions can be found in Refs.~\cite{Zhang:2017zap,DelleRose:2017xil,Fornal:2017msy,Jiang:2018uhs,Chen:2019ivz,Nam:2019osu,Pulice:2019xel,Krasnikov:2019dgh,Koch:2020ouk}. Besides, there exist some theoretical studies~\cite{Wong:2020hjc,Veselsky:2020ewb} at the QCD level using quarks and hadrons, which provide alternative assignments for the $X17$.

In this paper we study the $X17$ also at the QCD level. The idea originates from our previous QCD sum rule studies on light tetraquark states of $J^{P} = 0^{\pm}$~\cite{Chen:2006hy,Chen:2006zh,Chen:2007xr,Jiao:2009ra,Dong:2020okt}, based on which we find two interpolating currents,
\begin{eqnarray*}
J_{LL} &=& \bar{u}^a_L \gamma_\mu d^a_L~\bar{d}_L^b \gamma^\mu u_L^b \, ,
\\ J_{LR} &=& \bar{u}^a_L \gamma_\mu d^a_L~\bar{d}_R^b \gamma^\mu u_R^b \, ,
\end{eqnarray*}
as well as their chiral partners,
\begin{eqnarray*}
J_{RL} &=& \bar{u}^a_R \gamma_\mu d^a_R~\bar{d}_L^b \gamma^\mu u_L^b \, ,
\\ J_{RR} &=& \bar{u}^a_R \gamma_\mu d^a_R~\bar{d}_R^b \gamma^\mu u_R^b \, .
\end{eqnarray*}
Here $a$ and $b$ are color indices, and the sum over repeated indices is taken; $u_{L/R}$ and $d_{L/R}$ are the left/right-handed $up$ and $down$ quark fields, respectively.

In this paper we shall calculate their two-point correlation functions using the method of operator product expansion (OPE), and find that non-perturbative QCD effects do not contribute much to them. Hence, these tetraquark currents may behave as the ``singular points'' of the strong interaction. As we know, masses of hadrons are always significantly larger than masses of their valence quarks inside, and the extra masses are mainly responsible by the non-perturbative nature of strong interaction in the low energy region. Given the very limited non-perturbative QCD contributions, one naturally wonders whether the tetraquark states $X_{LL/LR/RL/RR}$ corresponding to $J_{LL/LR/RL/RR}$ can still have the normal hadron masses at the GeV level, if such states exist.

In this paper we shall assume the existence of $X_{LL/LR/RL/RR}$ and study them using the method of QCD sum rules. Our results suggest that these tetraquark states do have abnormally small masses, so they can be taken as combinations of bare quarks. This inspires us to investigate the possible assignment of the $X17$ as such a tetraquark state, based on which we shall study its possible production and decay mechanisms. We argue that the Atomki experiments~\cite{Krasznahorkay:2015iga,Krasznahorkay:2019lyl} might observe a new type of nuclear decay process, similar to the evaporation or the sublimation process in some aspects, where sea quarks take away the extra energy of the nucleus. We shall find the unique feature of this tetraquark assignment that we predict two almost degenerate states with significantly different widths.

This paper is organized as follows. In Sec.~\ref{sec:current} we construct the interpolating currents that are needed to study the $X17$. In Sec.~\ref{sec:quark} we investigate these currents at the quark-gluon level using the method of operator product expansion (OPE), where we pay special attention to non-perturbative QCD effects. The obtained QCD sum rule expressions are listed in Appendix~\ref{app:sumrule}. In Sec.~\ref{sec:phen} we investigate these currents at the hadron level, where we estimate their masses and study their possible production and decay mechanisms. In Sec.~\ref{sec:summary} we discuss the results and offer a summary.

%
%=====================================================================================
%=====================================================================================
\section{Tetraquark Currents}\label{sec:current}
%=====================================================================================
%=====================================================================================
%

There exist lots of tetraquark interpolating currents composed of two quark and two antiquark fields. To write them, Lorentz and color indices are contracted with suitable coefficients $C_{abcd}^{\mu\nu\rho\sigma}$ to provide necessary quantum numbers,
%
%%%%%%%%%%%%%%%%%%%%%%%%%%%%%%%%%%%%%%%%%%%%%%%%%%%%%%%%%%%%%%%%%%%%%%%%%%%%%%
\begin{equation}
J(x) = C_{abcd}^{\mu\nu\rho\sigma} ~ \bar{q}^a_\mu(x) ~ \bar{q}^b_\nu(x) ~ q^c_\rho(x) ~ q^d_\sigma(x) \, ,
\label{def:current}
\end{equation}
%%%%%%%%%%%%%%%%%%%%%%%%%%%%%%%%%%%%%%%%%%%%%%%%%%%%%%%%%%%%%%%%%%%%%%%%%%%%%%
%
where $q(x)$ represents an $up$ or $down$ quark field; $\mu, \nu, \rho, \sigma$ are Dirac spinor indices; $a, b, c, d$ are color indices; the sum over repeated indices is taken. Note that currents need not to have definite quantum numbers in principle. For examples, the vector meson current $J_\mu^{\rho^+} = \bar d \gamma_\mu u$ coupling to the $\rho^+$ meson does not have a $C$-parity, and the weak neutral current does not have a $P$-parity.

In the past years we have used many tetraquark currents to study tetraquark states through the method of QCD sum rules~\cite{Chen:2006hy,Chen:2006zh,Chen:2007xr,Jiao:2009ra,Dong:2020okt}. All of them have definite $P$-parities, either $P=+$ or $P=-$. A naive question is whether we can study some tetraquark current without a definite $P$-parity. The answer is yes in practice, {\it i.e.}, we can always use the method of operator product expansion to calculate its two-point correlation function, no matter whether it has a definite $P$-parity or not.

We have studied scalar tetraquark currents of $J^{P} = 0^{+}$ in Refs.~\cite{Chen:2006hy,Chen:2006zh,Chen:2007xr}, and pseudoscalar tetraquark currents of $J^{P} = 0^{-}$ in Refs.~\cite{Jiao:2009ra,Dong:2020okt}. Based on these results, we find the following two currents,
\begin{eqnarray}
J_{LL} &=& \bar{u}^a_L \gamma_\mu d^a_L~\bar{d}_L^b \gamma^\mu u_L^b
\label{def:JLL}
\\ \nonumber &=& \bar{u}^a \gamma_\mu {1-\gamma_5\over2} d^a~\bar{d}^b \gamma^\mu {1-\gamma_5\over2} u^b \, ,
\\ J_{LR} &=& \bar{u}^a_L \gamma_\mu d^a_L~\bar{d}_R^b \gamma^\mu u_R^b
\label{def:JLR}
\\ \nonumber &=& \bar{u}^a \gamma_\mu {1-\gamma_5\over2} d^a~\bar{d}^b \gamma^\mu {1+\gamma_5\over2} u^b \, ,
\end{eqnarray}
have the special property that non-perturbative QCD effects do not contribute much to them within the method of QCD sum rules. So do their chiral partners:
\begin{eqnarray}
J_{RL} &=& \bar{u}^a_R \gamma_\mu d^a_R~\bar{d}_L^b \gamma^\mu u_L^b
\label{def:JRL}
\\ \nonumber &=& \bar{u}^a \gamma_\mu {1+\gamma_5\over2} d^a~\bar{d}^b \gamma^\mu {1-\gamma_5\over2} u^b \, ,
\\ J_{RR} &=& \bar{u}^a_R \gamma_\mu d^a_R~\bar{d}_R^b \gamma^\mu u_R^b
\label{def:JRR}
\\ \nonumber &=& \bar{u}^a \gamma_\mu {1+\gamma_5\over2} d^a~\bar{d}^b \gamma^\mu {1+\gamma_5\over2} u^b \, .
\end{eqnarray}
We shall detailedly discuss them in the following sections.

For comparisons, we shall investigate several other tetraquark currents with definite $P$-parities:
\begin{eqnarray}
J_{VV} &=& \bar{u}^a \gamma_\mu d^a~\bar{d}^b \gamma^\mu u^b \, ,
\label{def:JVV}
\\ J_{AA} &=& \bar{u}^a \gamma_\mu \gamma_5 d^a~\bar{d}^b \gamma^\mu \gamma_5 u^b \, ,
\label{def:JAA}
\\ J_{VA} &=& \bar{u}^a \gamma_\mu d^a~\bar{d}^b \gamma^\mu \gamma_5 u^b \, ,
\label{def:JVA}
\\ J_{AV} &=& \bar{u}^a \gamma_\mu \gamma_5 d^a~\bar{d}^b \gamma^\mu u^b \, .
\label{def:JAV}
\end{eqnarray}
The former two $J_{VV}$ and $J_{AA}$ are scalar tetraquark currents of $J^{P} = 0^{+}$, and the latter two $J_{VA}$ and $J_{AV}$ are pseudoscalar tetraquark currents of $J^{P} = 0^{-}$.

We shall also separately investigate the positive- and negative-parity components of $J_{LL/LR/RL/RR}$:
\begin{eqnarray}
J_{LL} &=& J_{LL;+} + J_{LL;-} \, ,
\\ J_{LR} &=& J_{LR;+} + J_{LR;-} \, ,
\\ J_{RL} &=& J_{LR;+} - J_{LR;-} \, ,
\\ J_{RR} &=& J_{LL;+} - J_{LL;-} \, ,
\end{eqnarray}
where $J_{LL;+}$ and $J_{LR;+}$ have the positive parity $P=+$, and $J_{LL;-}$ and $J_{LR;-}$ have the negative parity $P=-$:
\begin{eqnarray}
J_{LL;+} &=& + {1\over4} ~ J_{VV} + {1\over4} ~ J_{AA} \, ,
\label{def:JLLP}
\\ J_{LL;-} &=& -{1\over4} ~ J_{VA} - {1\over4} ~ J_{AV} \, ,
\label{def:JLLN}
\\ J_{LR;+} &=& + {1\over4} ~ J_{VV} - {1\over4} ~ J_{AA} \, ,
\label{def:JLRP}
\\ J_{LR;-} &=& + {1\over4} ~ J_{VA} - {1\over4} ~ J_{AV} \, .
\label{def:JLRN}
\end{eqnarray}

The tetraquark currents $J_{LL/LR/RL/RR}$ can be separated into two color-singlet ``chiral'' quark-antiquark currents, which we shall study in Sec.~\ref{sec:binding}:
\begin{eqnarray}
J^\mu_{L} &=& \bar{u}^a_L \gamma^\mu d^a_L = \bar{u}^a \gamma^\mu {1-\gamma_5\over2} d^a \, ,
\label{def:JL}
\\ J^\mu_{R} &=& \bar{u}^a_R \gamma^\mu d^a_R = \bar{u}^a \gamma^\mu {1+\gamma_5\over2} d^a \, .
\label{def:JR}
\end{eqnarray}

In Sec.~\ref{sec:decay} we shall study possible decay processes of $J_{LL/LR/RL/RR}$. To do this, we apply the Fierz transformation on them to obtain:
\begin{eqnarray}
J_{LL} &=& \bar{u}^a_L \gamma_\mu d^a_L~\bar{d}_L^b \gamma^\mu u_L^b \rightarrow \bar{u}^a_L \gamma_\mu u_L^b~\bar{d}_L^b \gamma^\mu d^a_L \, ,
\label{fierzLL}
\\ J_{LR} &=& \bar{u}^a_L \gamma_\mu d^a_L~\bar{d}_R^b \gamma^\mu u_R^b \rightarrow -2~\bar{u}^a_L u_R^b~\bar{d}_R^b d^a_L \, ,
\label{fierzLR}
\\ J_{RL} &=& \bar{u}^a_R \gamma_\mu d^a_R~\bar{d}_L^b \gamma^\mu u_L^b \rightarrow -2~\bar{u}^a_R u_L^b~\bar{d}_L^b d^a_R \, ,
\label{fierzRL}
\\ J_{RR} &=& \bar{u}^a_R \gamma_\mu d^a_R~\bar{d}_R^b \gamma^\mu u_R^b \rightarrow \bar{u}^a_R \gamma_\mu u_R^b~\bar{d}_R^b \gamma^\mu d^a_R \, .
\label{fierzRR}
\end{eqnarray}

%
%=====================================================================================
%=====================================================================================
\section{Quark-Level Calculations}\label{sec:quark}
%=====================================================================================
%=====================================================================================
%

In the past decades the method of QCD sum rules has proven to be a very powerful and successful non-perturbative method~\cite{Shifman:1978bx,Reinders:1984sr}. In this method we consider the two-point correlation function,
%
%%%%%%%%%%%%%%%%%%%%%%%%%%%%%%%%%%%%%%%%%%%%%%%%%%%%%%%%%%%%%%%%%%%%%%%%%%%%%%
\begin{equation}
\Pi(q^2)\,\equiv\,i\int d^4x~e^{iqx}~\langle 0 | \mathbb{T}\left[ J(x){J^\dagger}(0) \right] | 0 \rangle \, ,
\label{eq:pi}
\end{equation}
%%%%%%%%%%%%%%%%%%%%%%%%%%%%%%%%%%%%%%%%%%%%%%%%%%%%%%%%%%%%%%%%%%%%%%%%%%%%%%
%
at both quark-gluon and hadron levels. In this section we study it at the quark-gluon level using the method of operator product expansion (OPE).

\subsection{Operator Product Expansion (OPE)}

Taking $J_{LL}$ as an example, we insert it into Eq.~(\ref{eq:pi}), and calculate its correlation function $\Pi_{LL}(q^2)$ using the method of operator product expansion (OPE). In the calculation we use the following quark propagator~\cite{Shifman:1978bx,Reinders:1984sr,Yang:1993bp,Hwang:1996pe}:
%
%%%%%%%%%%%%%%%%%%%%%%%%%%%%%%%%%%%%%%%%%%%%%%%%%%%%%%%%%%%%%%%%%%%%%%%%%%%%%%
\begin{eqnarray}
\label{eq:quark}
&& i S^{ab}(x) \equiv \langle0| \mathbb{T}\left[q^a(x)\bar{q}^b(0)\right]|0\rangle
\\ \nonumber &=& \frac{i \delta^{ab} \hat{x}}{2\pi^2 x^4}
-\frac{\delta^{ab}}{4\pi^2x^2}m_q
+\frac{i\delta^{ab}\hat{x}}{8\pi^2x^2}m_q^2
\\ \nonumber &&
+\frac{i}{32\pi^2}~g_c \frac{\lambda^n_{ab}}{2}G^n_{\mu\nu}~\frac{1}{x^2}(\sigma^{\mu\nu}\hat{x}+\hat{x}\sigma^{\mu\nu})
\\ \nonumber &&
-\frac{\delta^{ab}}{12}\langle\bar{q}q\rangle
+\frac{\delta^{ab} x^2}{192}\langle g_c\bar{q}\sigma Gq\rangle
\\ \nonumber &&
+\frac{i\delta^{ab}\hat{x}}{48}m_q \langle\bar{q}q\rangle
-\frac{i\delta^{ab}x^2 \hat{x}}{1152}m_q \langle g_c\bar{q}\sigma Gq\rangle \, ,
\end{eqnarray}
%%%%%%%%%%%%%%%%%%%%%%%%%%%%%%%%%%%%%%%%%%%%%%%%%%%%%%%%%%%%%%%%%%%%%%%%%%%%%%
%
where $\hat{x} = \gamma_\mu x^\mu$; $m_q$ is the current quark mass; $\langle\bar{q}q\rangle$ and $\langle g_c\bar{q}\sigma Gq\rangle$ are the $[{\rm D(imenion)}=3]$ quark condensate and $[{\rm D}=5]$ quark-gluon mixed condensate, respectively; the $[{\rm D}=4]$ and $[{\rm D}=6]$ gluon condensates $\langle g_c^2 GG\rangle \equiv \langle g_c^2 G^a_{\mu\nu} G^a_{\mu\nu} \rangle$ and $\langle g_c^3 fGGG\rangle \equiv \langle g_c^3 f^{abc} G^a_{\mu\nu} G^b_{\nu\rho} G^c_{\rho\mu}\rangle$ can be obtained by contracting two and three gluon fields, respectively.

\begin{figure*}[hbtp]
\begin{center}
\subfigure[($a$)]{
\scalebox{0.15}{\includegraphics{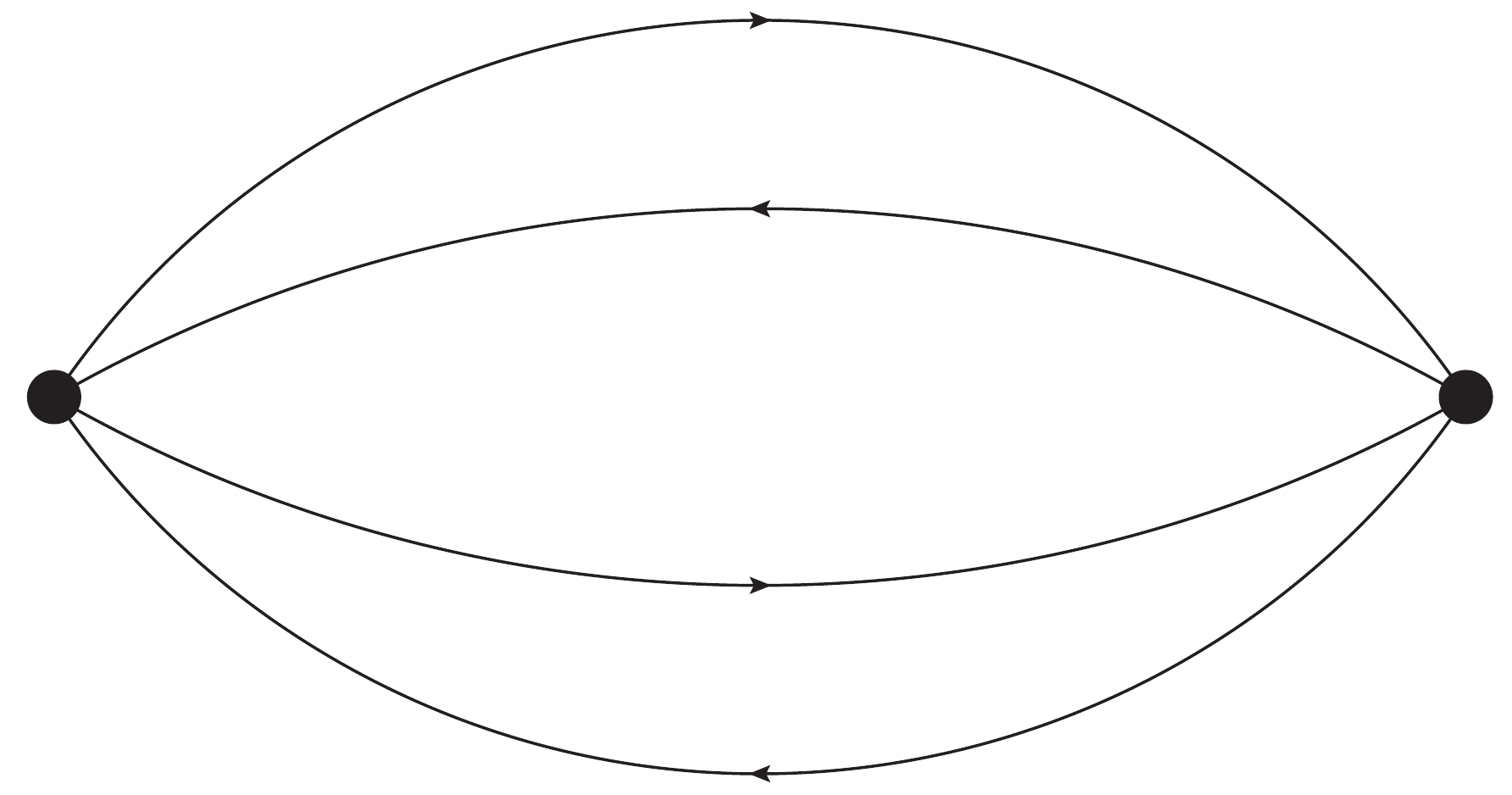}}}
\\[3mm]
\subfigure[($b{\rm-}1$)]{
\scalebox{0.15}{\includegraphics{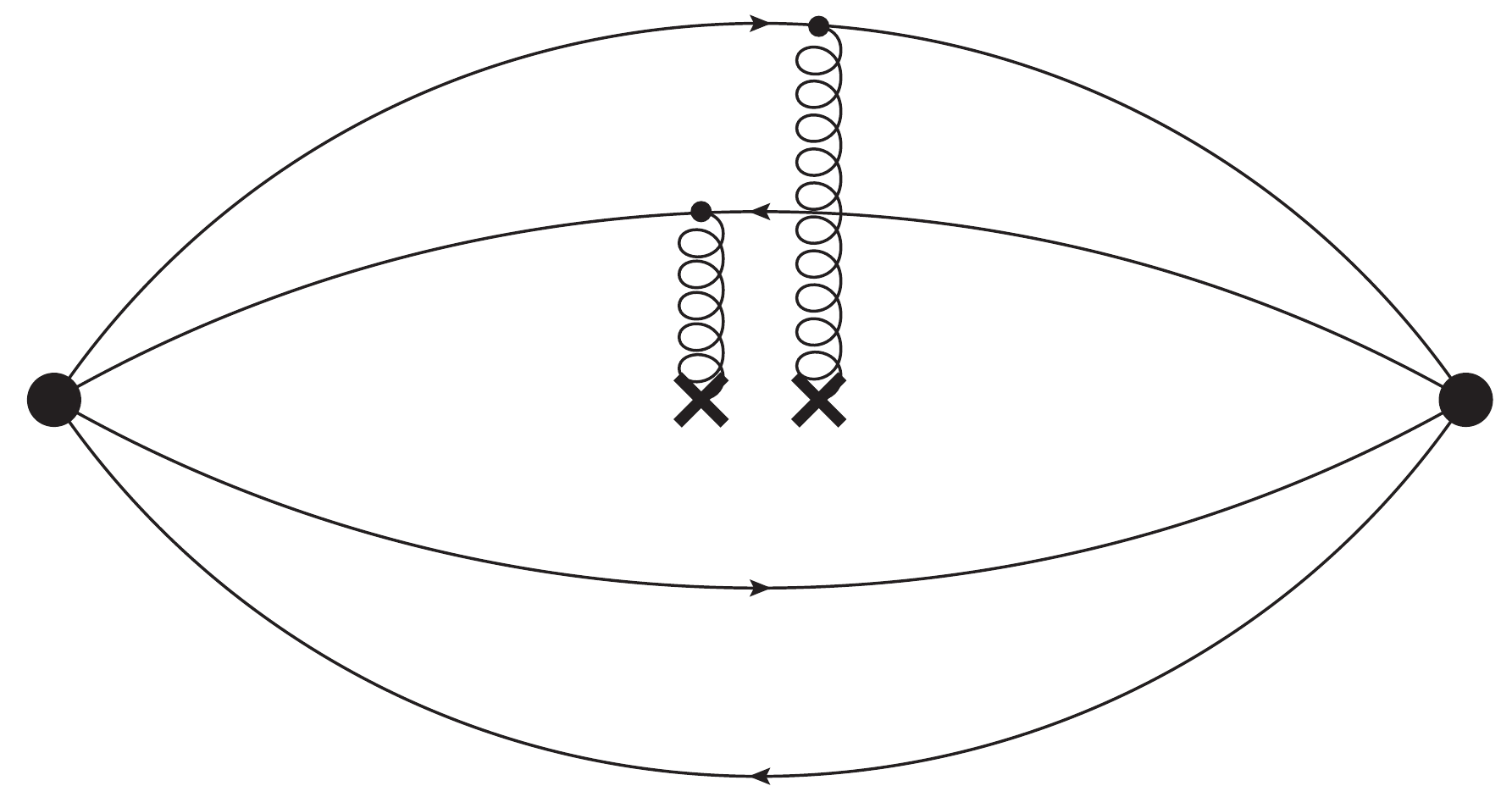}}}~~~~~
\subfigure[($b{\rm-}2$)]{
\scalebox{0.15}{\includegraphics{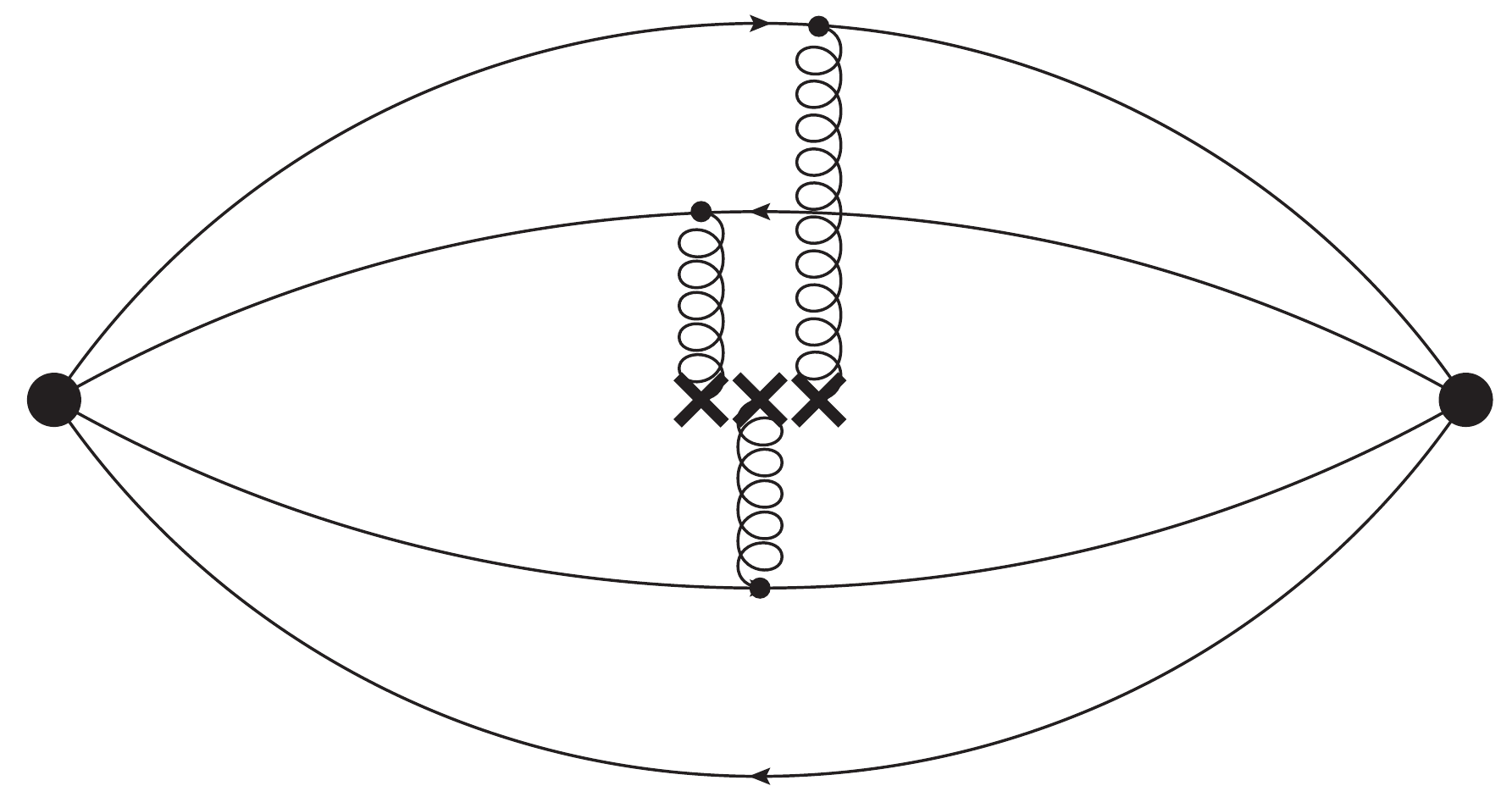}}}
\\[3mm]
\subfigure[($c{\rm-}1$)]{
\scalebox{0.15}{\includegraphics{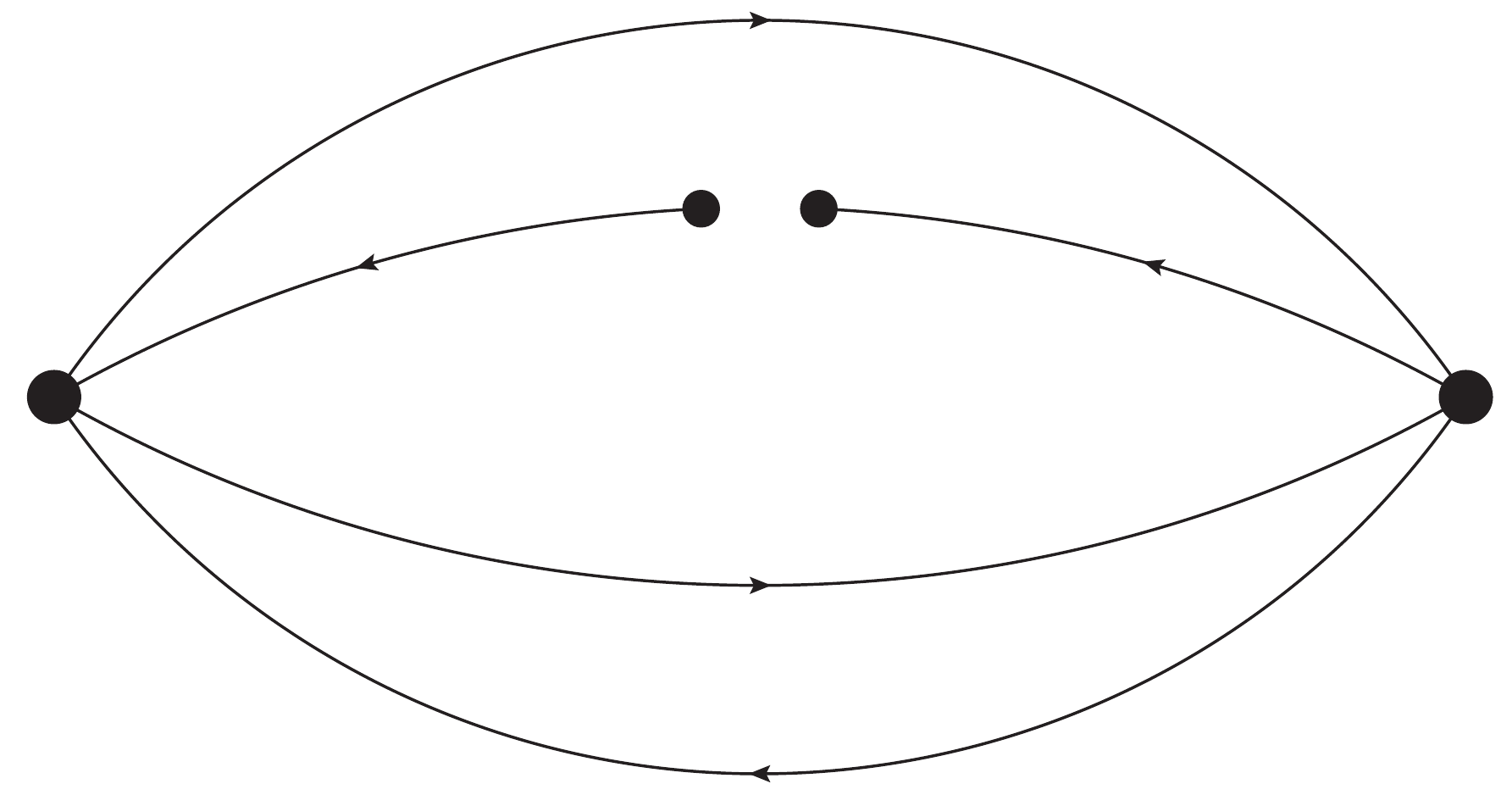}}}~~~~~
\subfigure[($c{\rm-}2$)]{
\scalebox{0.15}{\includegraphics{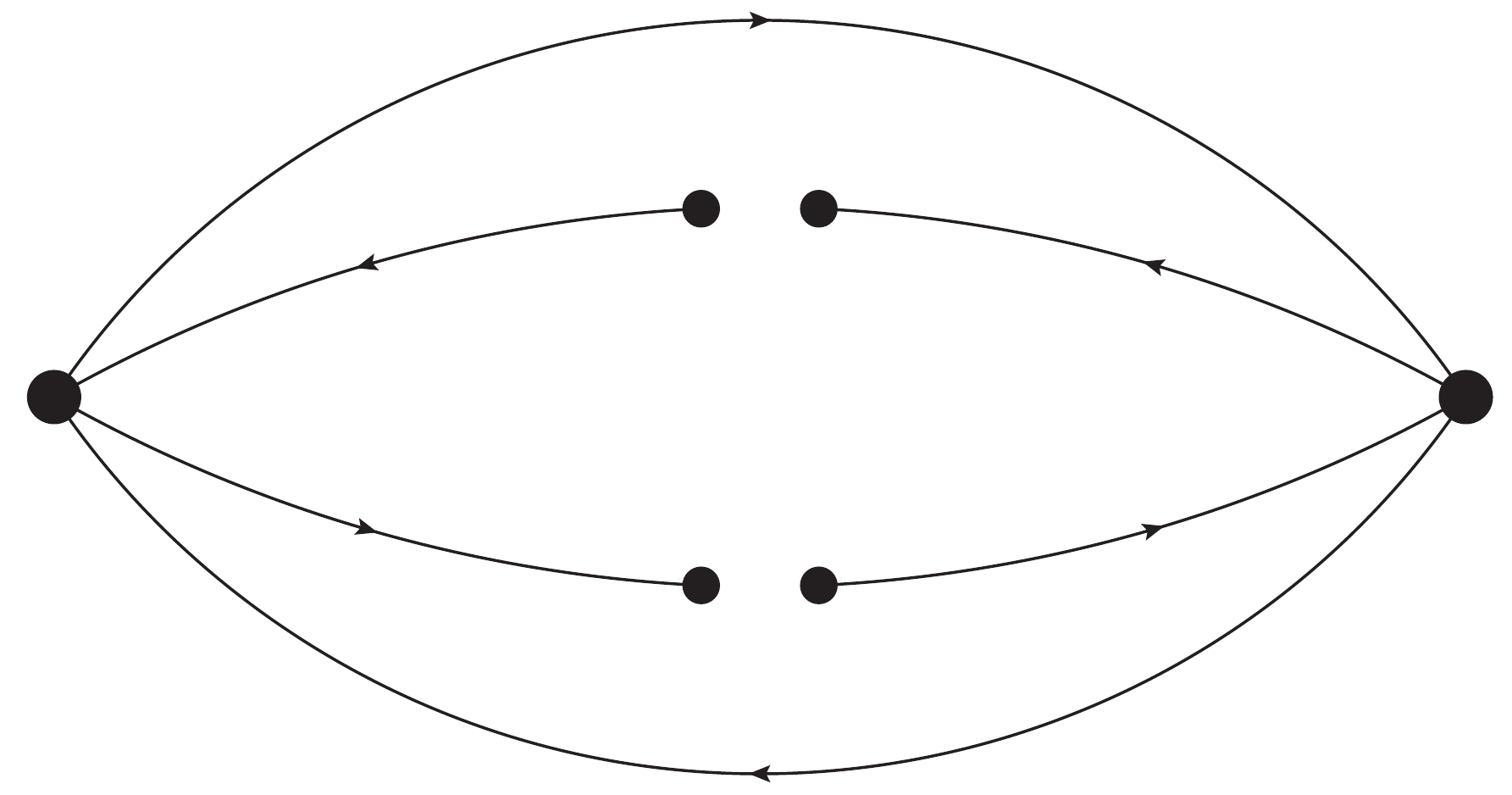}}}~~~~~
\subfigure[($c{\rm-}3$)]{
\scalebox{0.15}{\includegraphics{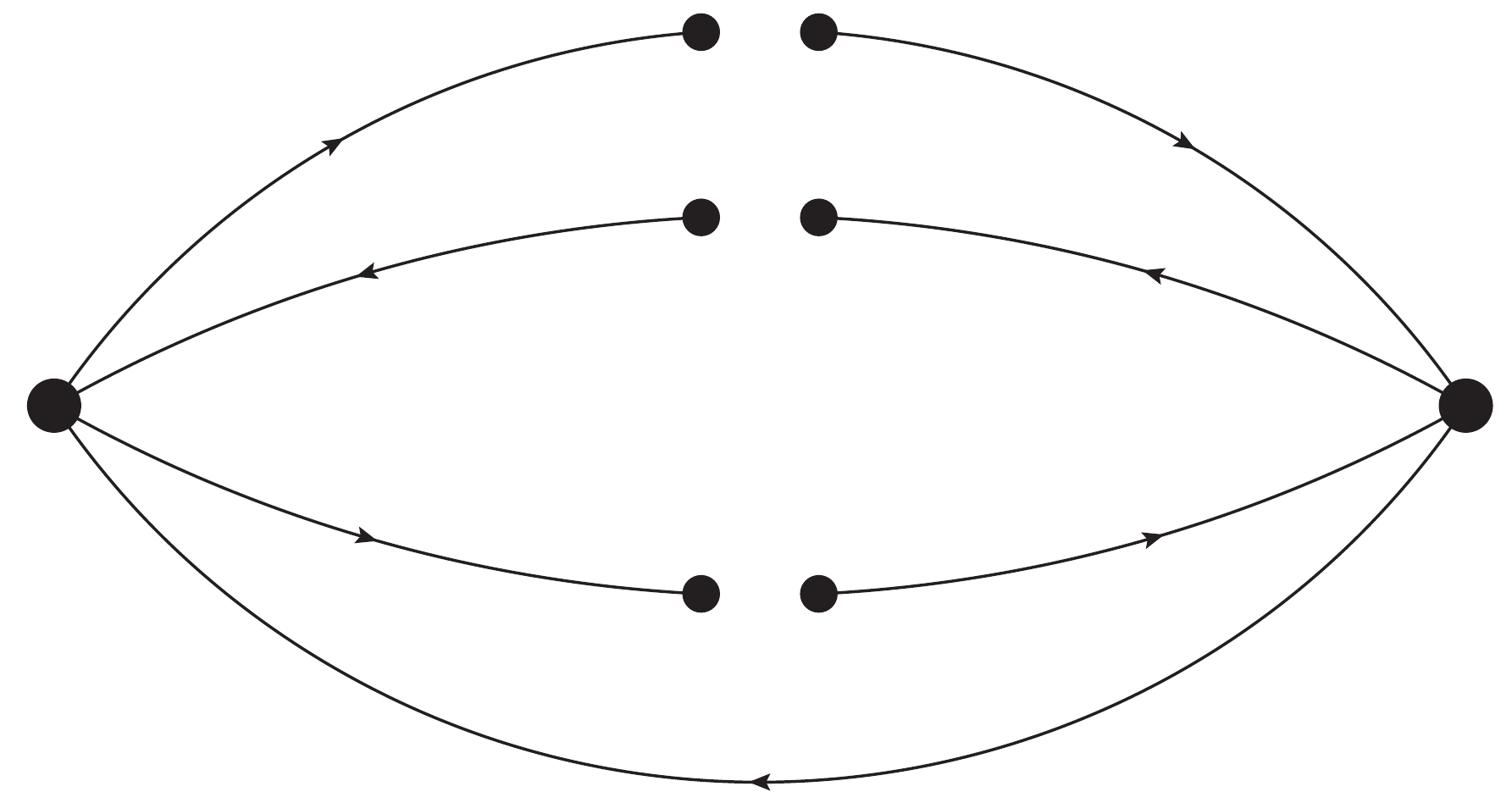}}}~~~~~
\subfigure[($c{\rm-}4$)]{
\scalebox{0.15}{\includegraphics{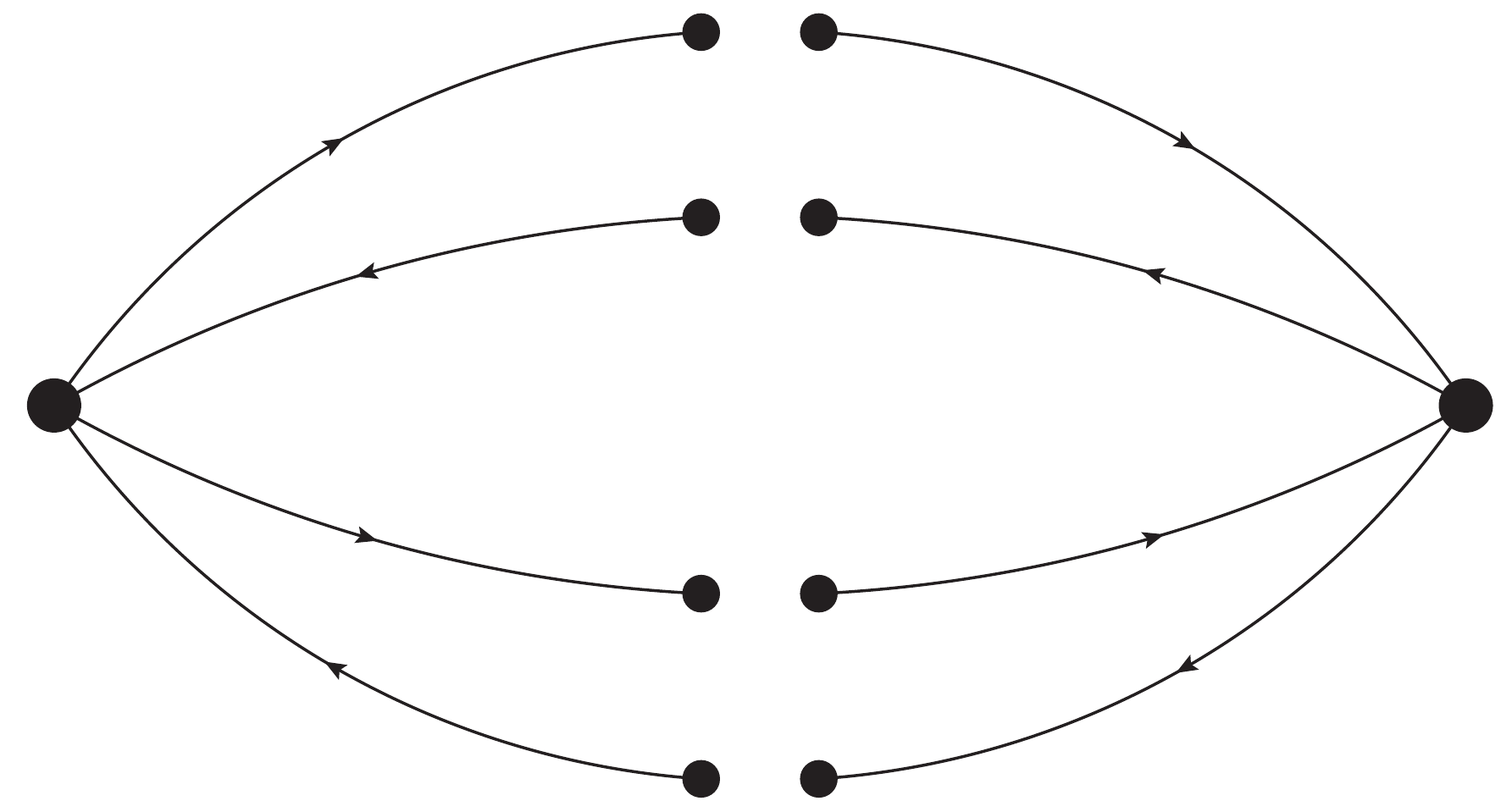}}}
\\[3mm]
\subfigure[($d{\rm-}1$)]{
\scalebox{0.15}{\includegraphics{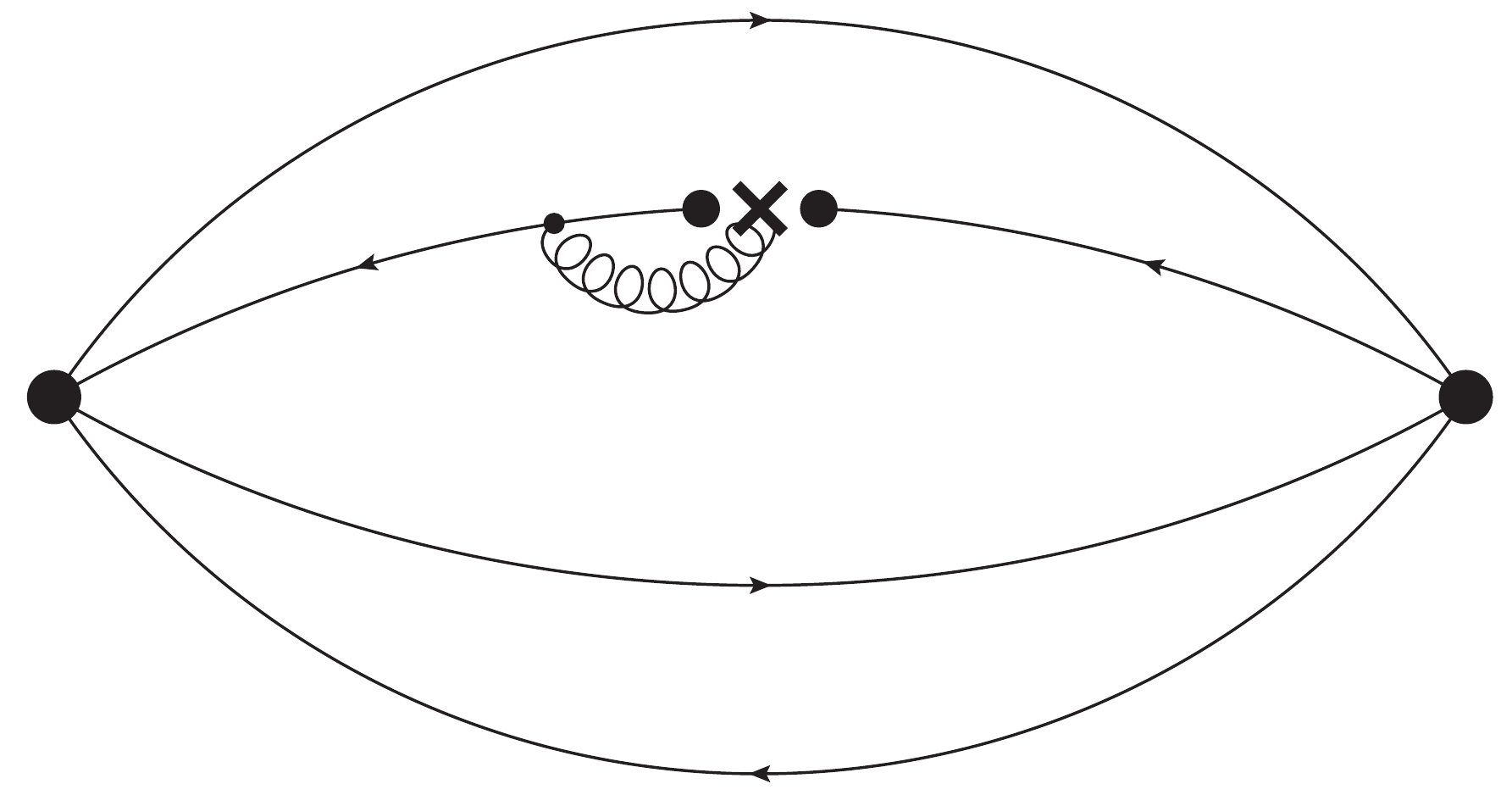}}}~~~~~
\subfigure[($d{\rm-}2$)]{
\scalebox{0.15}{\includegraphics{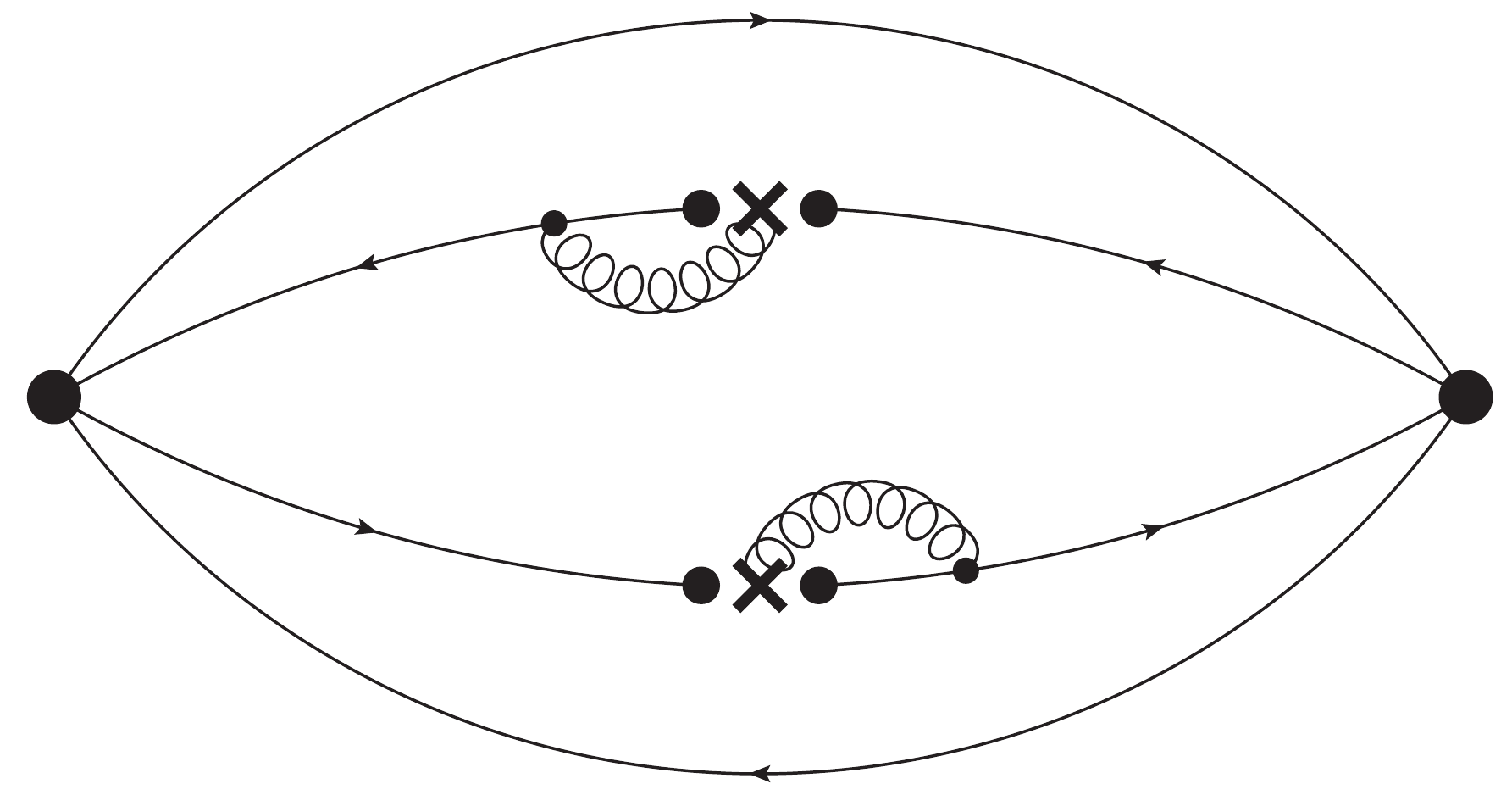}}}~~~~~
\subfigure[($d{\rm-}3$)]{
\scalebox{0.15}{\includegraphics{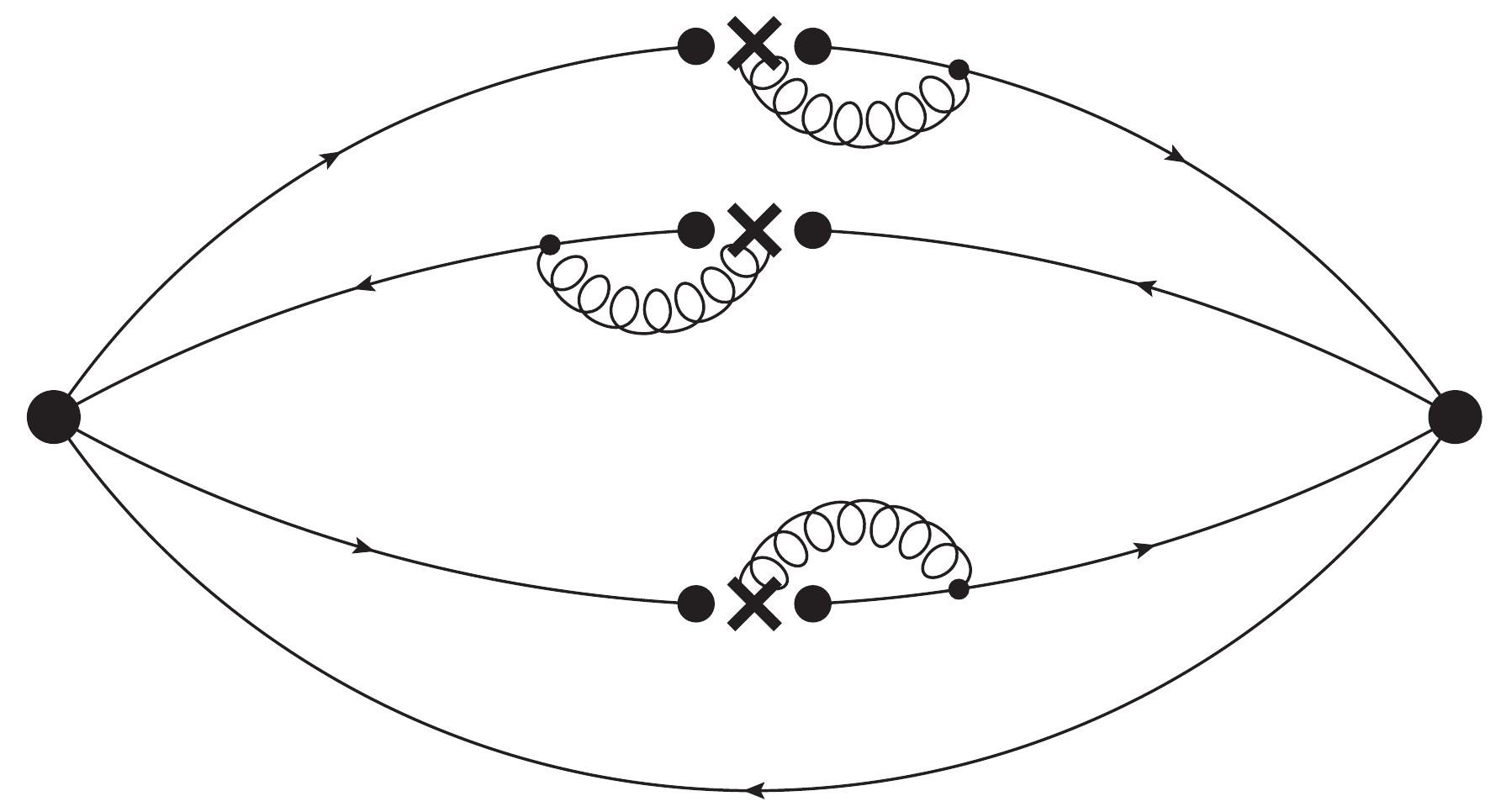}}}~~~~~
\subfigure[($d{\rm-}4$)]{
\scalebox{0.15}{\includegraphics{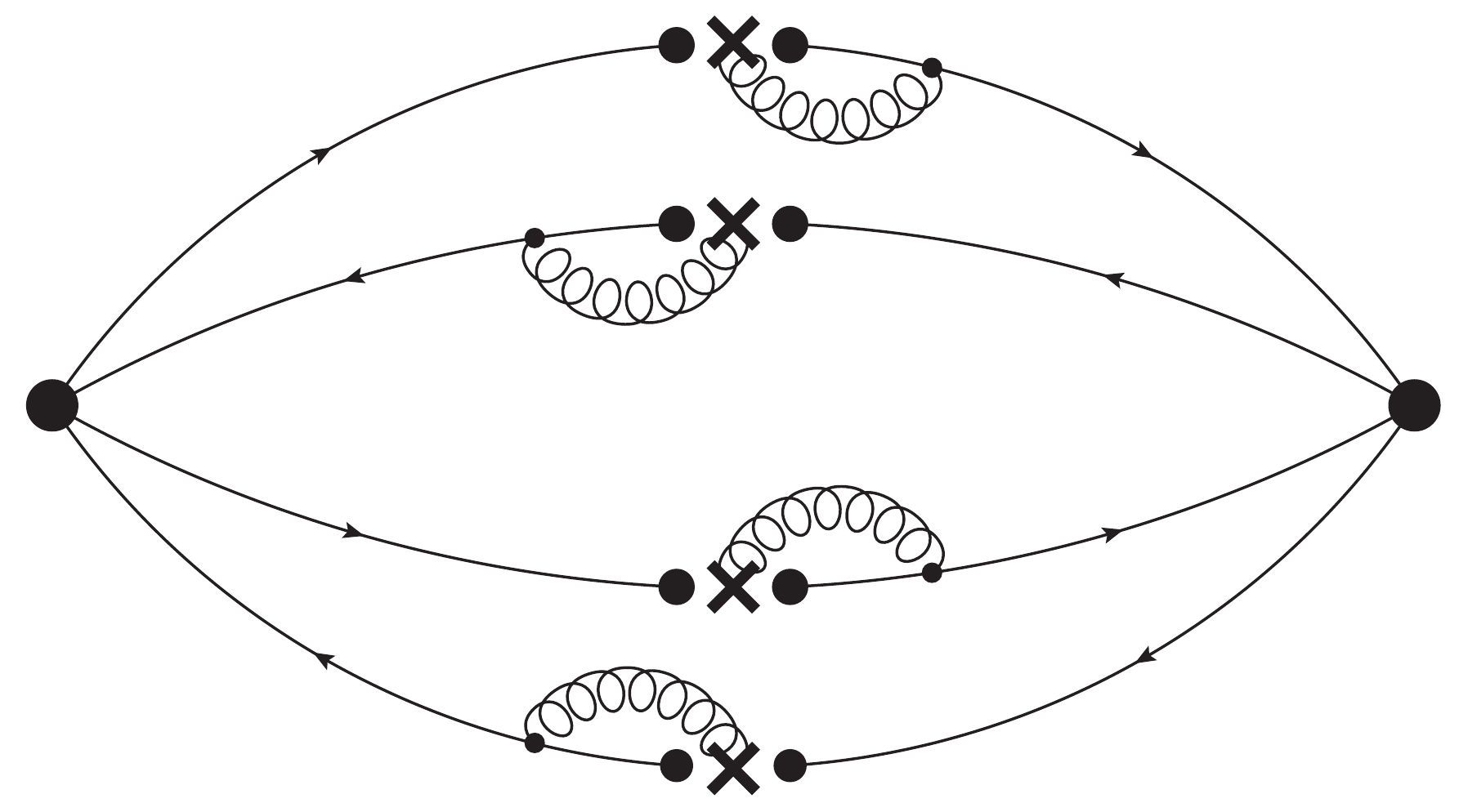}}}
\\
\subfigure[($d{\rm-}5$)]{
\scalebox{0.15}{\includegraphics{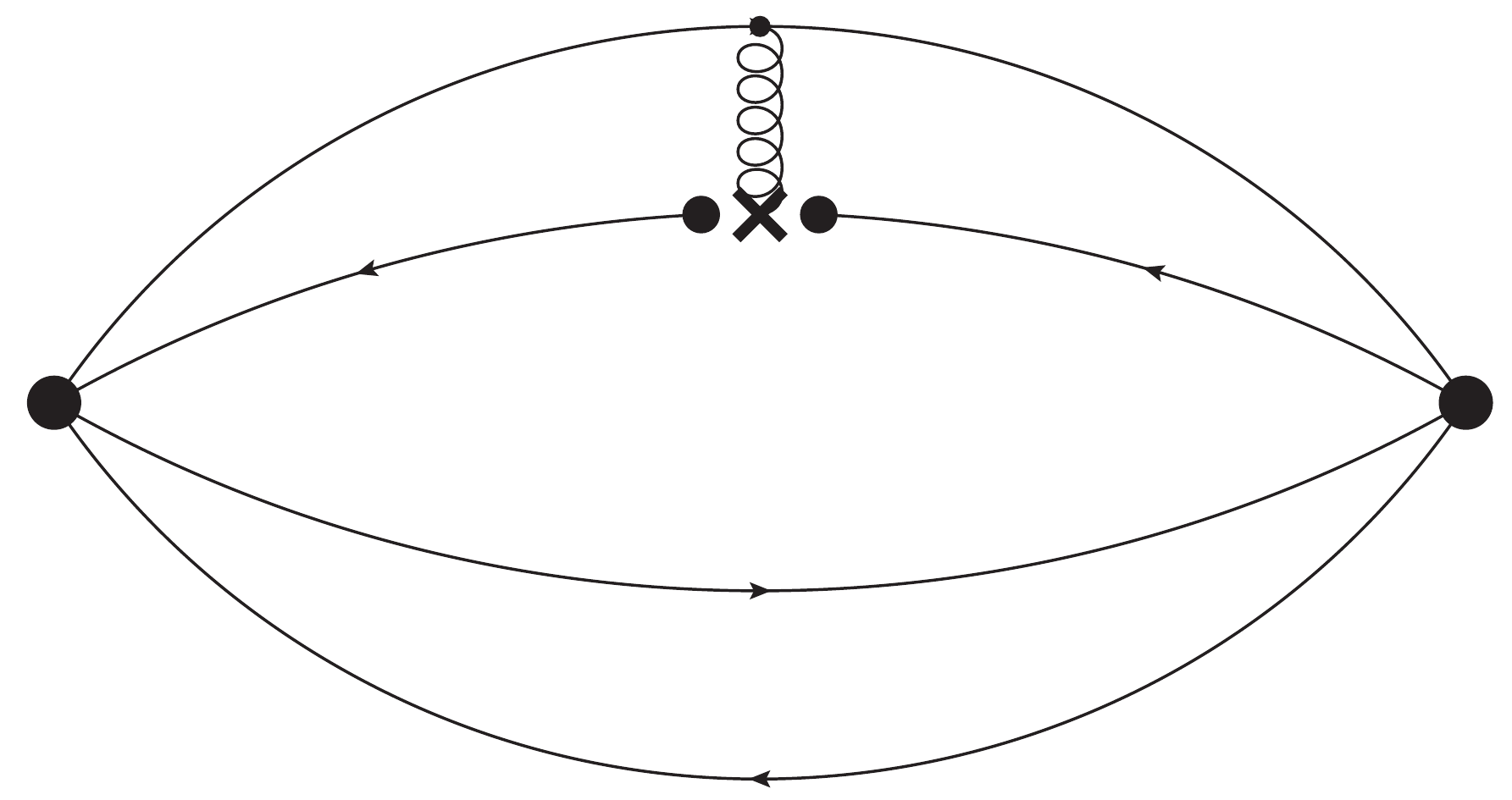}}}~~~~~
\subfigure[($d{\rm-}6$)]{
\scalebox{0.15}{\includegraphics{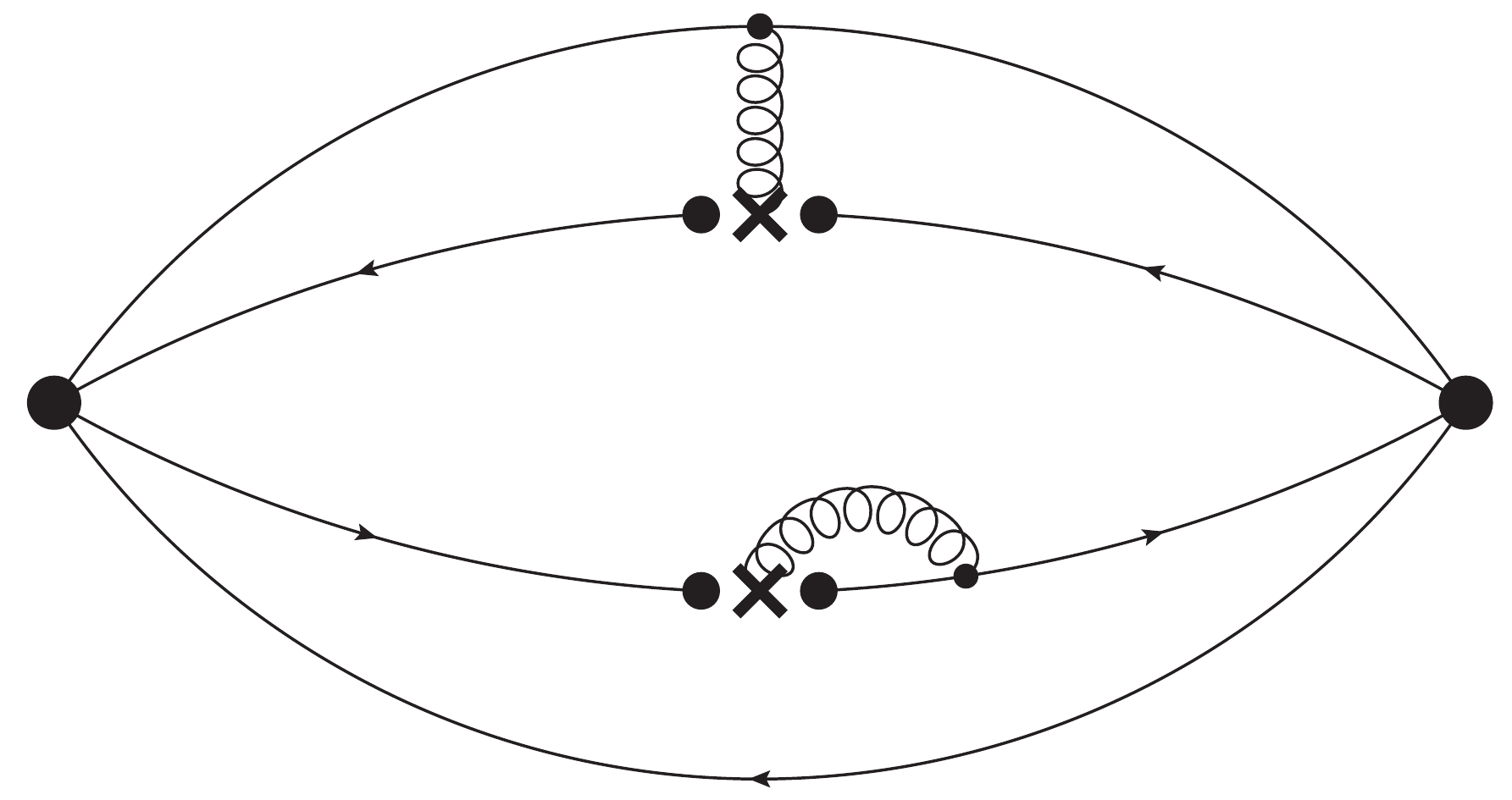}}}~~~~~
\subfigure[($d{\rm-}7$)]{
\scalebox{0.15}{\includegraphics{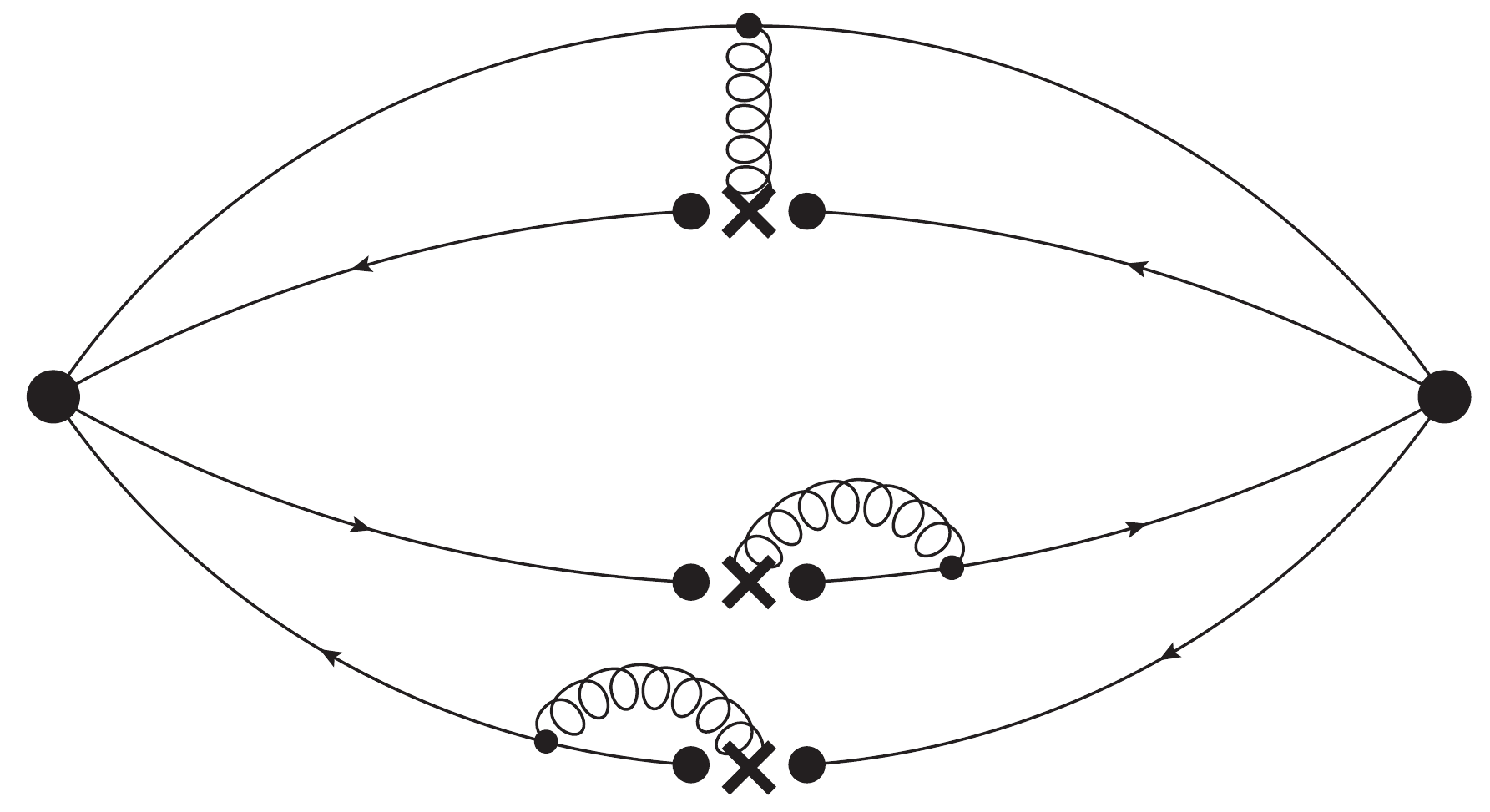}}}
\\[3mm]
\subfigure[($e{\rm-}1$)]{
\scalebox{0.15}{\includegraphics{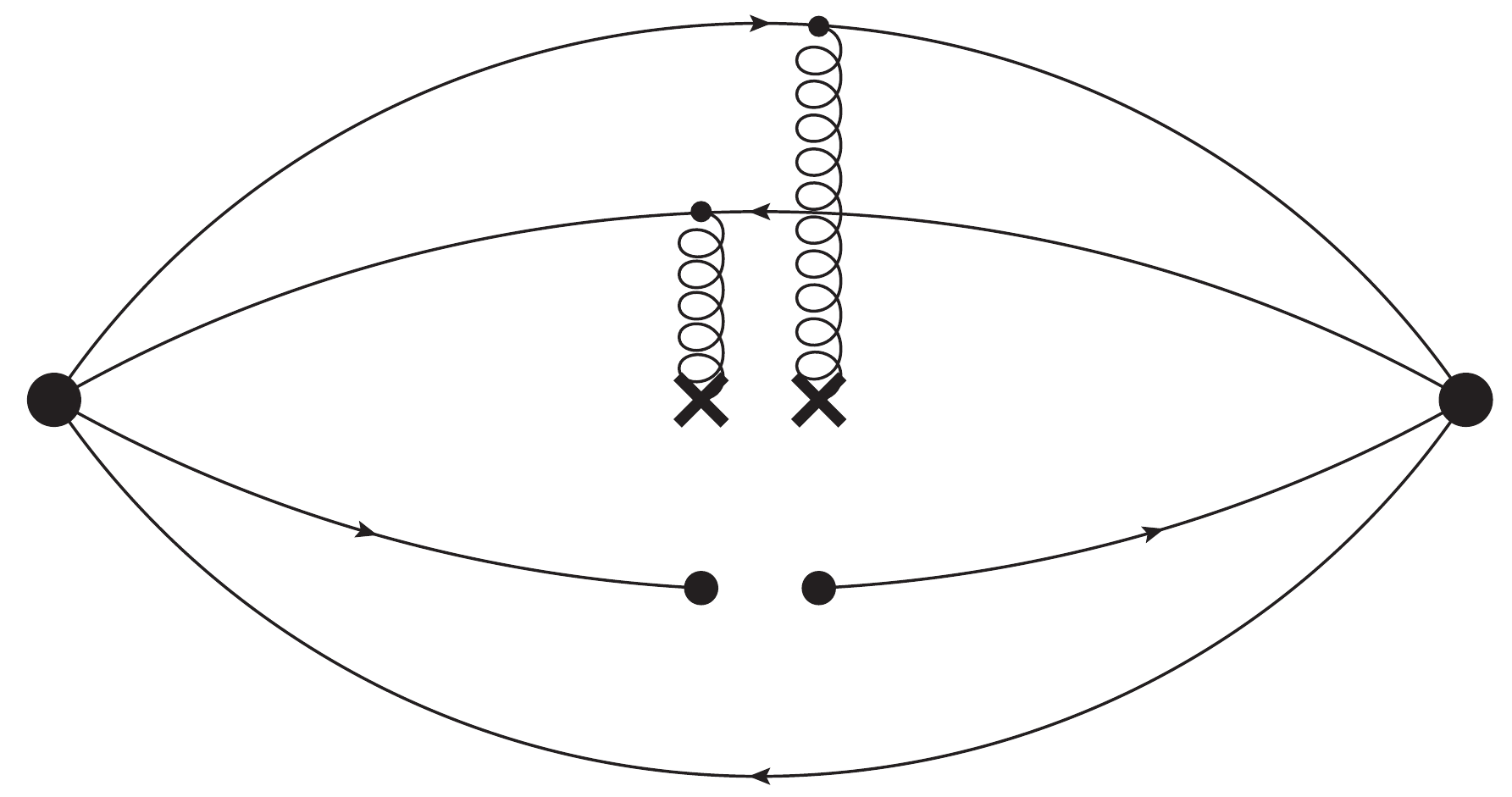}}}~~~~~
\subfigure[($e{\rm-}2$)]{
\scalebox{0.15}{\includegraphics{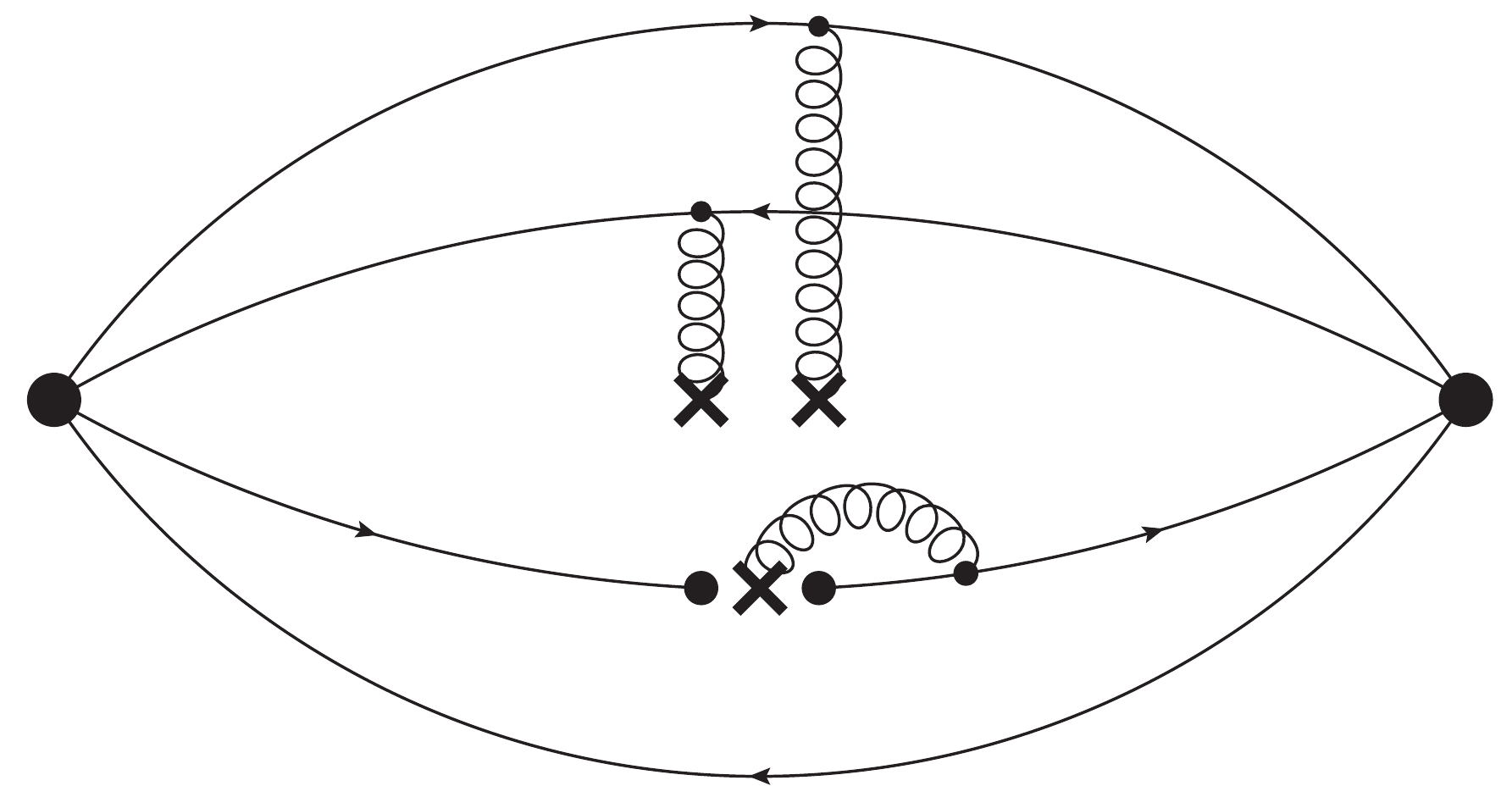}}}~~~~~
\subfigure[($e{\rm-}3$)]{
\scalebox{0.15}{\includegraphics{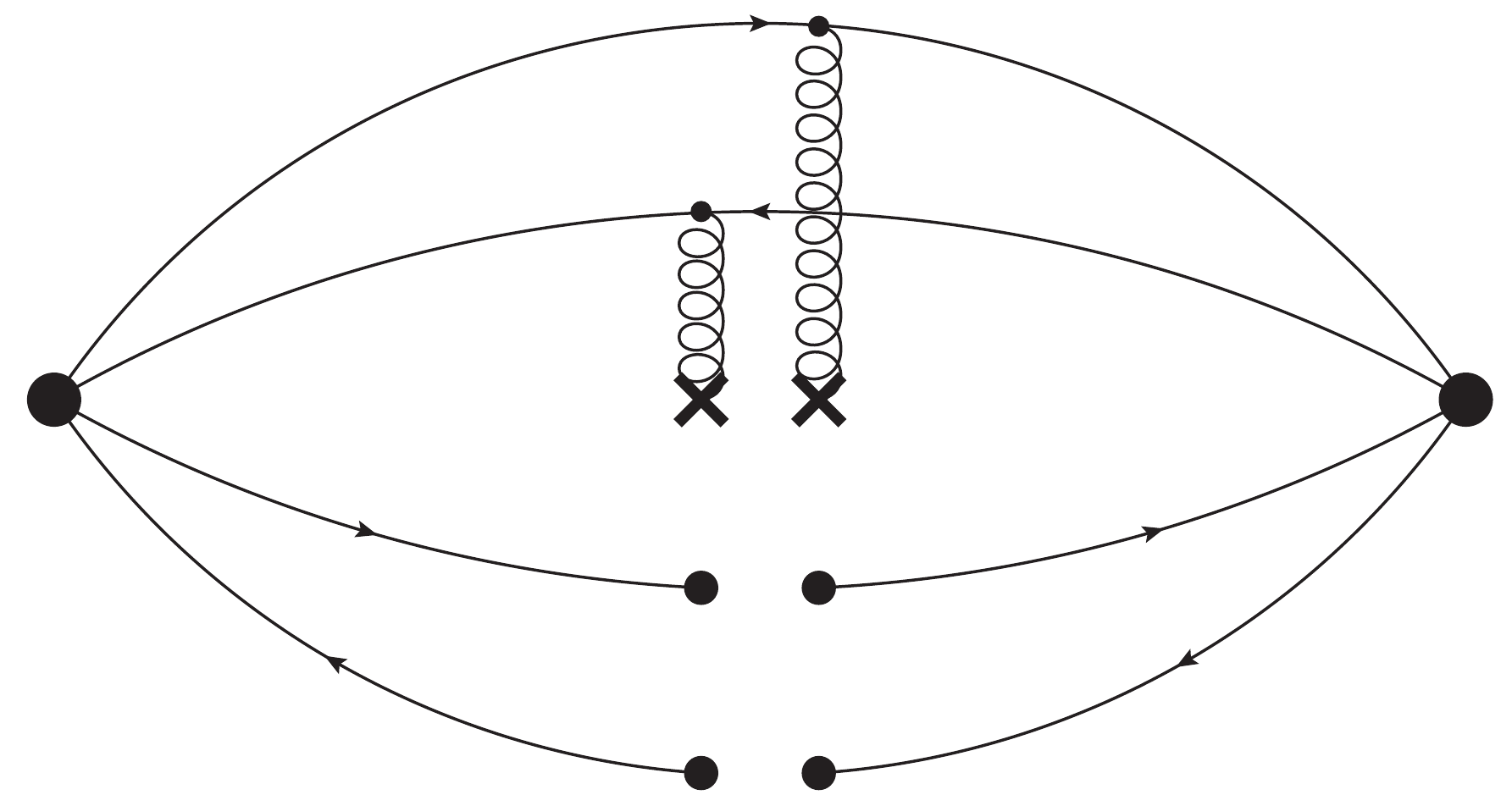}}}~~~~~
\subfigure[($e{\rm-}4$)]{
\scalebox{0.15}{\includegraphics{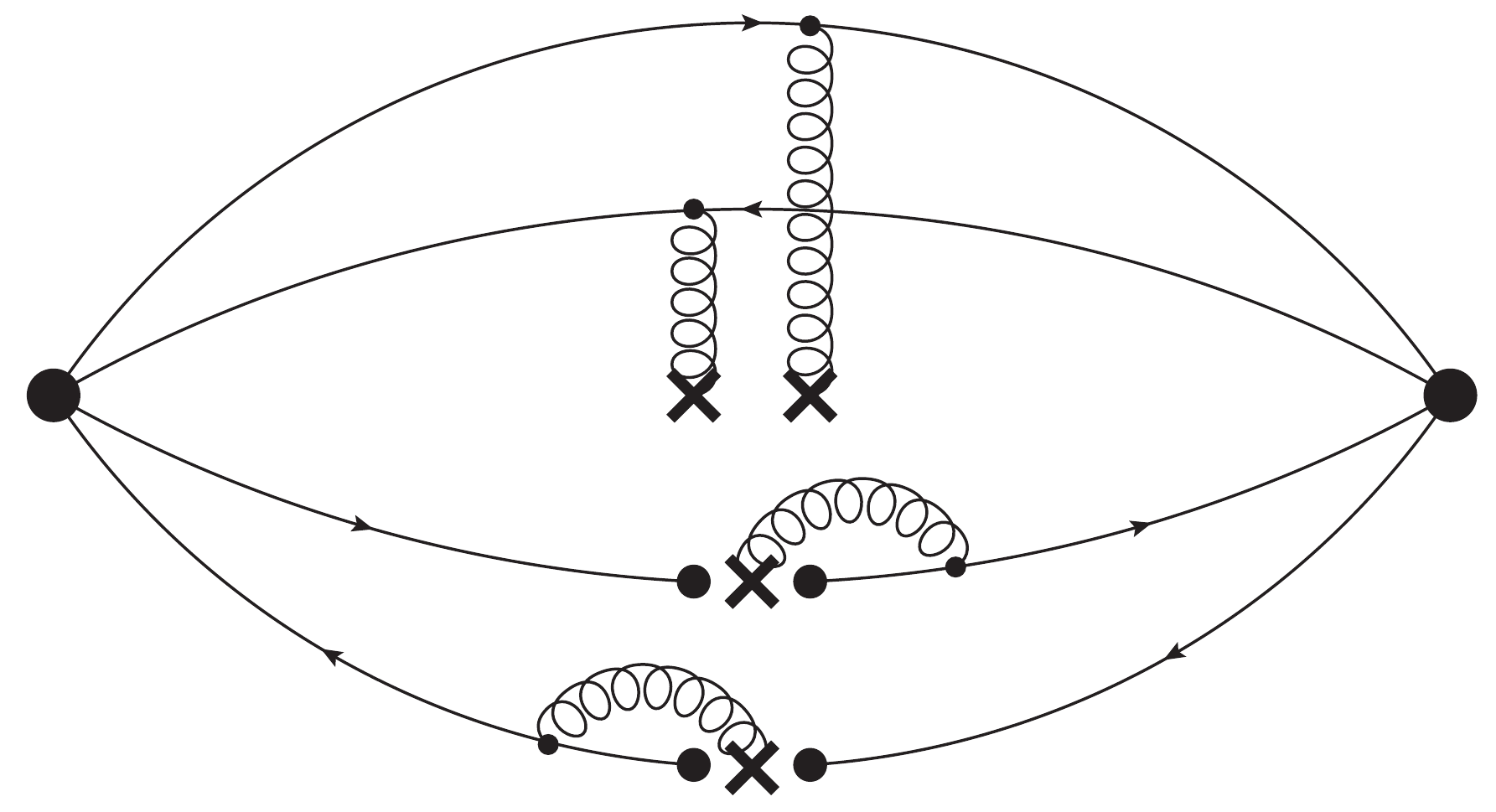}}}
\\
\subfigure[($e{\rm-}5$)]{
\scalebox{0.15}{\includegraphics{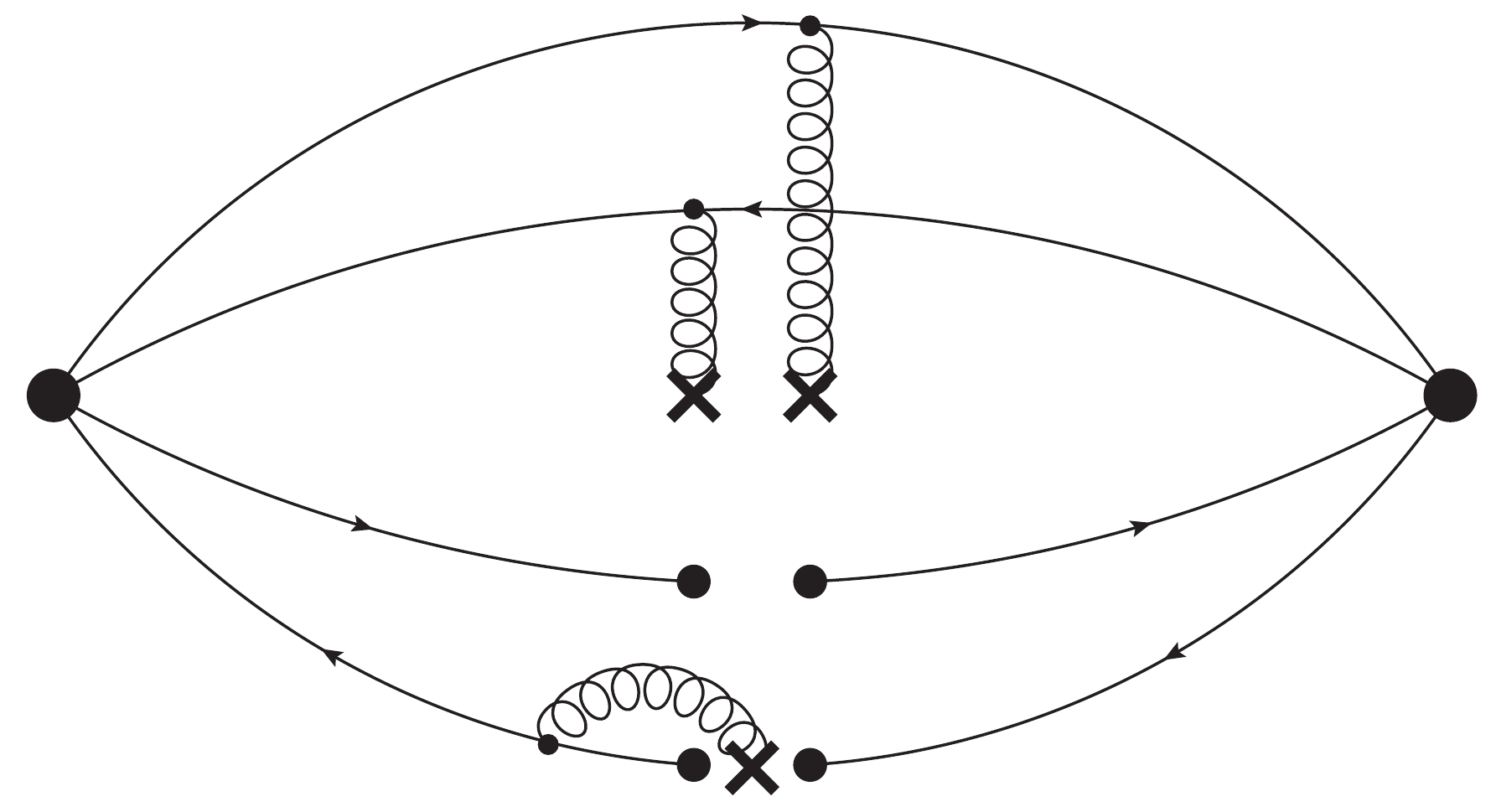}}}~~~~~
\subfigure[($e{\rm-}6$)]{
\scalebox{0.15}{\includegraphics{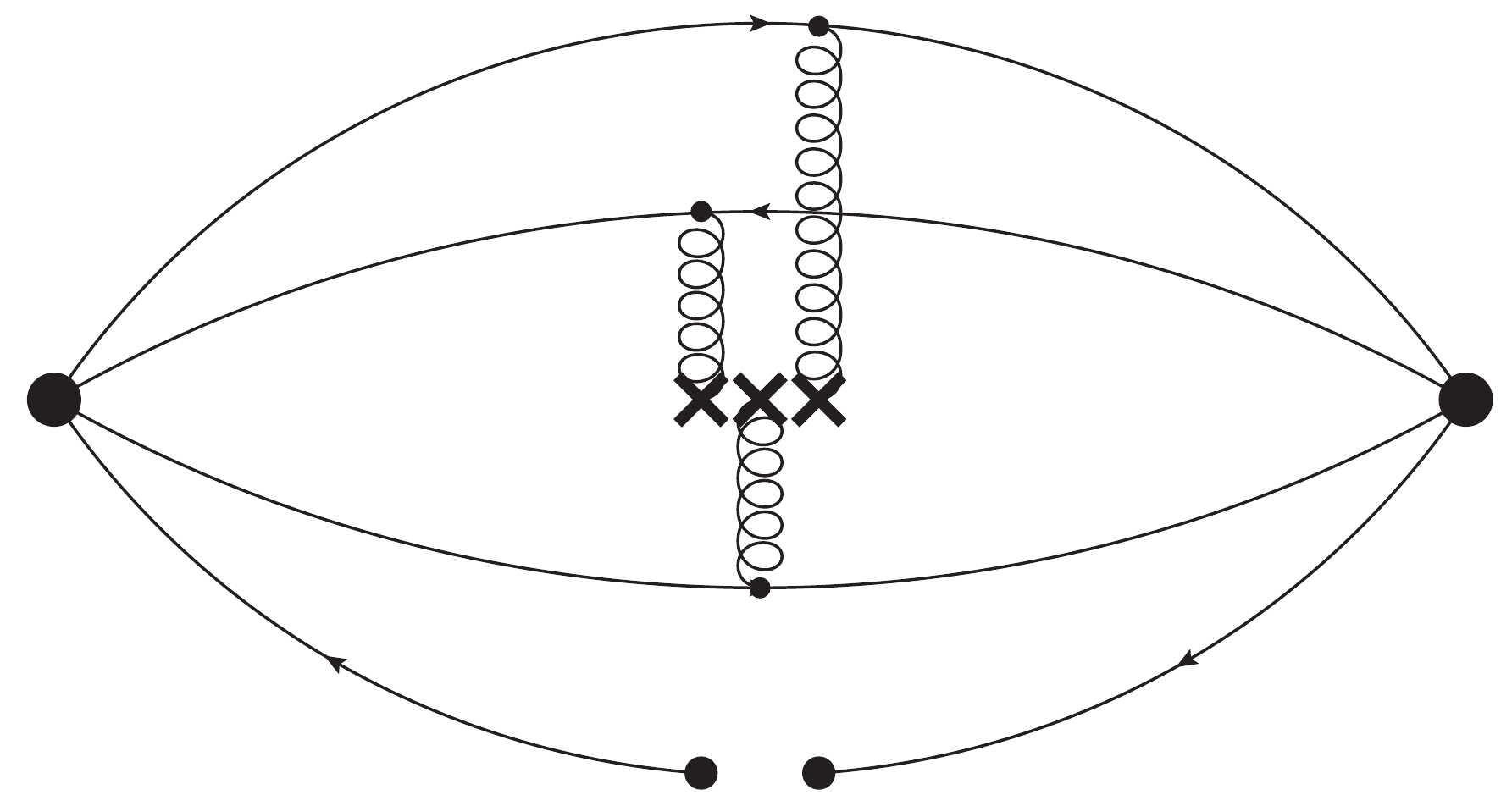}}}~~~~~
\subfigure[($e{\rm-}7$)]{
\scalebox{0.15}{\includegraphics{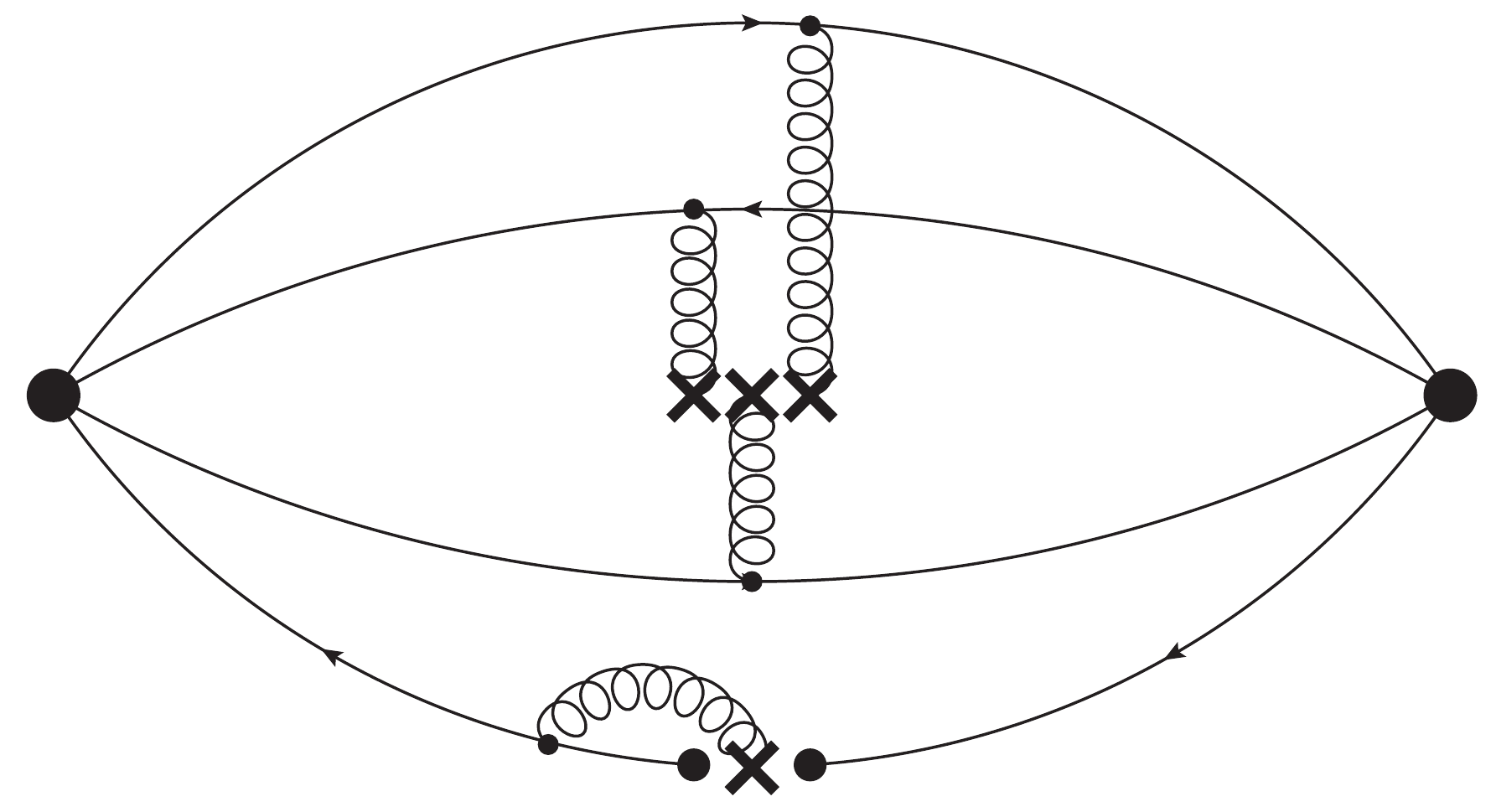}}}
\\[3mm]
\subfigure[($f{\rm-}1$)]{
\scalebox{0.15}{\includegraphics{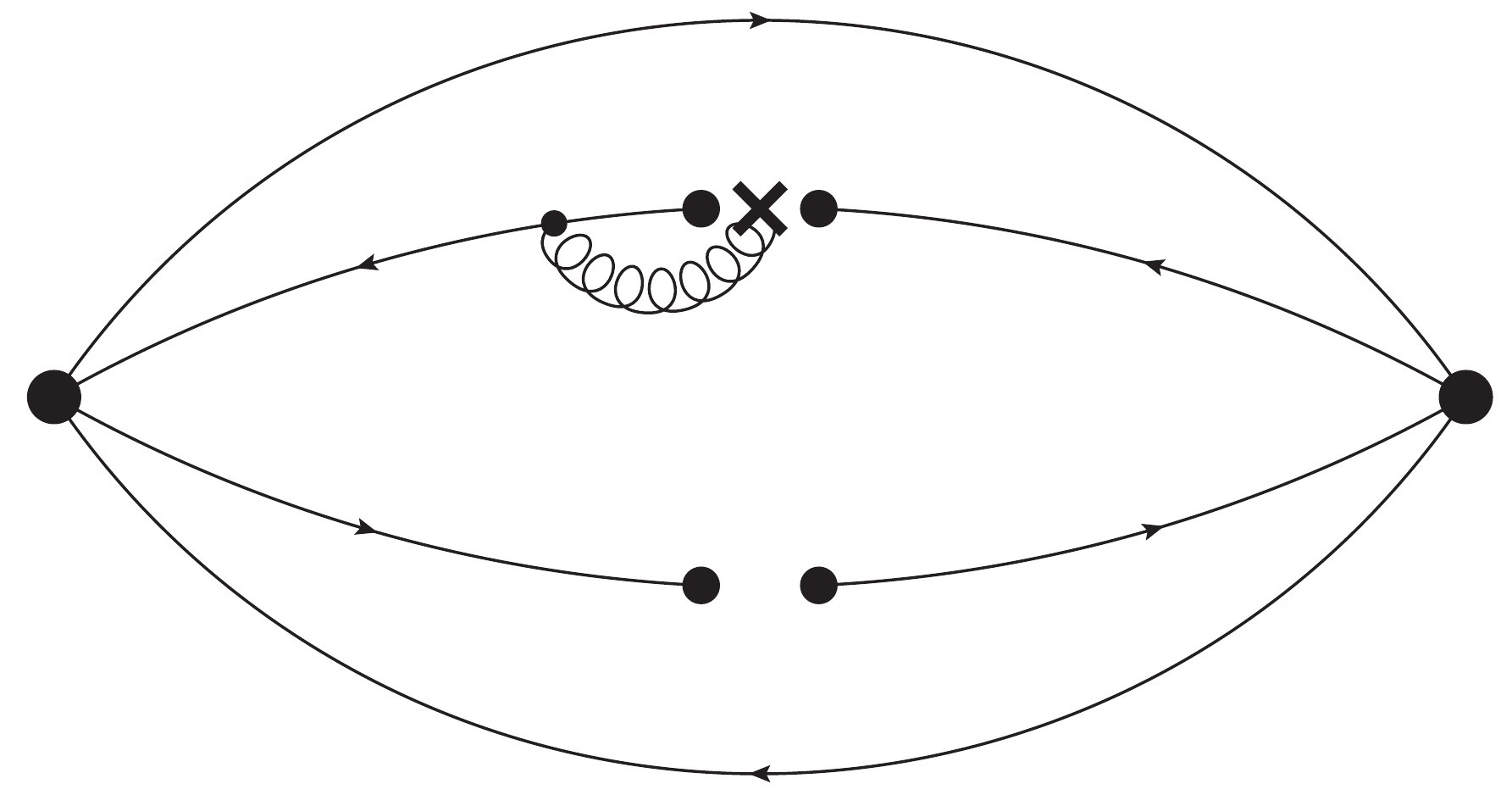}}}~~~~~
\subfigure[($f{\rm-}2$)]{
\scalebox{0.15}{\includegraphics{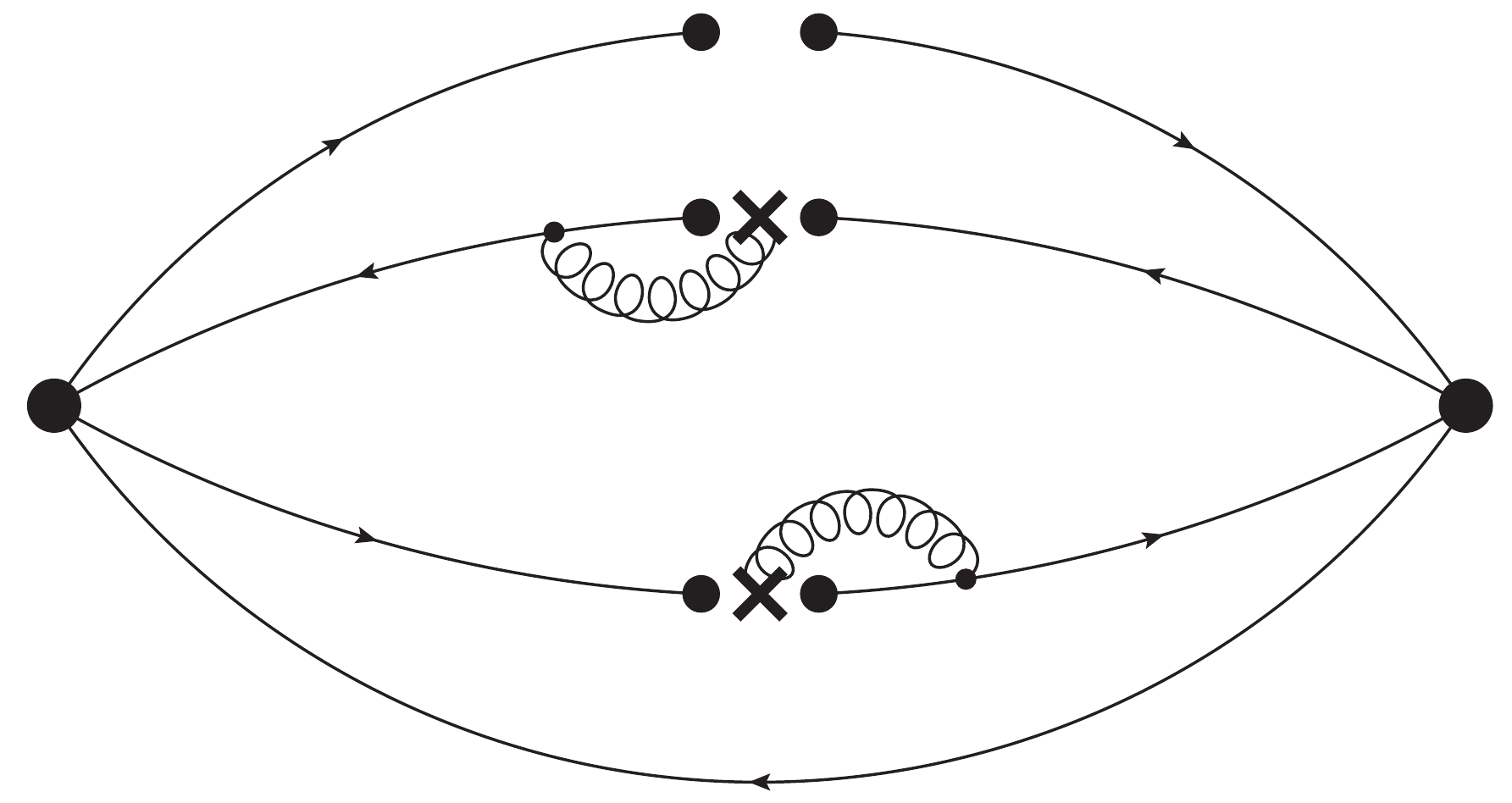}}}~~~~~
\subfigure[($f{\rm-}3$)]{
\scalebox{0.15}{\includegraphics{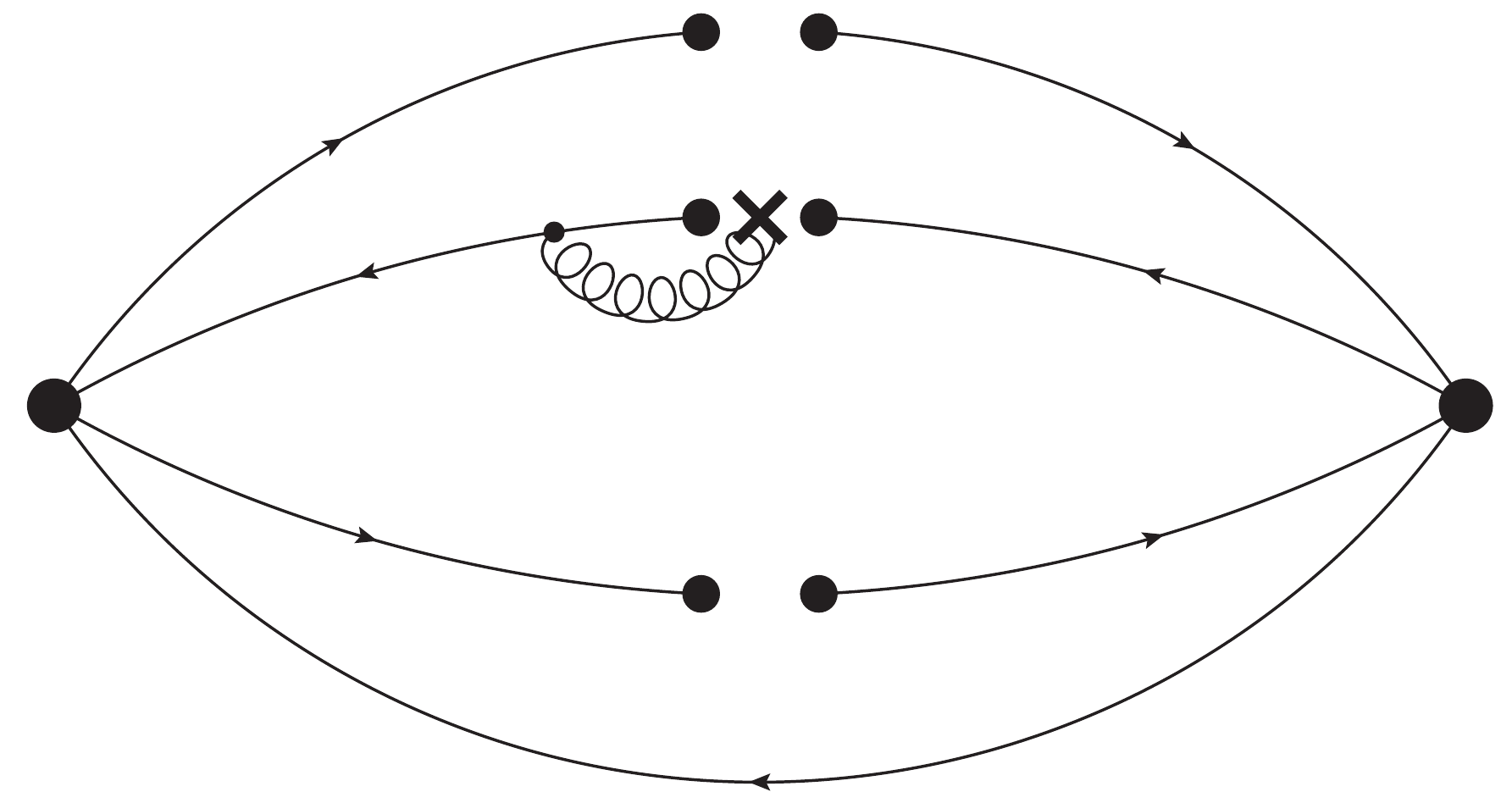}}}
\\
\subfigure[($f{\rm-}4$)]{
\scalebox{0.15}{\includegraphics{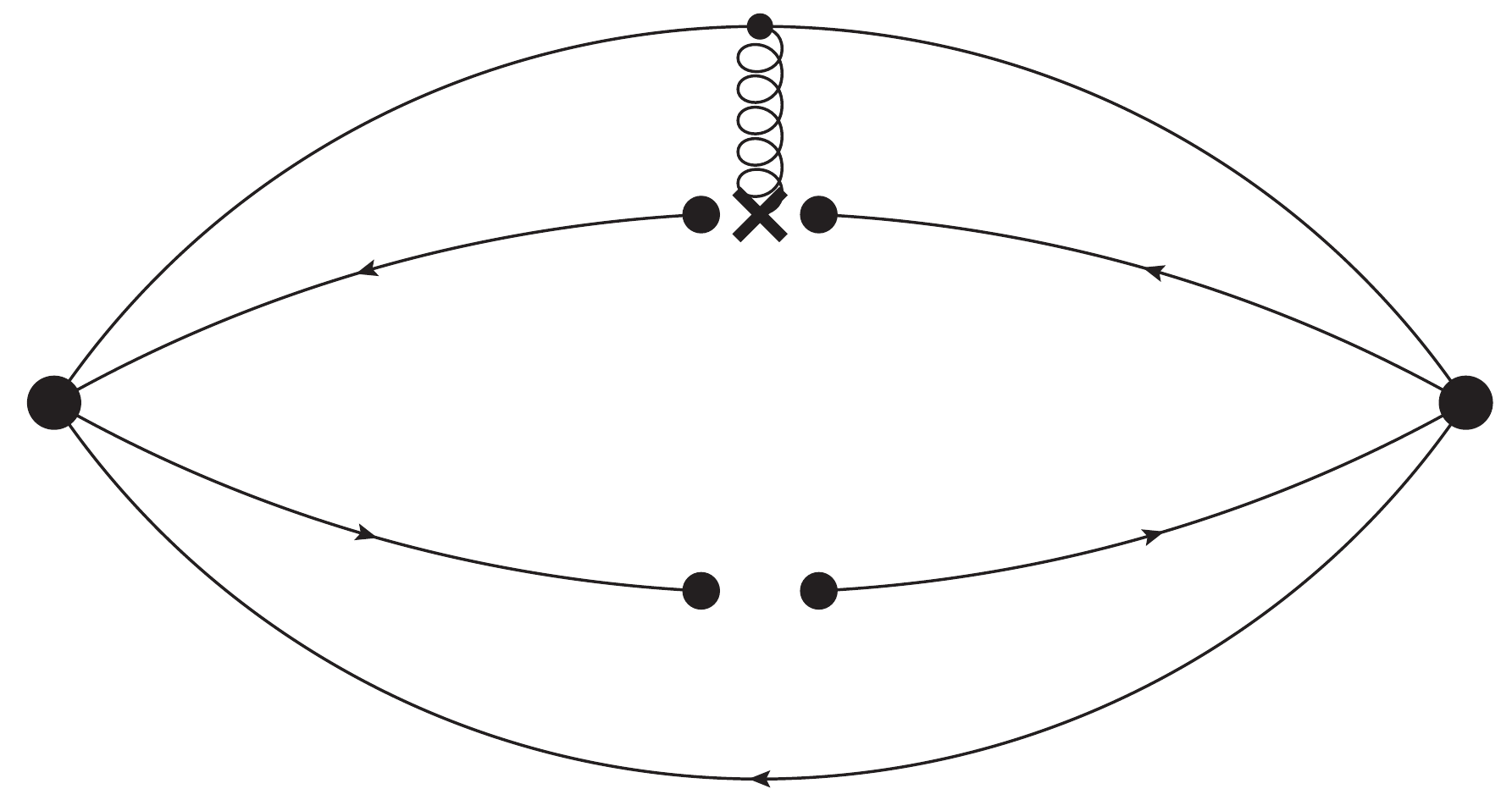}}}~~~~~
\subfigure[($f{\rm-}5$)]{
\scalebox{0.15}{\includegraphics{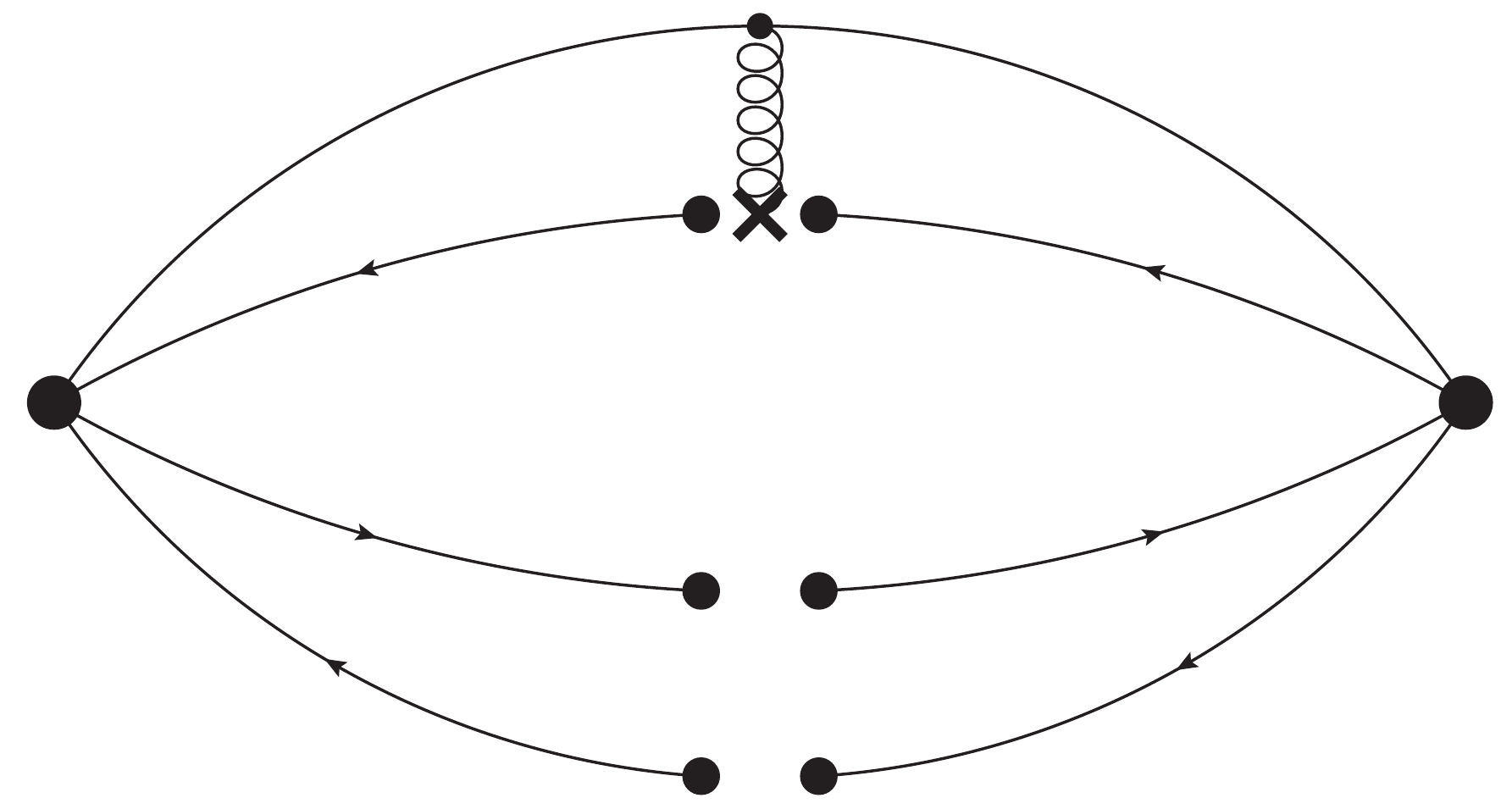}}}~~~~~
\subfigure[($f{\rm-}6$)]{
\scalebox{0.15}{\includegraphics{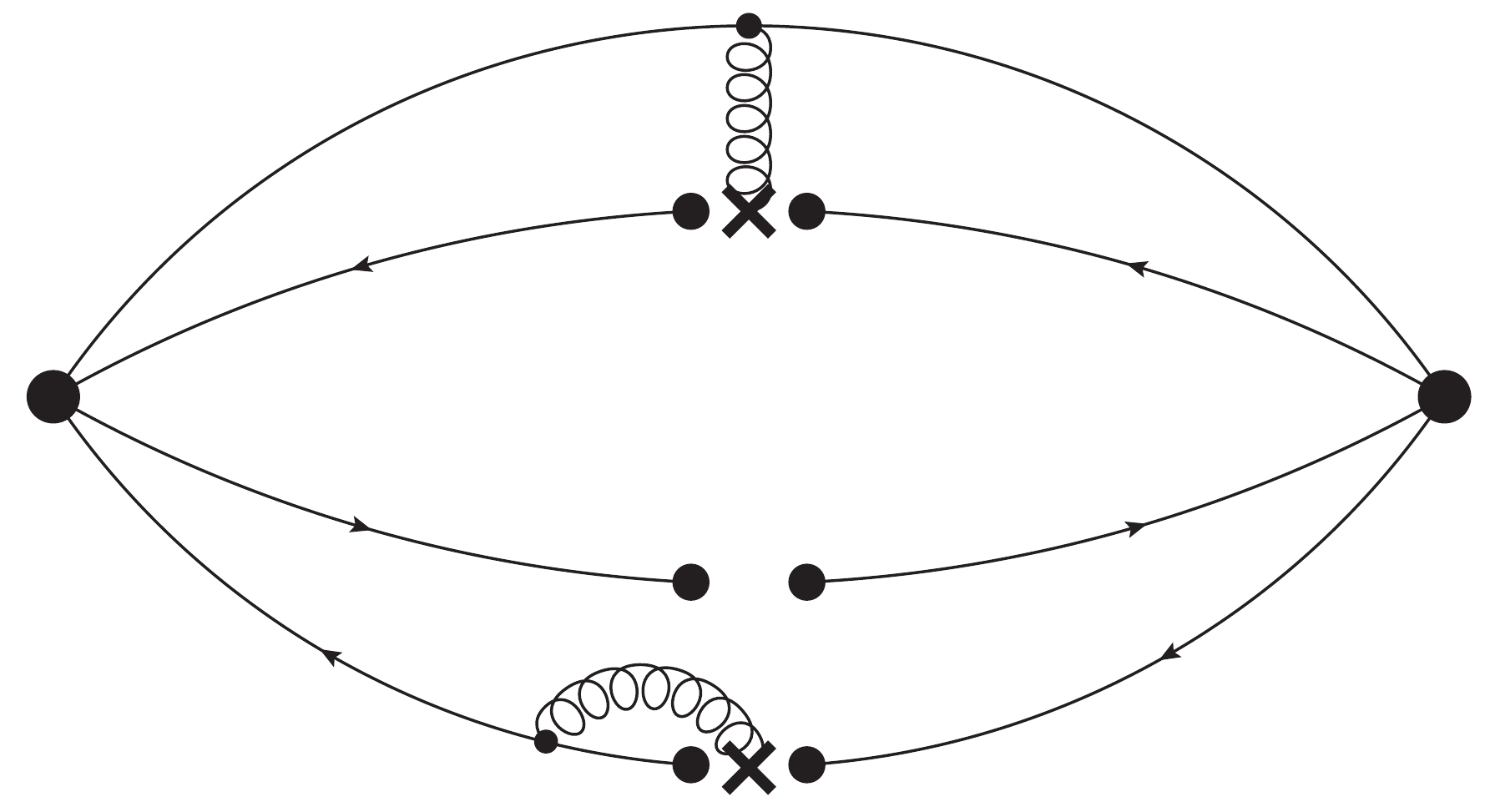}}}
\\[3mm]
$\Huge \bullet~\bullet~\bullet~~~\bullet~\bullet~\bullet$
\end{center}
\caption{Leading-order Feynman diagrams for tetraquark interpolating currents, including the perturbative term ($a$), gluon condensates ($b$), quark condensates ($c$), quark-gluon mixed condensates ($d$), and their combinations ($e$ and $f$).}
\label{fig:feynman1}
\end{figure*}

\begin{figure*}[hbtp]
\begin{center}
\subfigure[($g$)]{
\scalebox{0.15}{\includegraphics{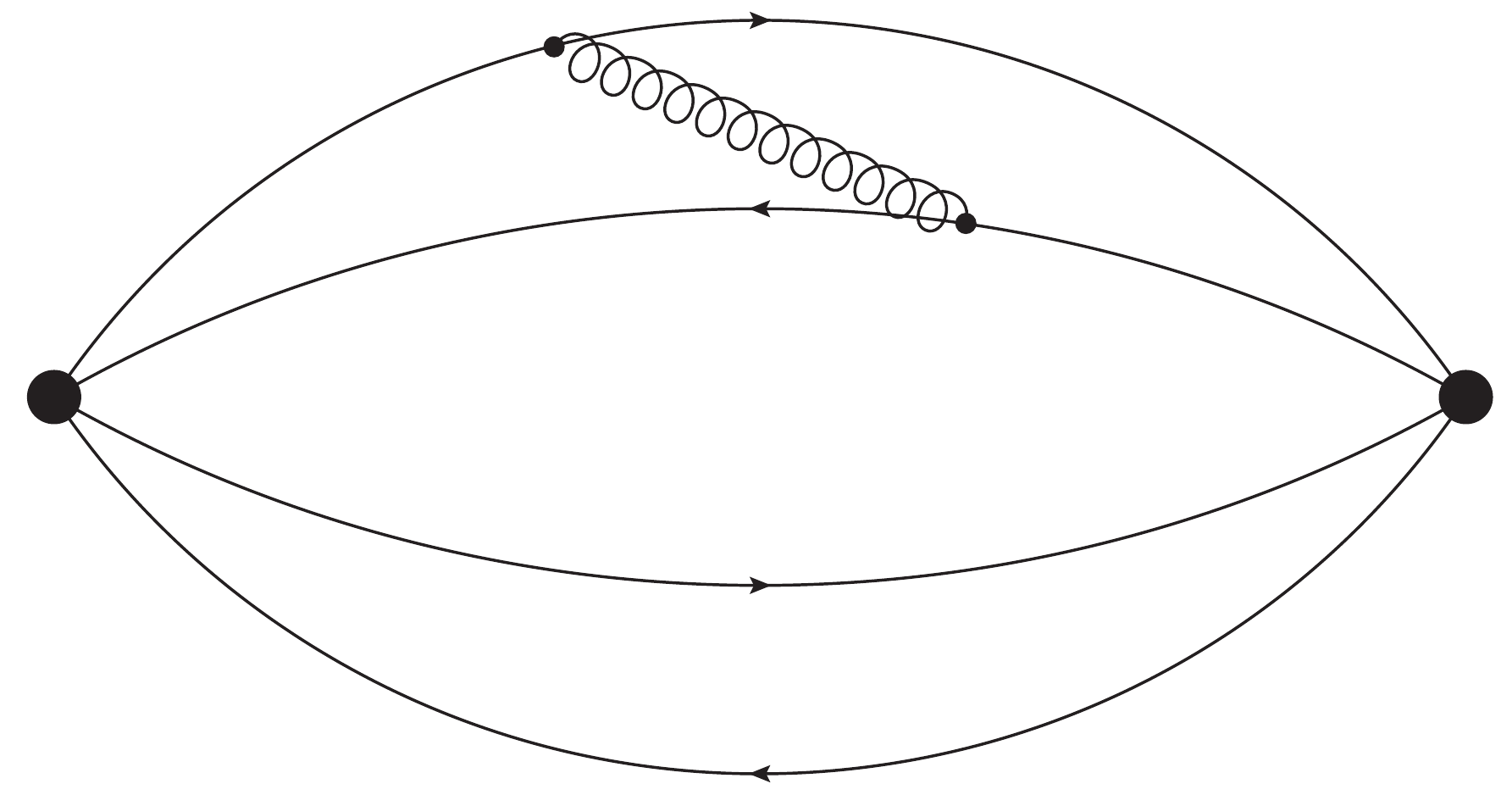}}}
\\[5mm]
\subfigure[($h{\rm-}1$)]{
\scalebox{0.15}{\includegraphics{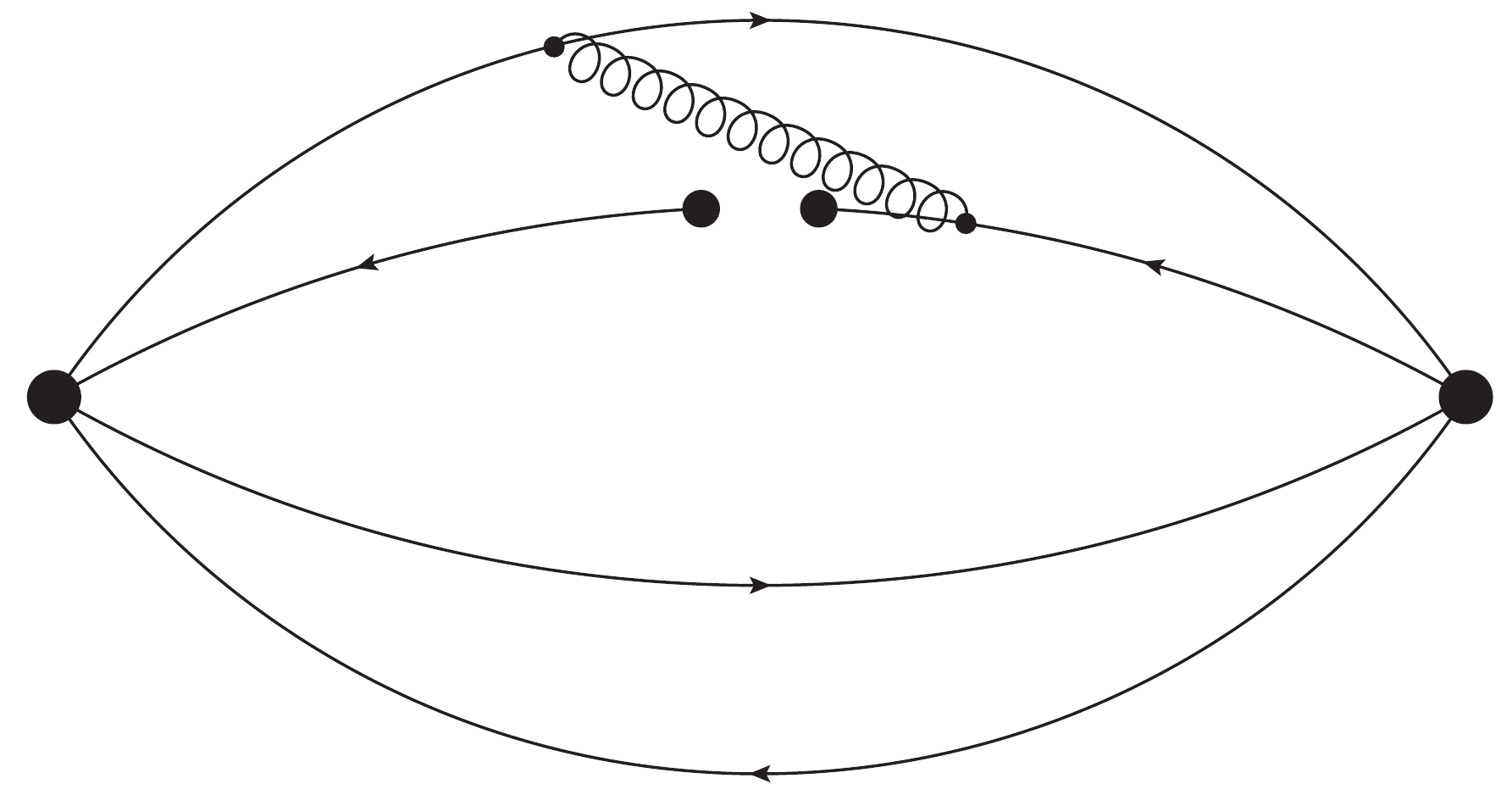}}}~~~~~
\subfigure[($h{\rm-}2$)]{
\scalebox{0.15}{\includegraphics{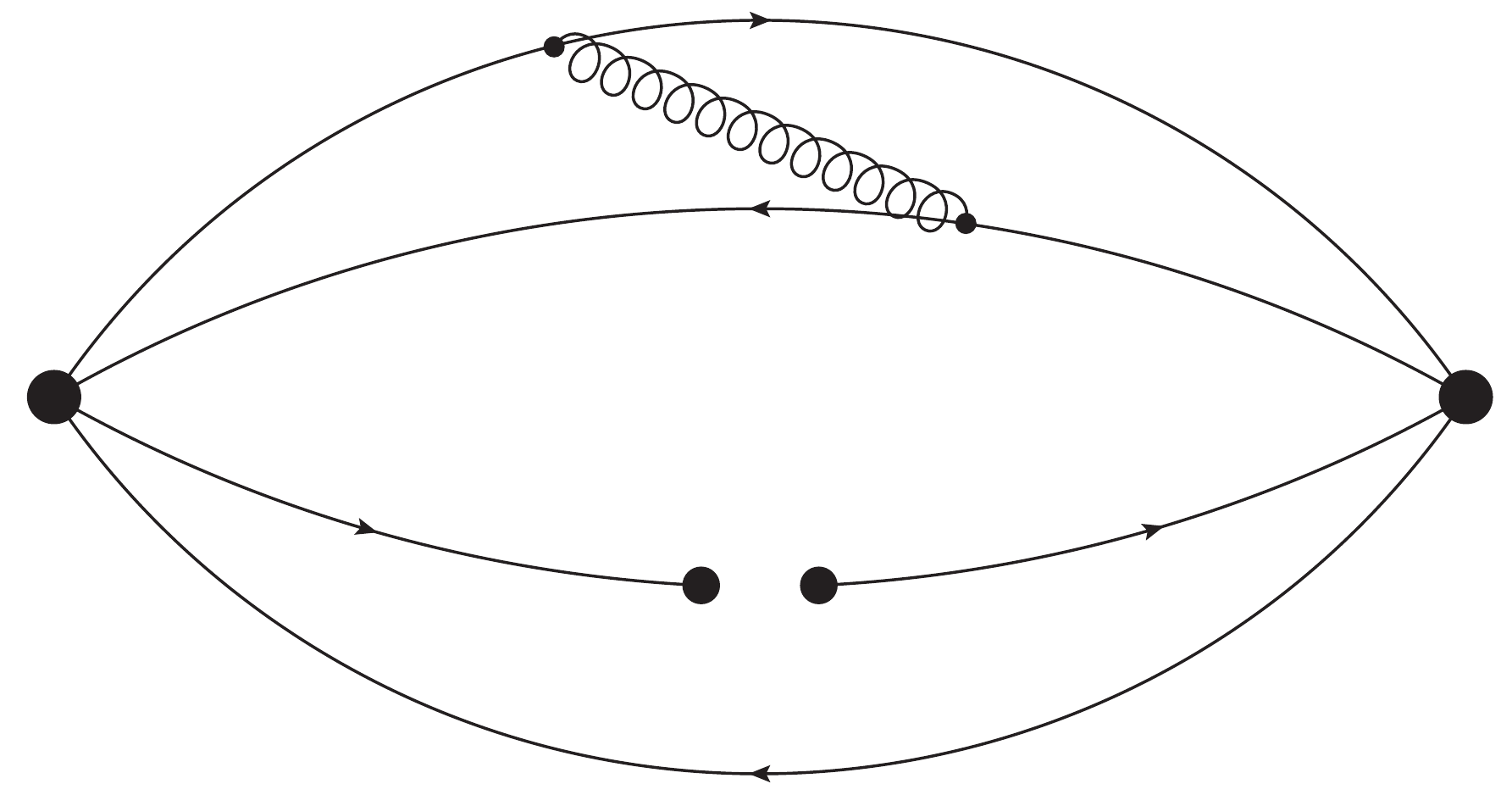}}}~~~~~
\subfigure[($h{\rm-}3$)]{
\scalebox{0.15}{\includegraphics{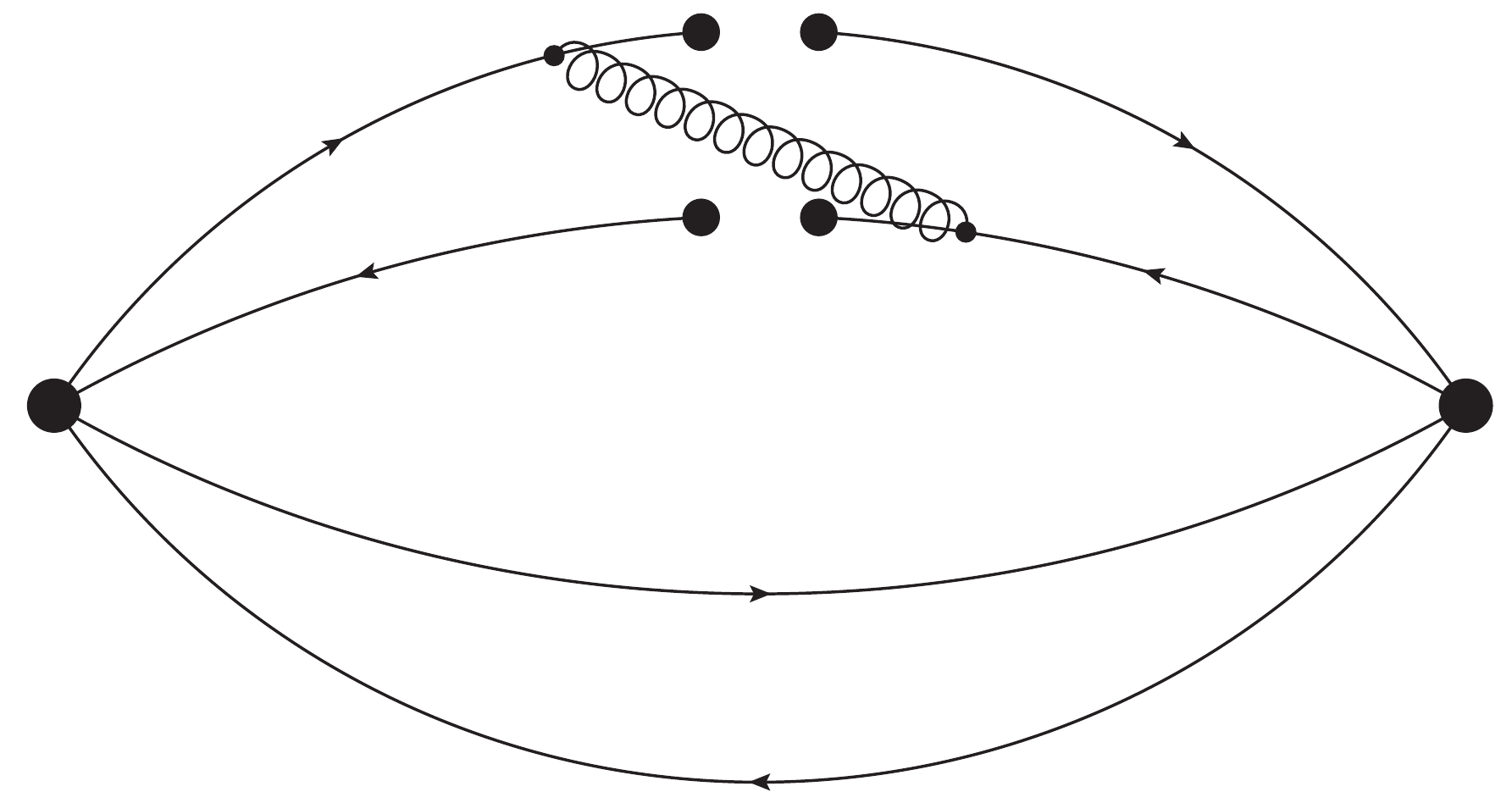}}}~~~~~
\subfigure[($h{\rm-}4$)]{
\scalebox{0.15}{\includegraphics{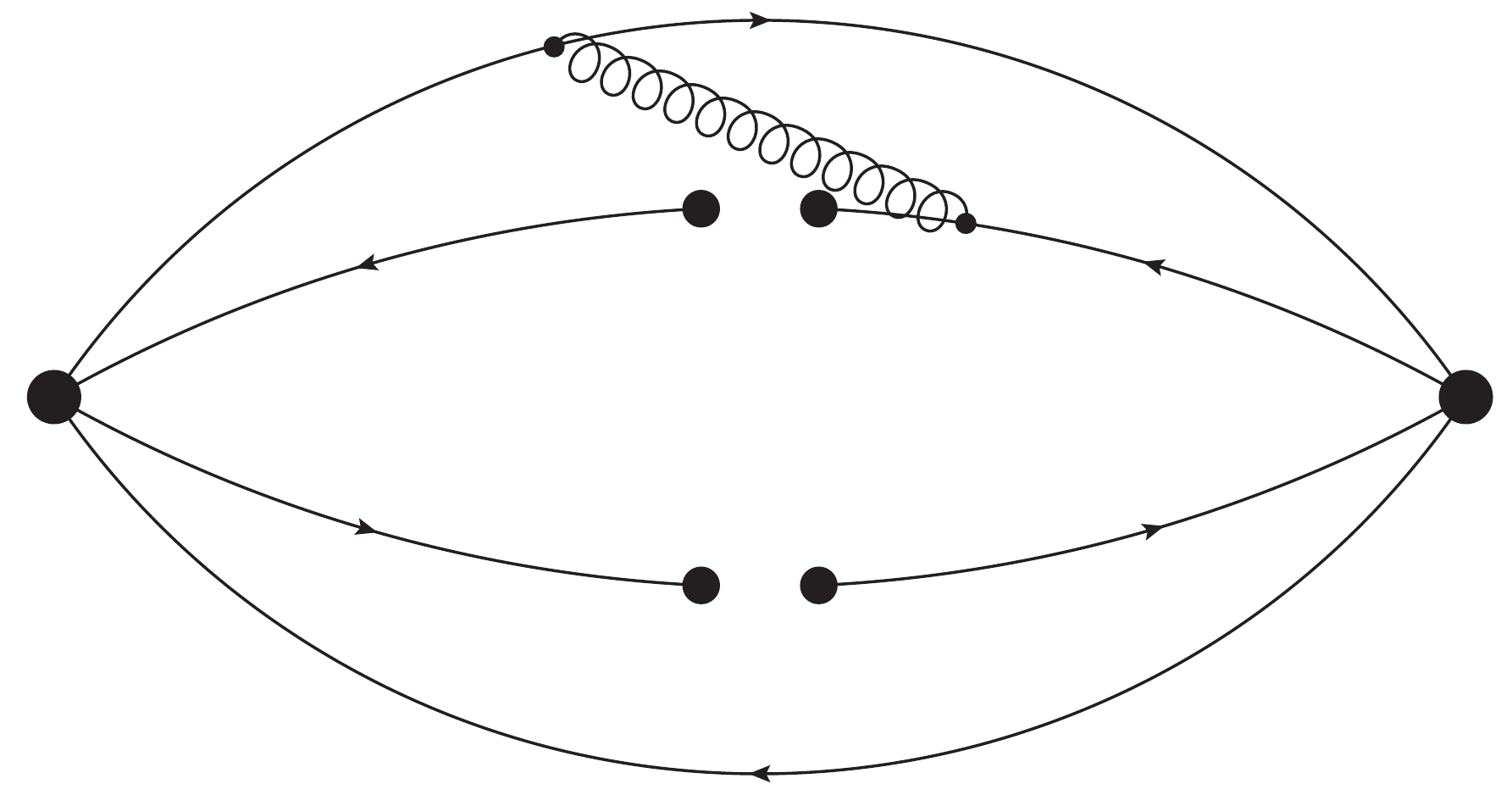}}}
\\
\subfigure[($h{\rm-}5$)]{
\scalebox{0.15}{\includegraphics{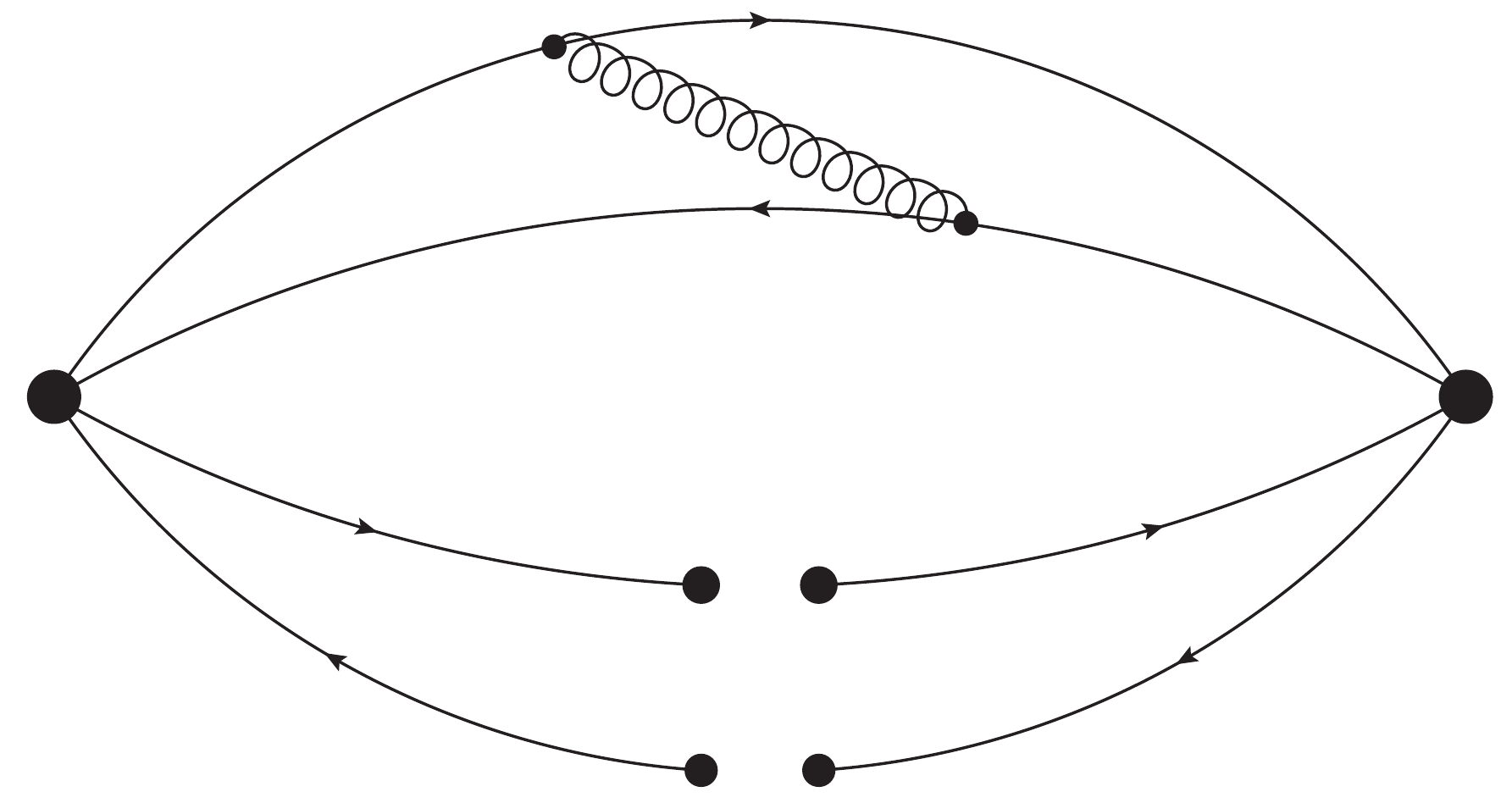}}}~~~~~
\subfigure[($h{\rm-}6$)]{
\scalebox{0.15}{\includegraphics{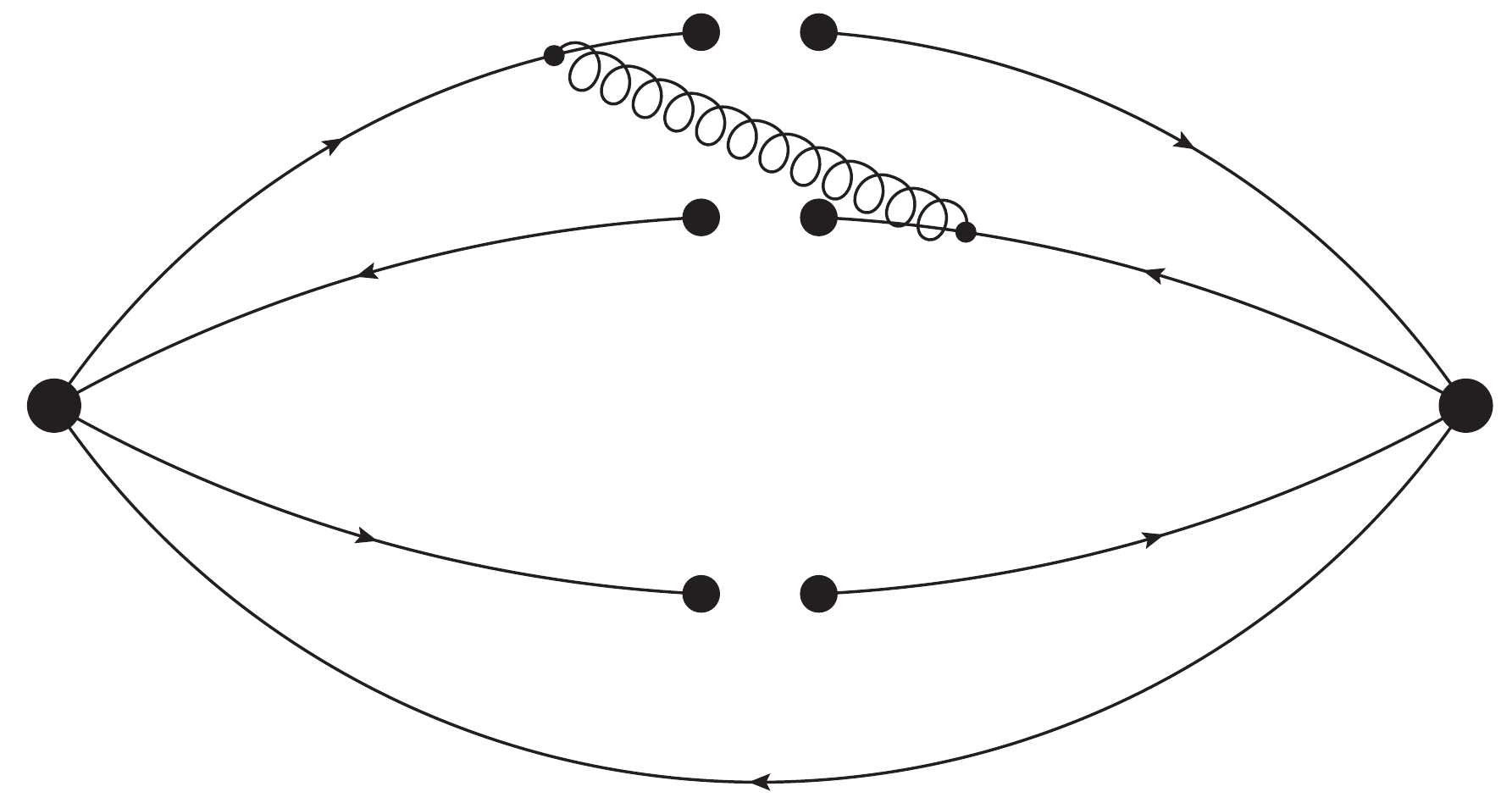}}}~~~~~
\subfigure[($h{\rm-}7$)]{
\scalebox{0.15}{\includegraphics{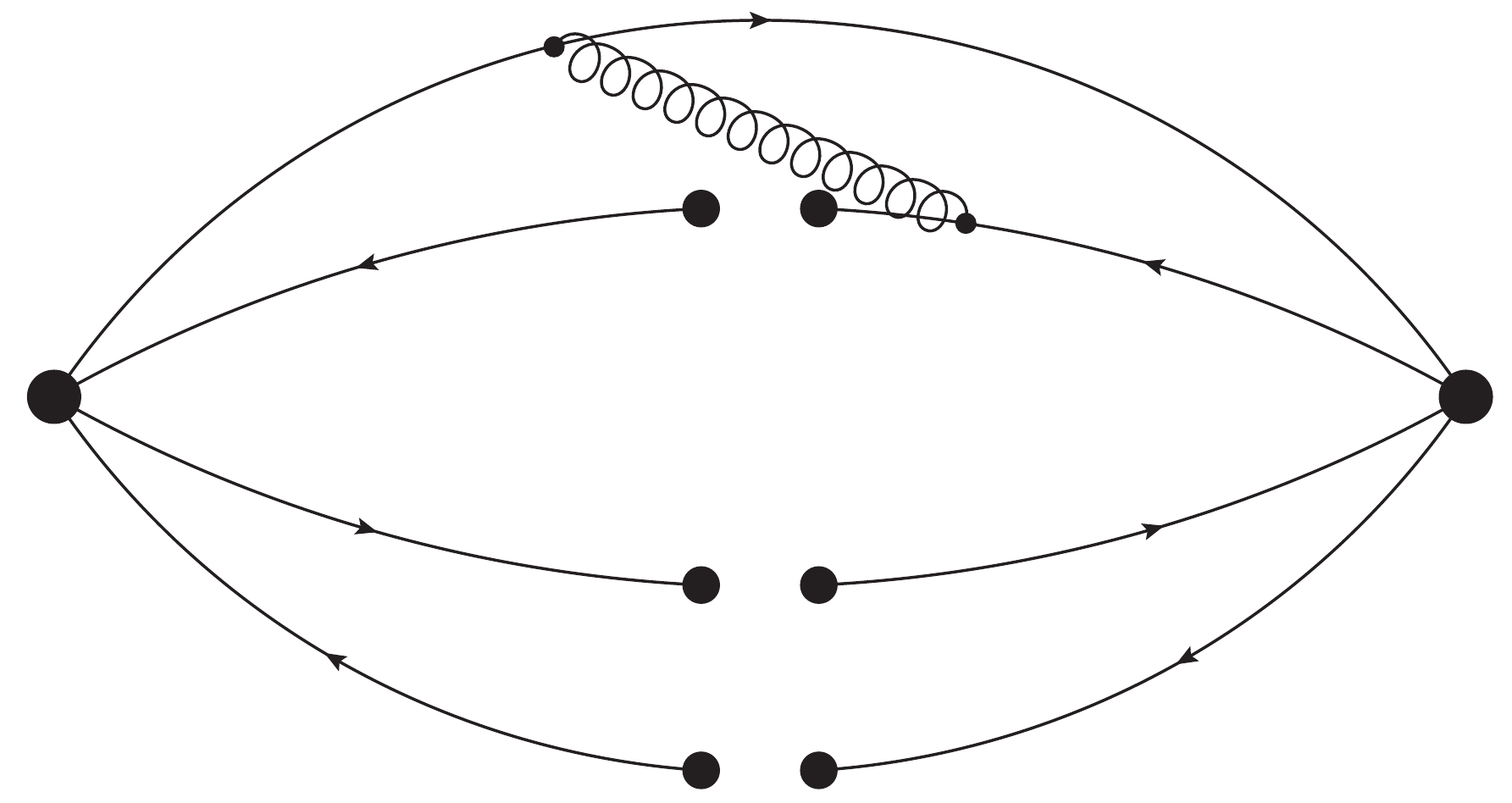}}}~~~~~
\subfigure[($h{\rm-}8$)]{
\scalebox{0.15}{\includegraphics{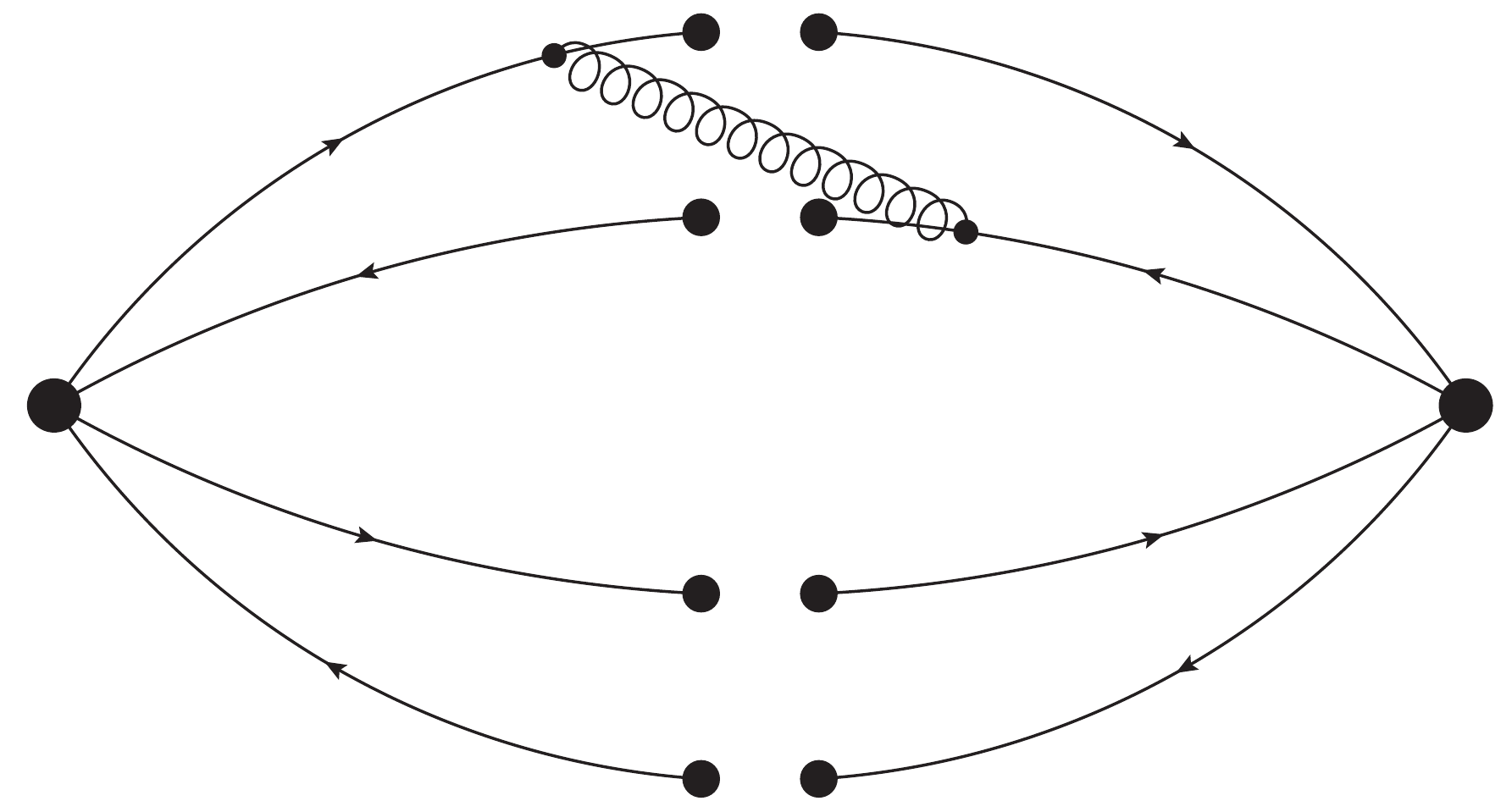}}}
\\[5mm]
\subfigure[($i{\rm-}1$)]{
\scalebox{0.15}{\includegraphics{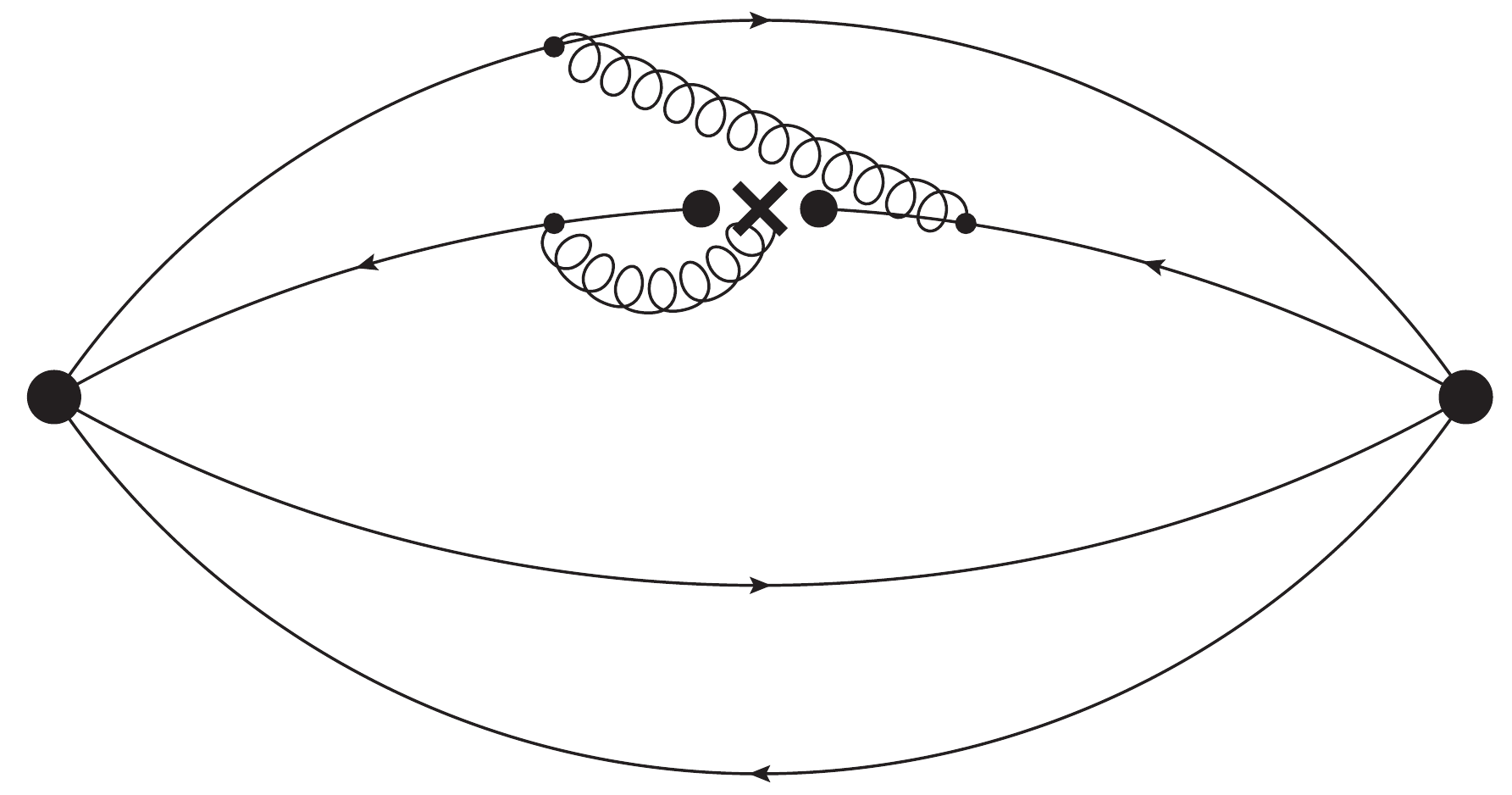}}}~~~~~
\subfigure[($i{\rm-}2$)]{
\scalebox{0.15}{\includegraphics{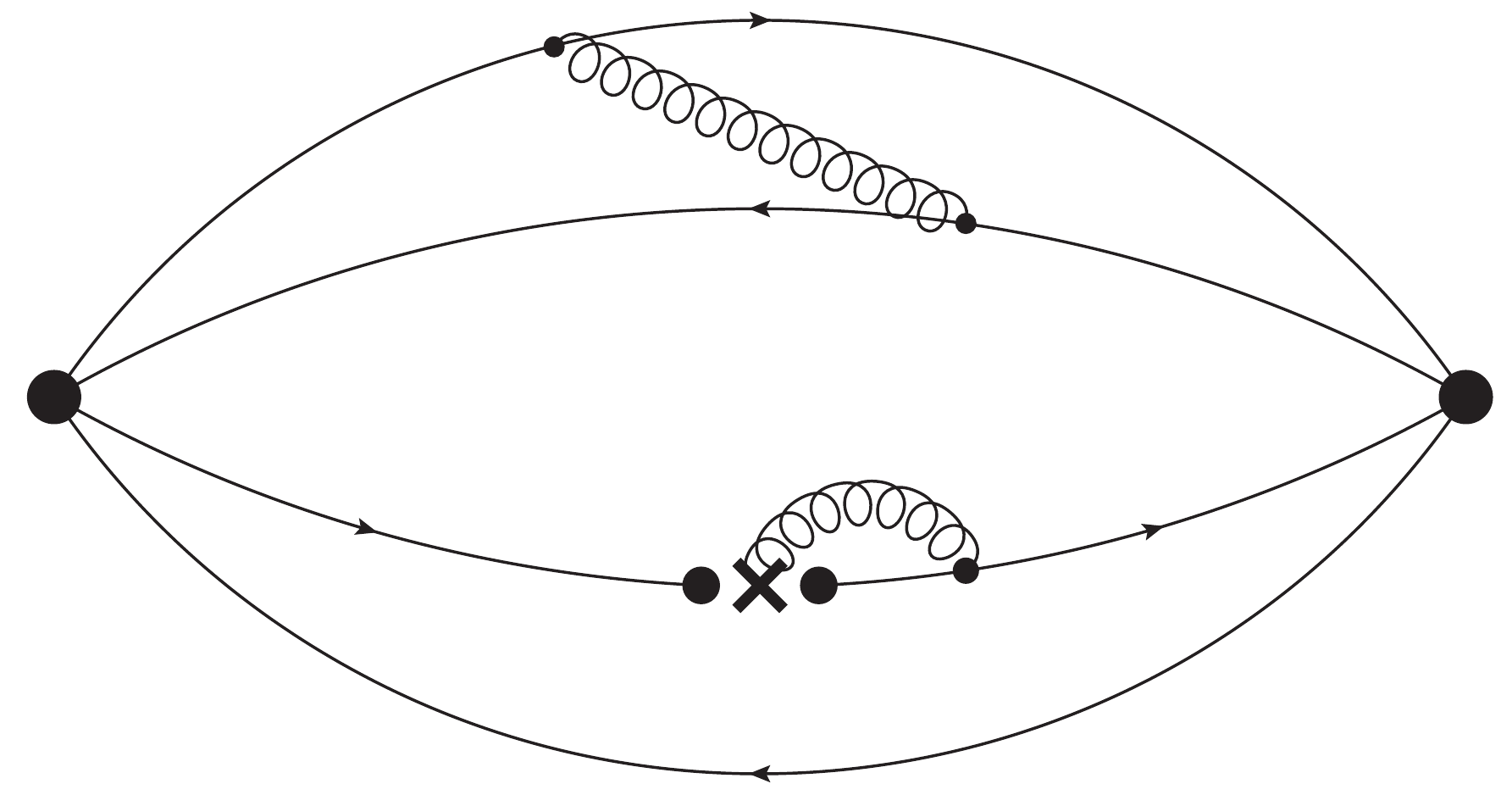}}}~~~~~
\subfigure[($i{\rm-}3$)]{
\scalebox{0.15}{\includegraphics{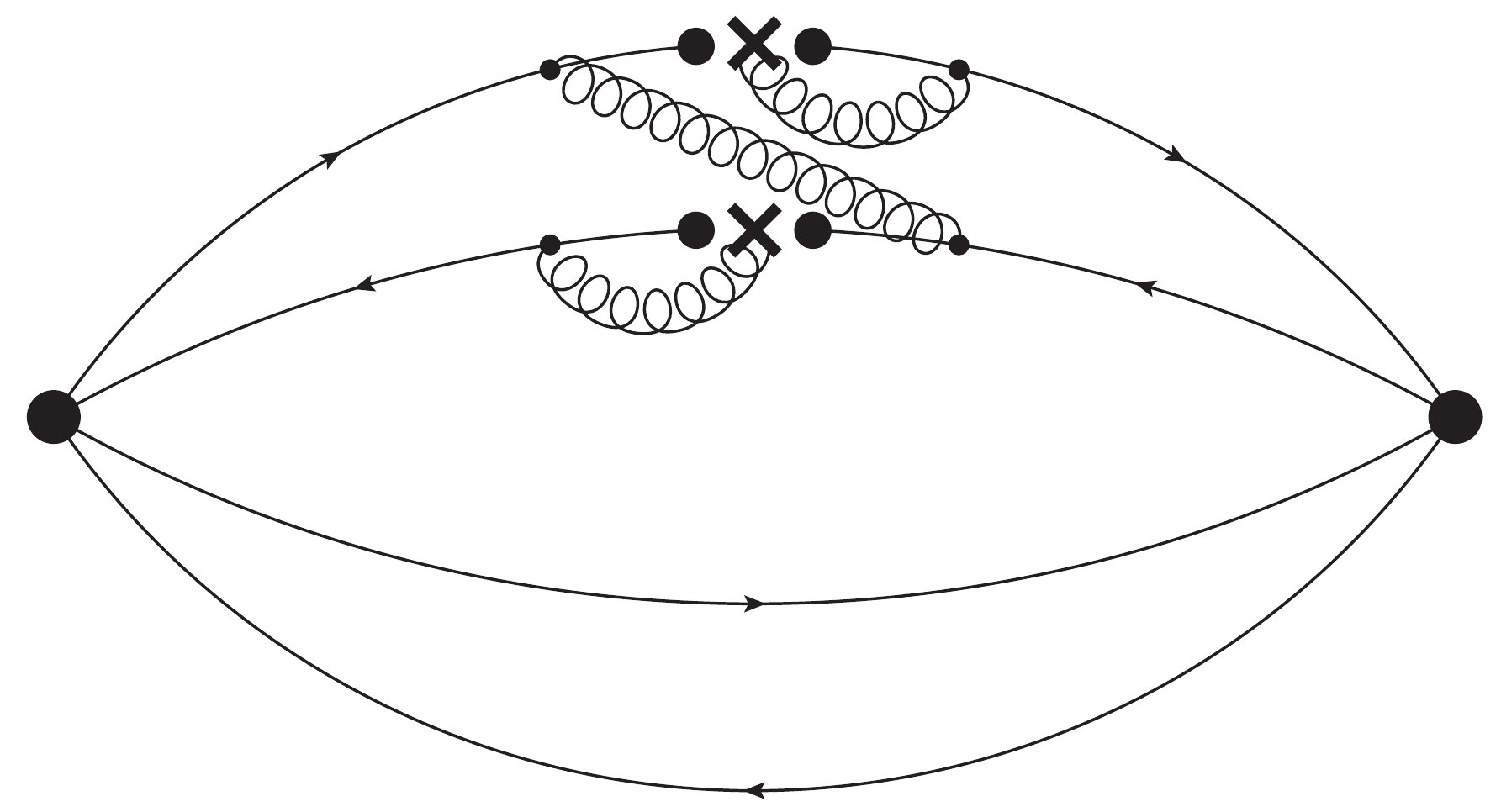}}}~~~~~
\subfigure[($i{\rm-}4$)]{
\scalebox{0.15}{\includegraphics{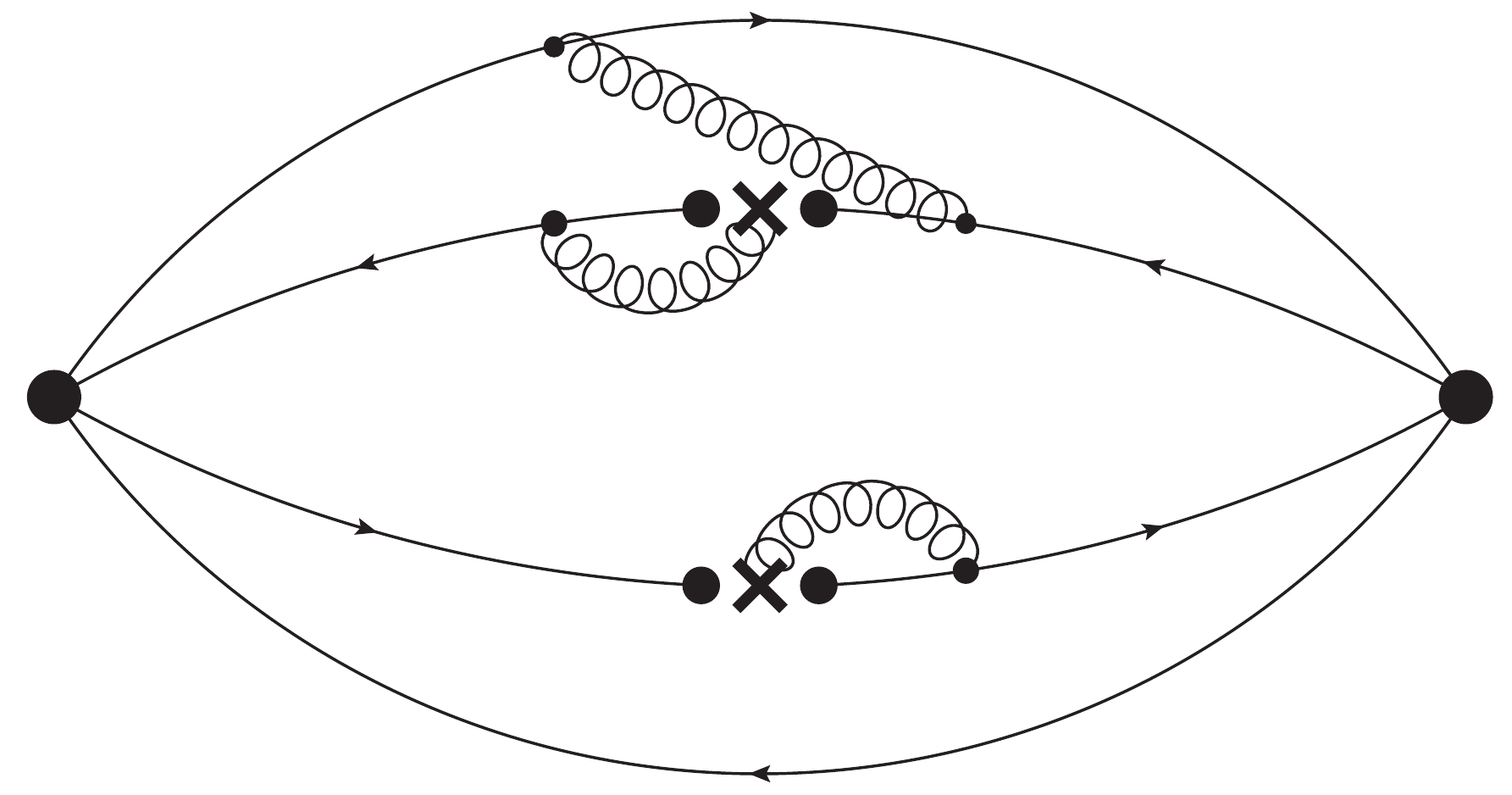}}}
\\
\subfigure[($i{\rm-}5$)]{
\scalebox{0.15}{\includegraphics{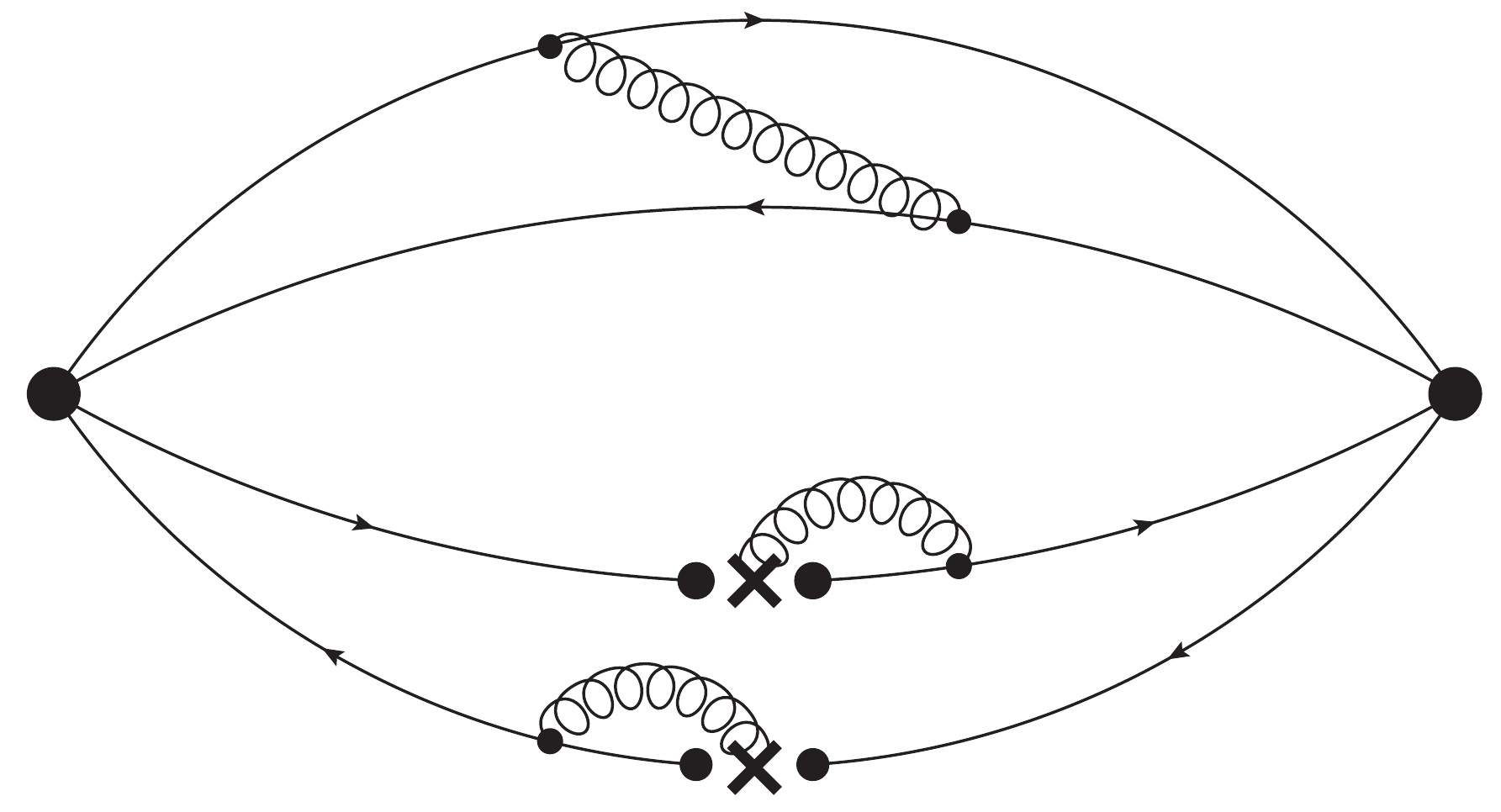}}}~~~~~
\subfigure[($i{\rm-}6$)]{
\scalebox{0.15}{\includegraphics{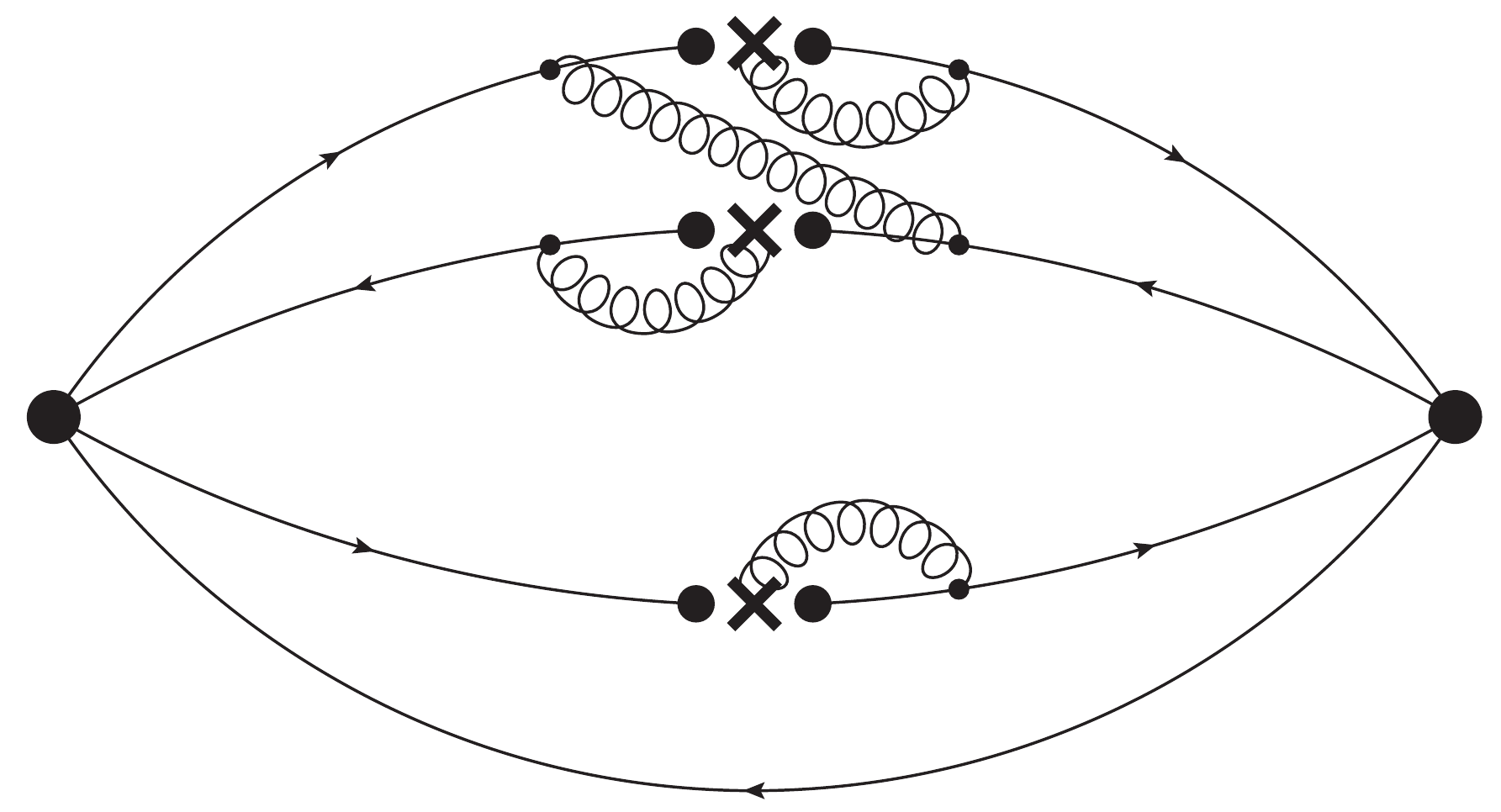}}}~~~~~
\subfigure[($i{\rm-}7$)]{
\scalebox{0.15}{\includegraphics{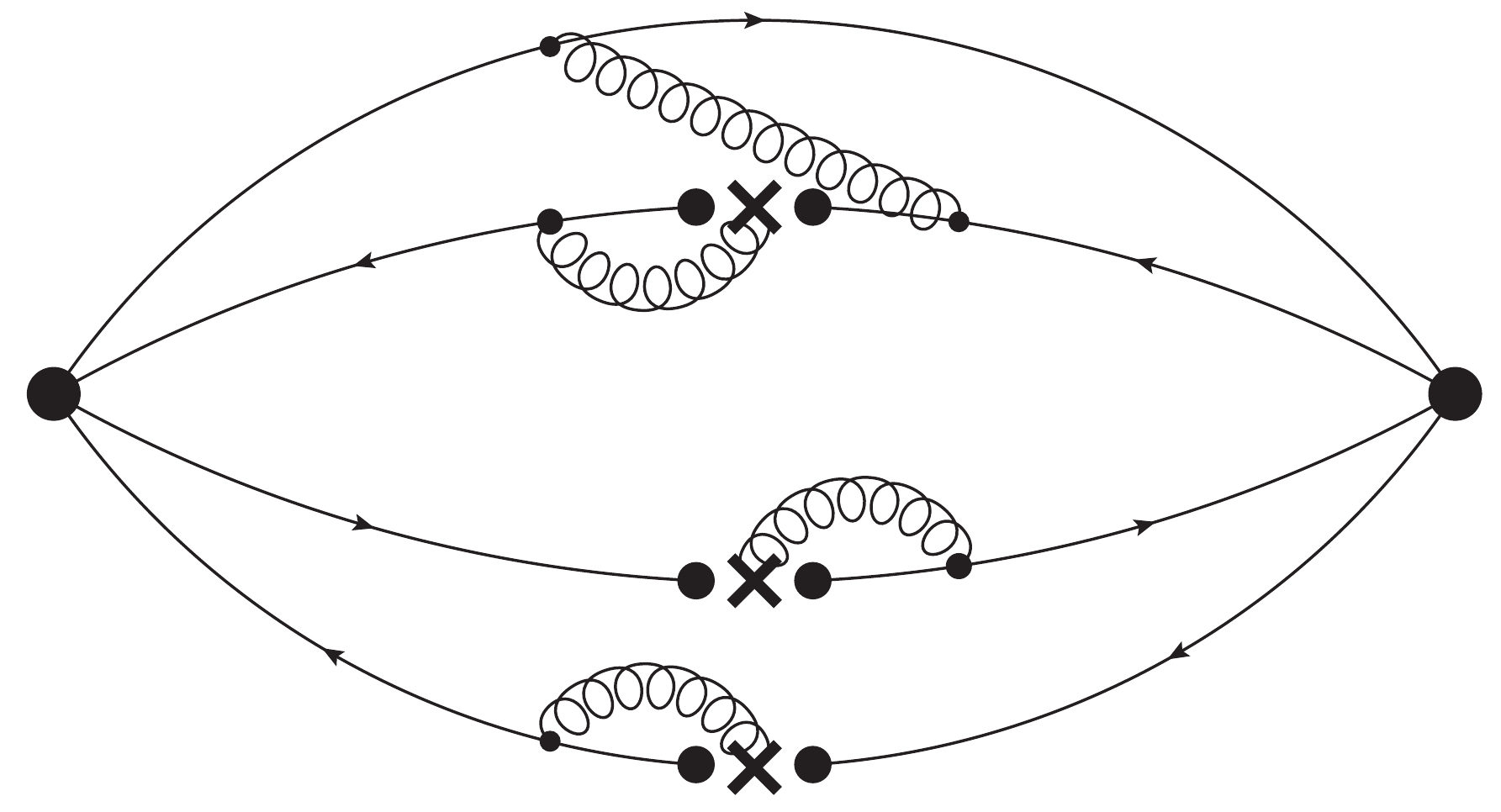}}}~~~~~
\subfigure[($i{\rm-}8$)]{
\scalebox{0.15}{\includegraphics{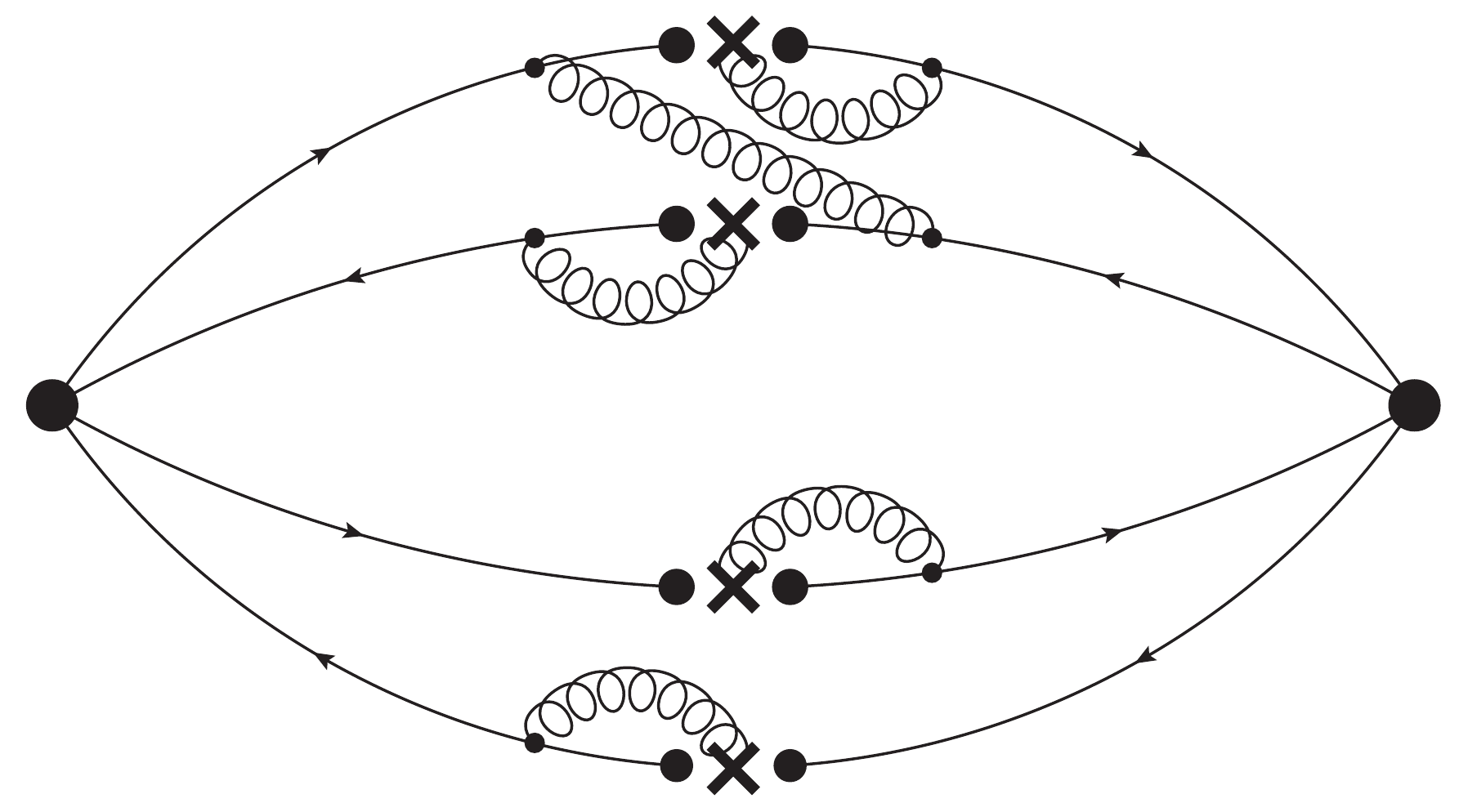}}}
\\[5mm]
\subfigure[($j{\rm-}1$)]{
\scalebox{0.15}{\includegraphics{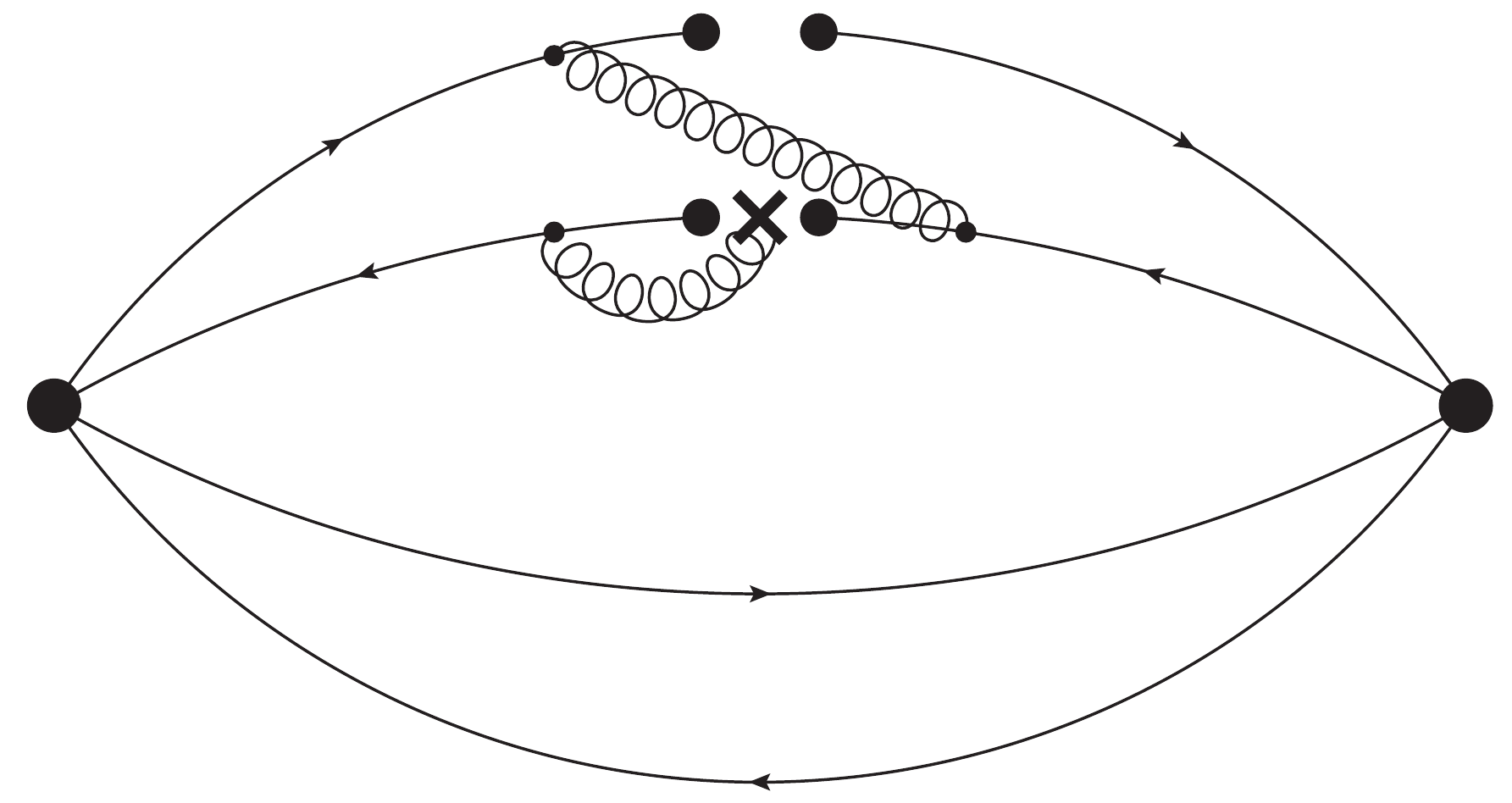}}}~~~~~
\subfigure[($j{\rm-}2$)]{
\scalebox{0.15}{\includegraphics{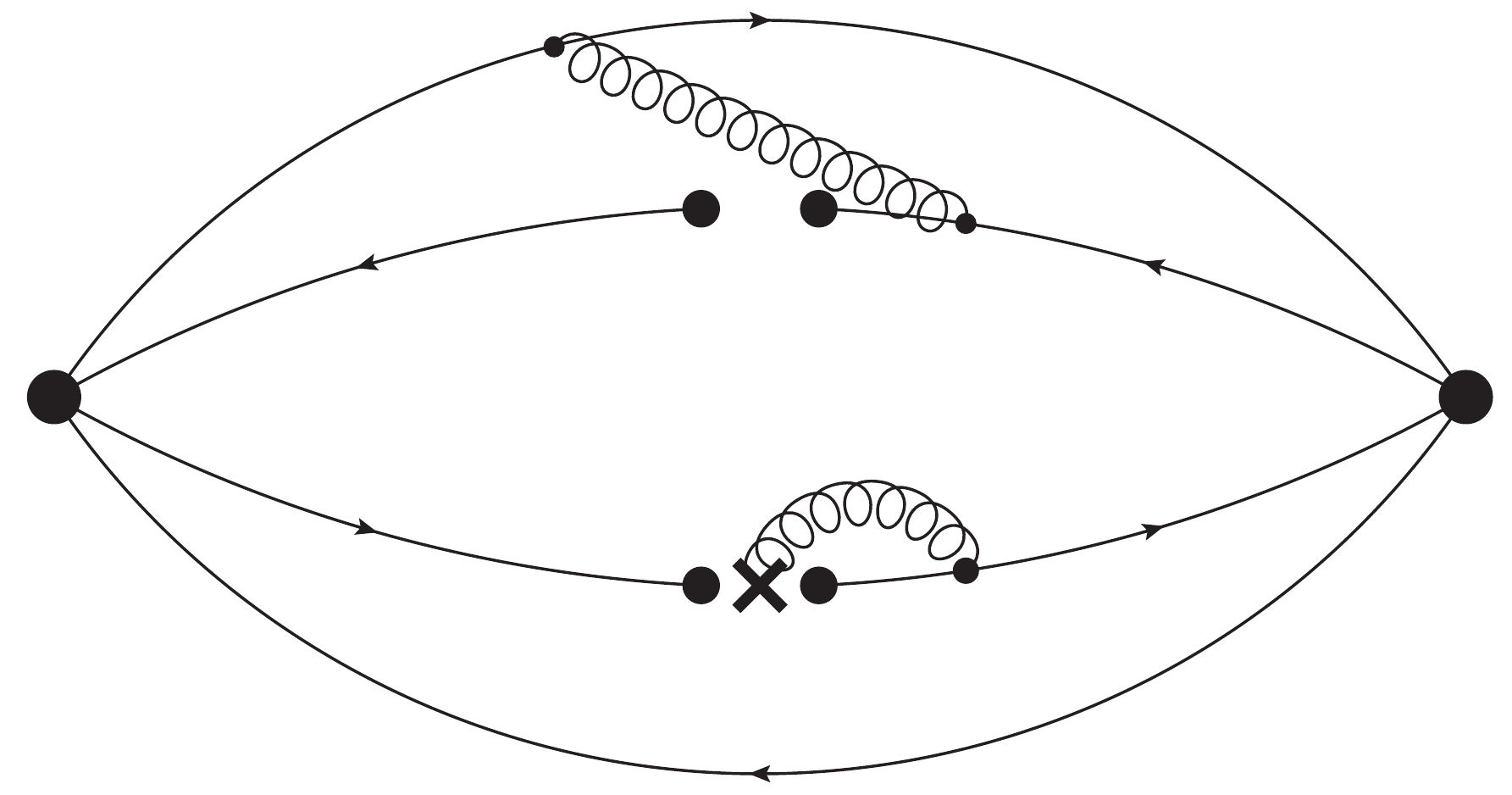}}}~~~~~
\subfigure[($j{\rm-}3$)]{
\scalebox{0.15}{\includegraphics{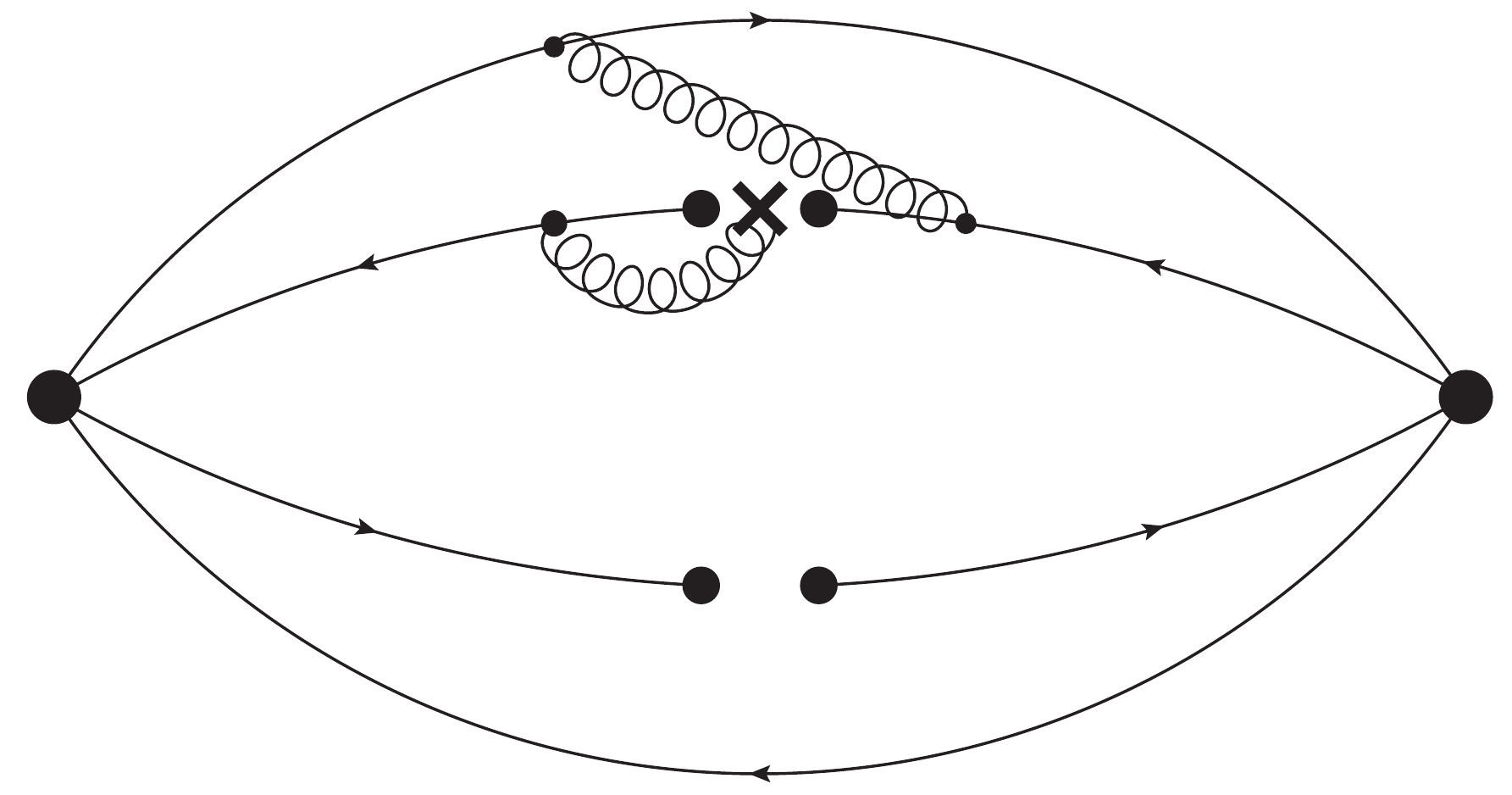}}}~~~~~
\subfigure[($j{\rm-}4$)]{
\scalebox{0.15}{\includegraphics{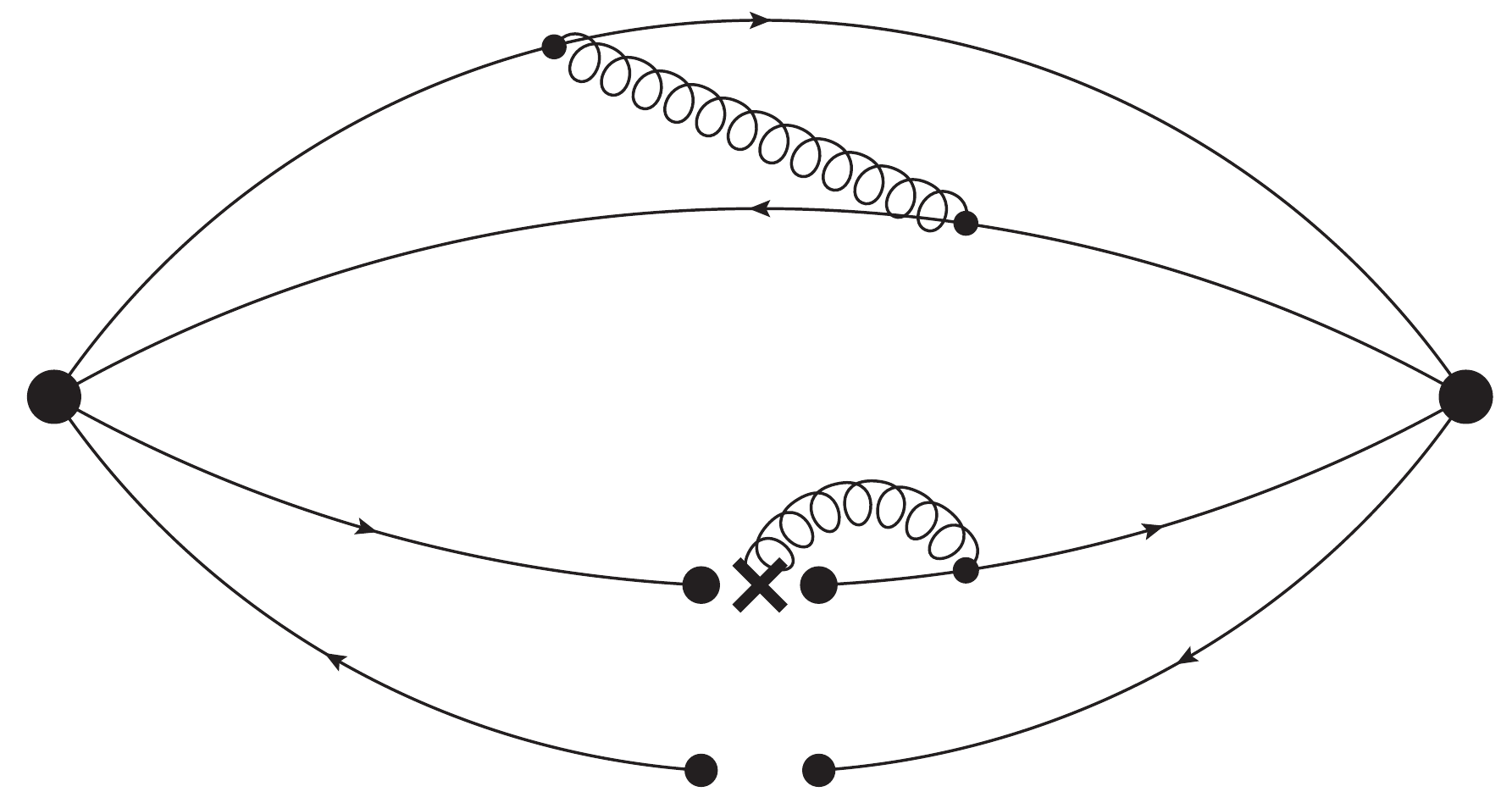}}}
\\
\subfigure[($j{\rm-}5$)]{
\scalebox{0.15}{\includegraphics{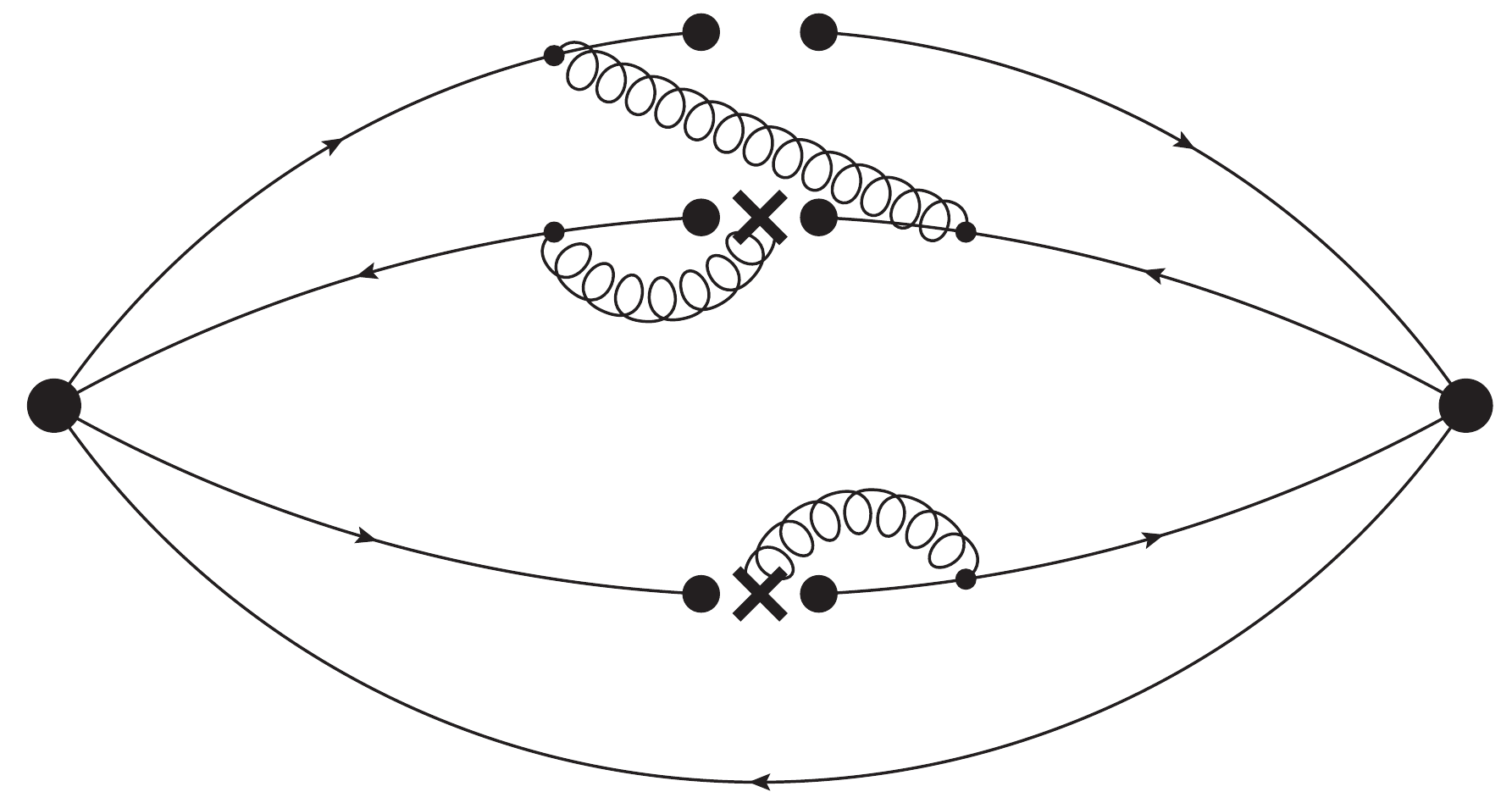}}}~~~~~
\subfigure[($j{\rm-}6$)]{
\scalebox{0.15}{\includegraphics{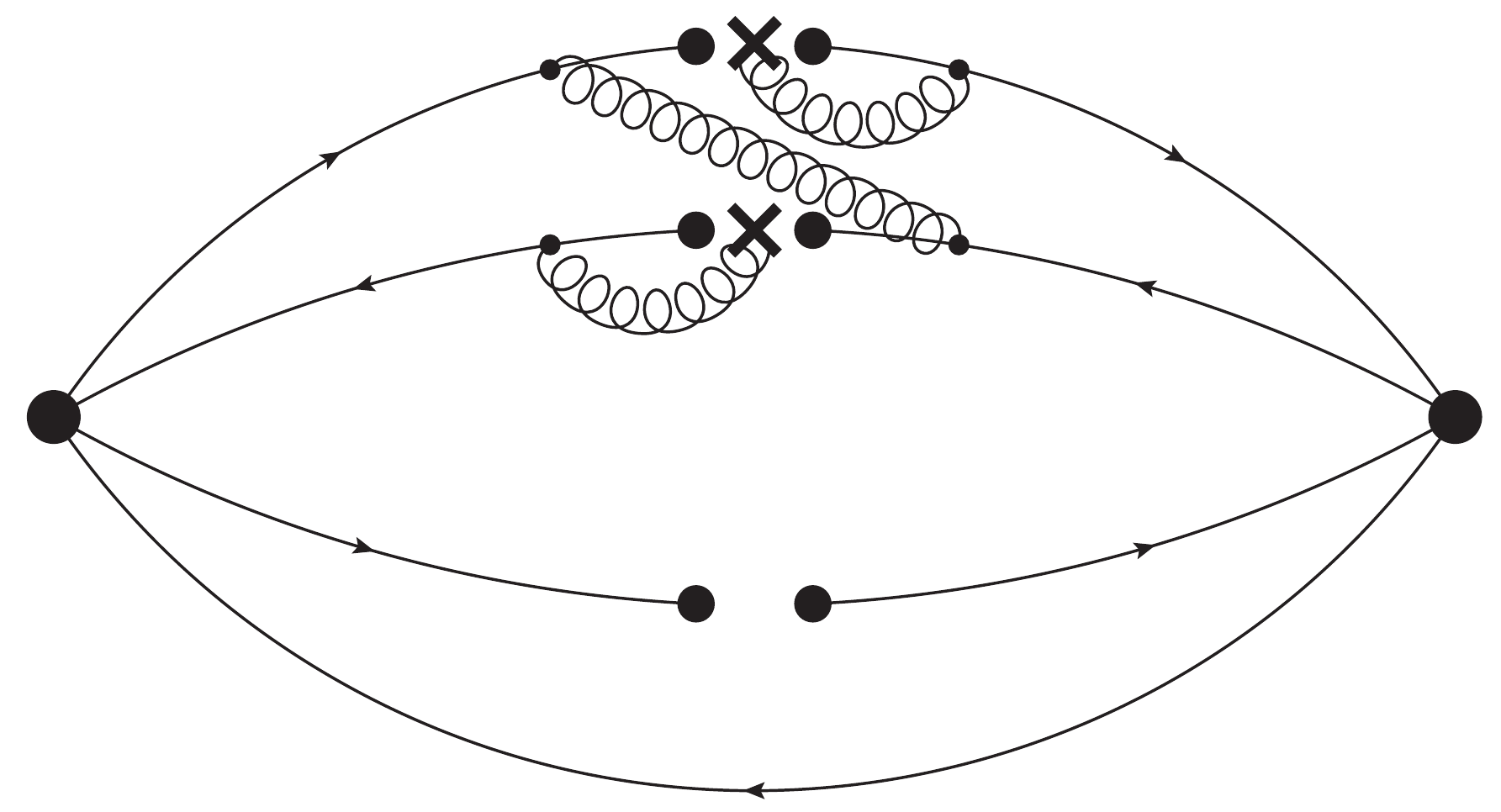}}}~~~~~
\subfigure[($j{\rm-}7$)]{
\scalebox{0.15}{\includegraphics{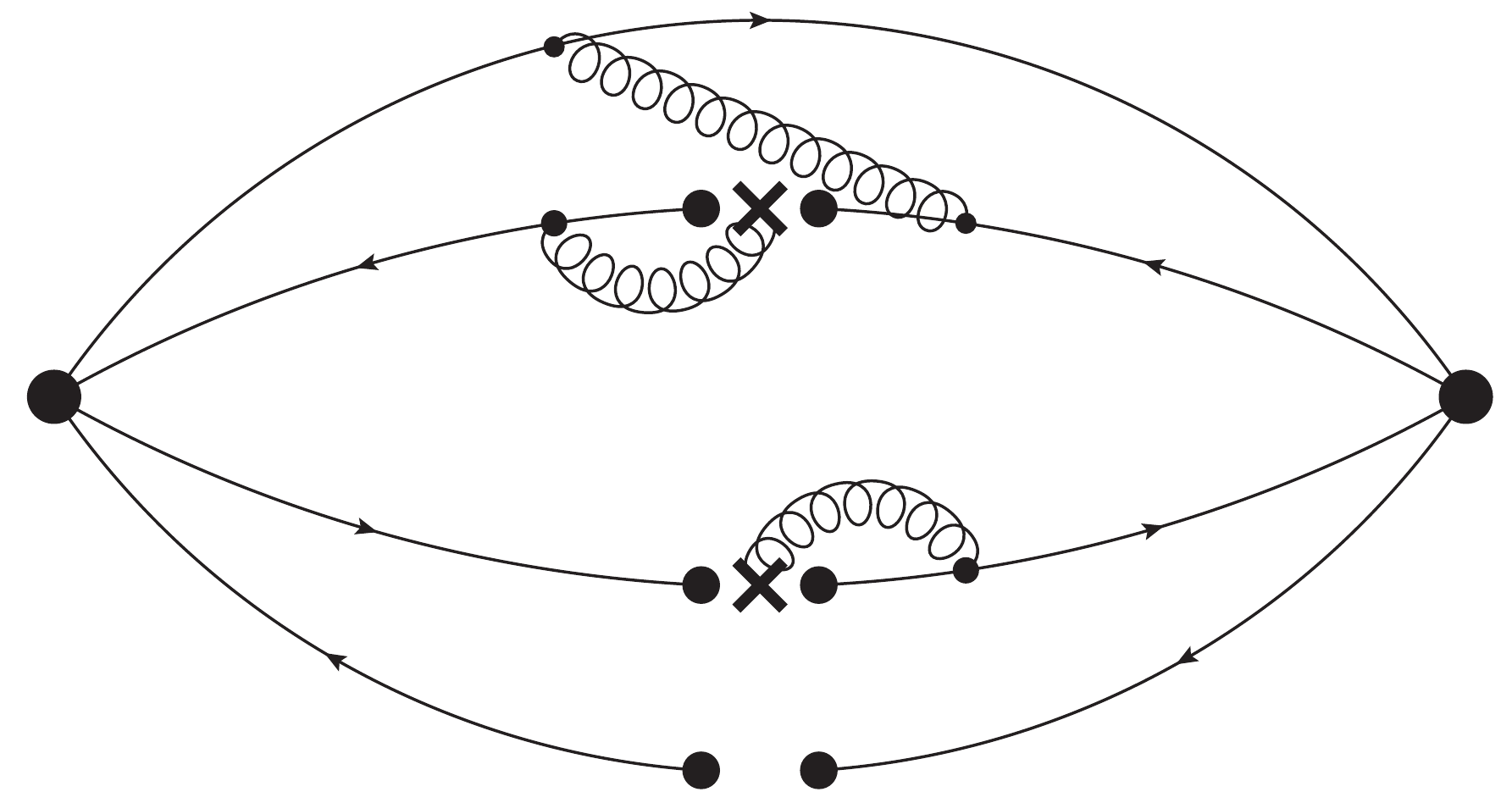}}}~~~~~
\subfigure[($j{\rm-}8$)]{
\scalebox{0.15}{\includegraphics{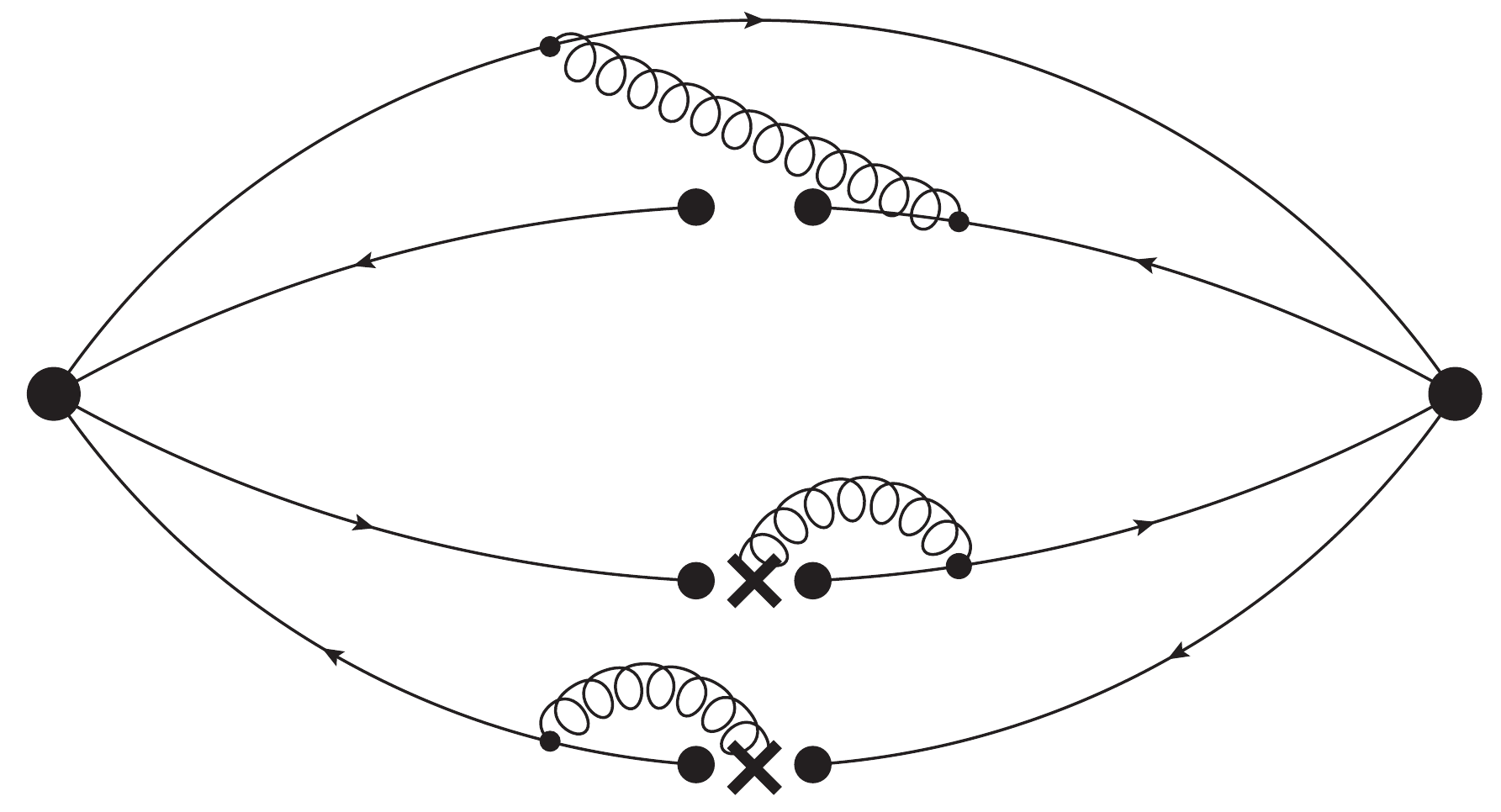}}}
\\
\subfigure[($j{\rm-}9$)]{
\scalebox{0.15}{\includegraphics{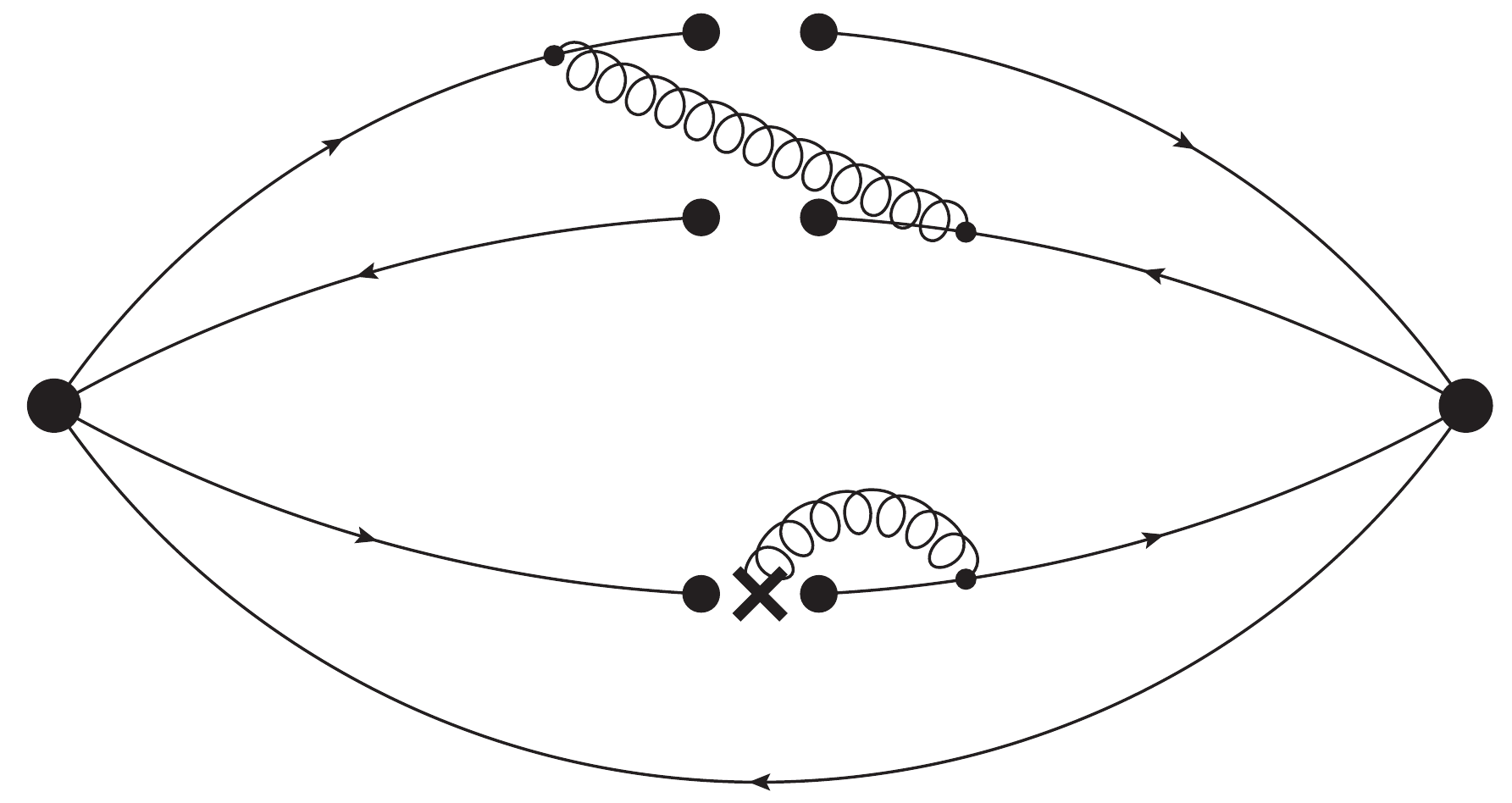}}}~~~~~
\subfigure[($j{\rm-}10$)]{
\scalebox{0.15}{\includegraphics{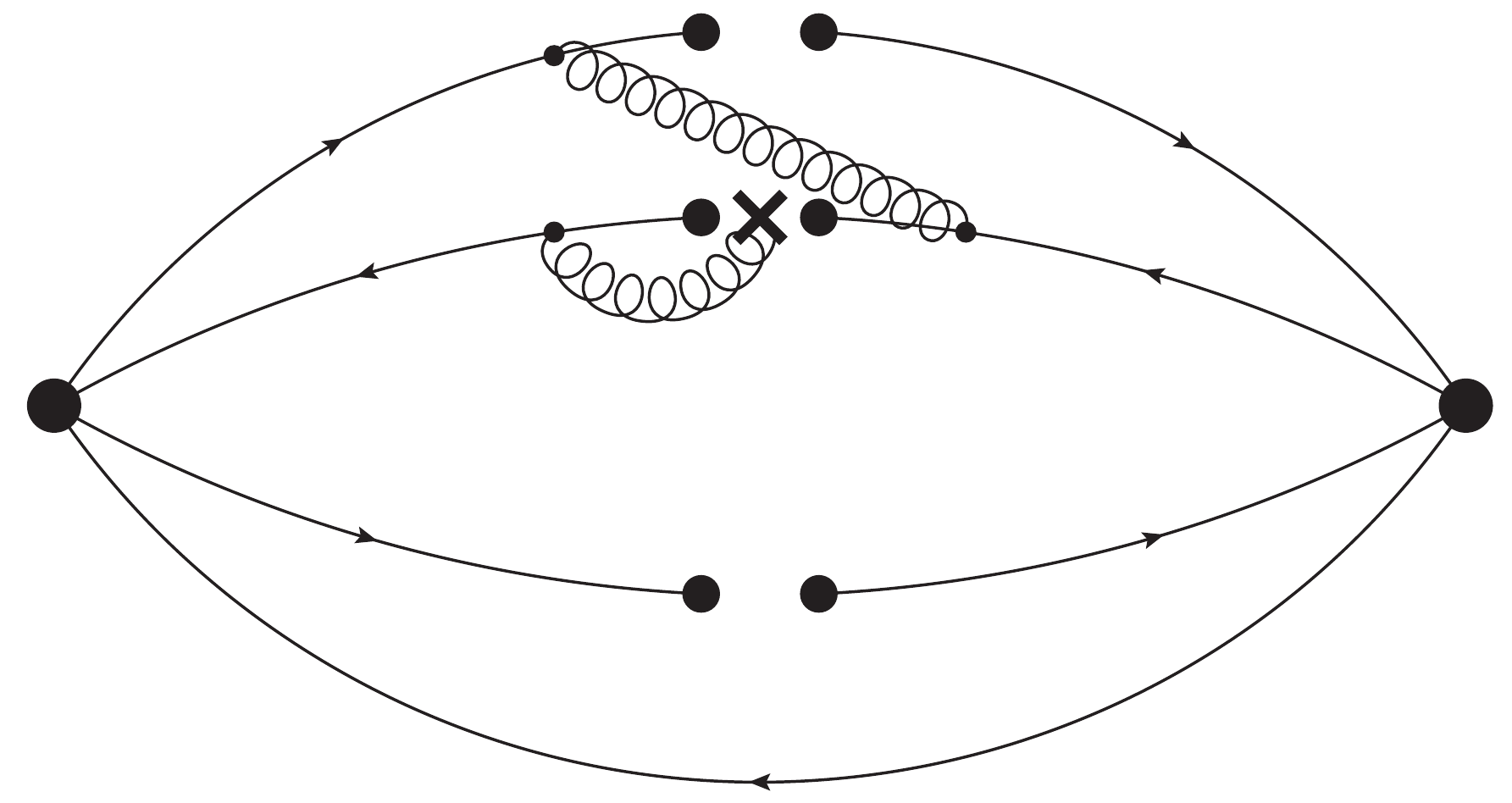}}}~~~~~
\subfigure[($j{\rm-}11$)]{
\scalebox{0.15}{\includegraphics{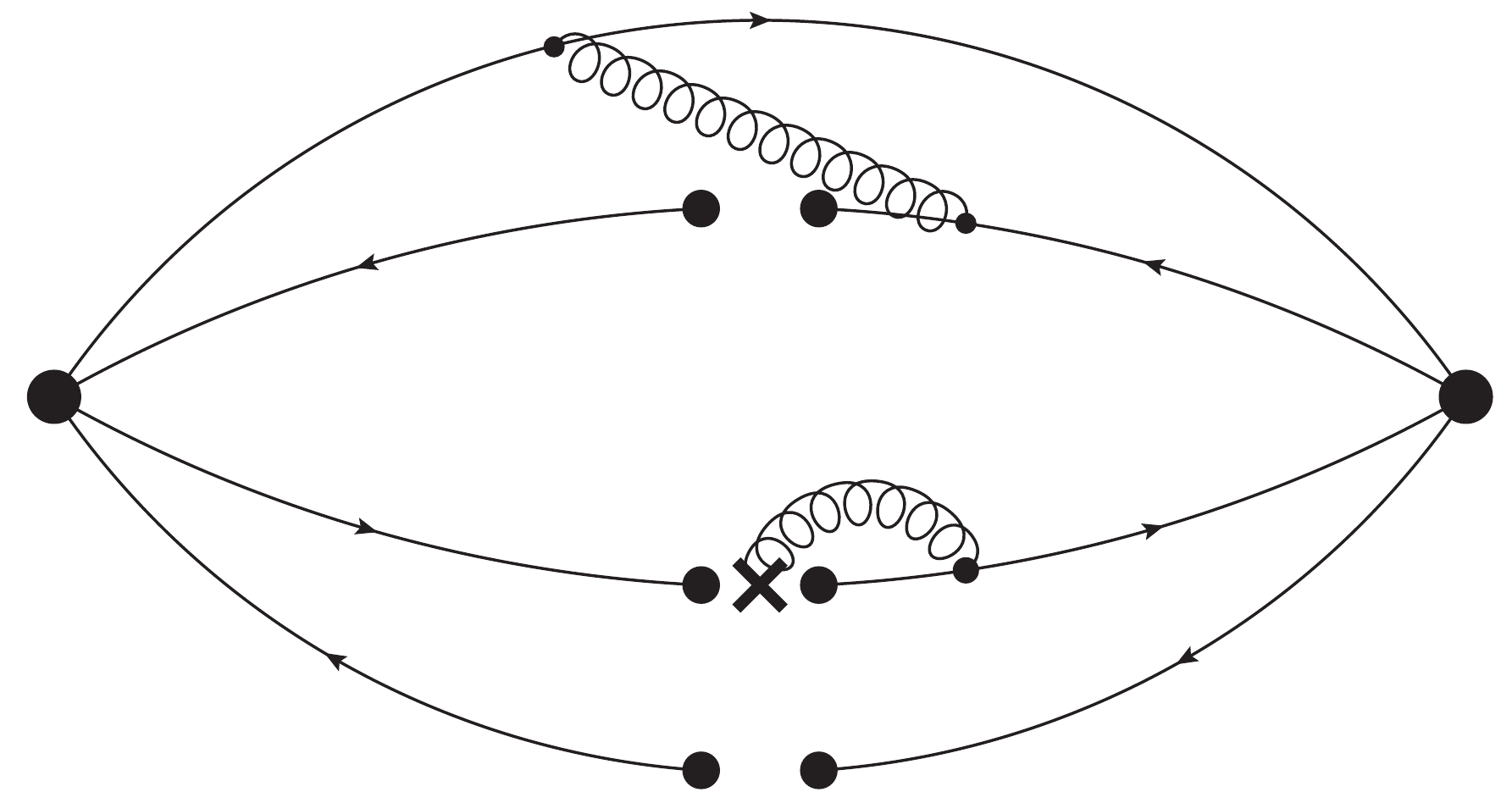}}}~~~~~
\subfigure[($j{\rm-}12$)]{
\scalebox{0.15}{\includegraphics{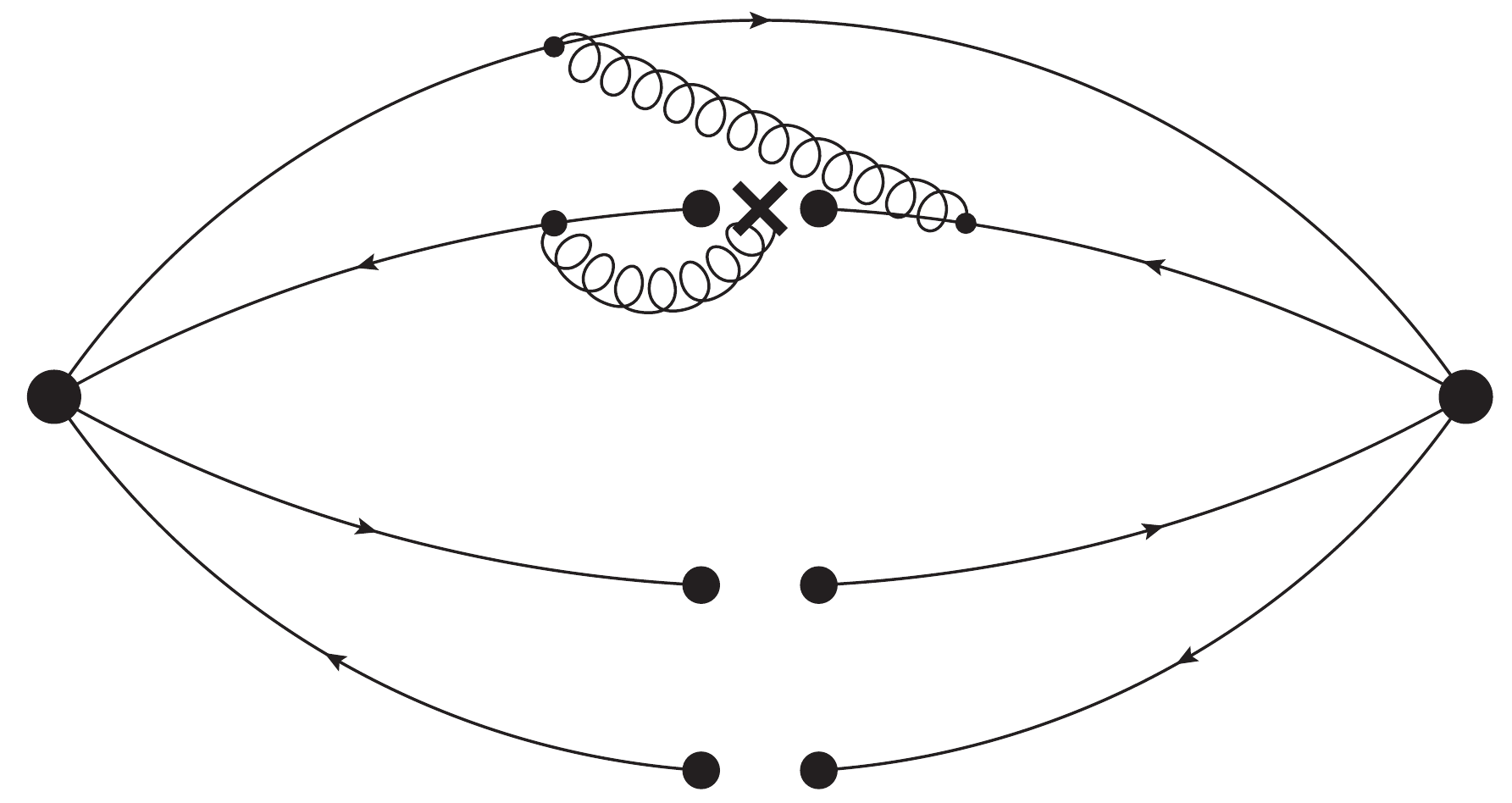}}}
\\[5mm]
$\Huge \bullet~\bullet~\bullet~~~\bullet~\bullet~\bullet$
\\[5mm]
\end{center}
\caption{Next-to-leading-order Feynman diagrams for tetraquark interpolating currents, including the $\mathcal{O}(\alpha_s)$ corrections to the perturbative term ($g$), to quark condensates ($h$), to quark-gluon mixed condensates ($i$), and to their combinations ($j$).}
\label{fig:feynman2}
\end{figure*}

For the tetraquark currents $J_{LL/LR/RL/RR}$, we have calculated all the leading-order Feynman diagrams, including:
\begin{itemize}

\item the perturbative term, as shown in Fig.~\ref{fig:feynman1}($a$);

\item the gluon condensates $\langle g_c^2 GG\rangle$ and $\langle g_c^3 f GGG\rangle$, as shown in Figs.~\ref{fig:feynman1}($b$);

\item the quark condensates $\langle\bar{q}q\rangle$, $\langle\bar{q}q\rangle^2$, $\langle\bar{q}q\rangle^3$, and $\langle\bar{q}q\rangle^4$ (disconnected), as shown in Figs.~\ref{fig:feynman1}($c$);

\item the quark-gluon mixed condensates $\langle g_c\bar{q}\sigma Gq\rangle$, $\langle g_c\bar{q}\sigma Gq\rangle^2$, $\langle g_c\bar{q}\sigma Gq\rangle^3$, and $\langle g_c\bar{q}\sigma Gq\rangle^4$ (disconnected), as shown in Figs.~\ref{fig:feynman1}($d$);

\item their combinations, as shown in Figs.~\ref{fig:feynman1}($e$) and Figs.~\ref{fig:feynman1}($f$).

\end{itemize}
For these diagrams, we use non-zero current quark masses ($m_{u/d} \neq 0$).

Besides them, we have calculated the following next-to-leading-order Feynman diagrams:
\begin{itemize}

\item the $\mathcal{O}(\alpha_s)$ corrections to the perturbative term, as shown in Fig.~\ref{fig:feynman2}($g$);

\item the $\mathcal{O}(\alpha_s)$ corrections to quark condensates, as shown in Figs.~\ref{fig:feynman2}($h$);

\item the $\mathcal{O}(\alpha_s)$ corrections to quark-gluon mixed condensates, as shown in Figs.~\ref{fig:feynman2}($i$);

\item the $\mathcal{O}(\alpha_s)$ corrections to their combinations, as shown in Figs.~\ref{fig:feynman2}($j$).

\end{itemize}
For these diagrams, we use zero current quark masses ($m_{u/d} \rightarrow 0$). The results for Fig.~\ref{fig:feynman2}($g$) are partly taken from Ref.~\cite{Groote:2014pva}, where the $\mathcal{O}(\alpha_s)$ corrections to the perturbative term were systematically studied for all light scalar tetraquark currents.

Still taking $J_{LL}$ as an example, we use the method of operator product expansion (OPE) to calculate its correlation function $\Pi_{LL}(q^2)$, and then use the Borel transformation with the truncation at the threshold value $s_0$ to obtain
\begin{eqnarray}
%%%%%%%%%%%%%%%%%%%%%%%%%%%%%%%%%%%%%%%%%%%%%%%%%%%%%%%%%%%%%%%%%%%%%%%%%%%%%%
&& \Pi_{LL}(s_0, M_B^2) = \mathcal{B}_{q^2 \rightarrow M_B^2}\Pi_{LL}(q^2)
\label{eq:piLL}
\\ \nonumber &=& \int^{s_0}_{0} \Bigg( {s^4 \over 81920 \pi^6} \left( 1 + {4\alpha_s \over \pi} \right)
\\ \nonumber && ~~~~~~ + {s^2 \over 256 \pi^4}~\Big( m_u + m_d \Big)~\langle\bar{q}q\rangle
\\ \nonumber && ~~~~~~ + {s \over 256 \pi^4}~\Big( m_u + m_d \Big)~\langle g_c\bar{q}\sigma Gq\rangle \Bigg)~e^{-s/M_B^2}~ds \, .
\end{eqnarray}
We also extract its spectral density $\rho^{\rm OPE}_{LL}(s)$ to be:
\begin{eqnarray}
%%%%%%%%%%%%%%%%%%%%%%%%%%%%%%%%%%%%%%%%%%%%%%%%%%%%%%%%%%%%%%%%%%%%%%%%%%%%%%
\nonumber \rho^{\rm OPE}_{LL}(s) &=& {s^4 \over 81920 \pi^6}  \left( 1 + {4\alpha_s \over \pi} \right)
\\ \nonumber && + {s^2 \over 256 \pi^4}~\Big( m_u + m_d \Big)~\langle\bar{q}q\rangle
\\ && + {s \over 256 \pi^4}~\Big( m_u + m_d \Big)~\langle g_c\bar{q}\sigma Gq\rangle \, ,
\label{eq:rhoLL}
\end{eqnarray}
where $\Pi_{LL}(q^2)$ and $\rho^{\rm OPE}_{LL}(s)$ are related to each other through the dispersion relation:
\begin{equation}
\Pi_{LL}(q^2)=\int^\infty_0\frac{\rho^{\rm OPE}_{LL}(s)}{s-q^2-i\varepsilon}ds \, .
\label{eq:disper}
\end{equation}

In Eq.~(\ref{eq:piLL}) we have performed the OPE calculation up to the twentieth dimension $({\rm D}=20)$, during which we try to calculate as many terms as we can. Here we only keep the terms up to the $\mathcal{O}(m_q)$ order, while a more detailed expression is given in Appendix~\ref{app:sumrule}. Sum rules obtained using the other three tetraquark currents $J_{LR/RL/RR}$ are just the same:
\begin{equation}
\Pi_{LL}(q^2) = \Pi_{LR}(q^2) = \Pi_{RL}(q^2) = \Pi_{RR}(q^2) \, .
\end{equation}
%We also extract the off-diagonal correlation functions among the tetraquark currents $J_{LL/LR/RL/RR}$. The results are given in Appendix~\ref{app:sumrule}.

We clearly see from Eq.~(\ref{eq:piLL}) that it only contains the perturbative terms (Fig.~\ref{fig:feynman1}($a$) and Fig.~\ref{fig:feynman2}($g$)) and two non-perturbative terms depending on current quark masses (Fig.~\ref{fig:feynman1}($c{\rm-}1$) and Fig.~\ref{fig:feynman1}($d{\rm-}1$)). The leading-order perturbative term can not give a normal hadron mass by itself, while it is highly doubtful that the $\mathcal{O}(\alpha_s)$ corrections to the perturbative term could do this. Besides, the contributions of the latter two terms, which contain the current $up$ and $down$ quark masses, to hadron masses are usually thought to be at the $\left({m_d + m_u \over m_d - m_u} \approx  3~{\rm times}\right)$ isospin breaking (MeV) level~\cite{Yang:1993bp,Hwang:1996pe}. Therefore, Eq.~(\ref{eq:piLL}) indicates an abnormal mass significantly smaller than the normal hadron mass at the GeV level.%, given such a tetraquark state exists.

For comparisons, we also perform QCD sum rule analyses using the other eight tetraquark currents $J_{VV}$, $J_{AA}$, $J_{VA}$, $J_{AV}$, $J_{LL;+}$, $J_{LL;-}$, $J_{LR;+}$, and $J_{LR;-}$. For these currents, we calculate all the leading-order terms up to the twelfth dimension $({\rm D}=12)$ and up to the $\mathcal{O}(m_q)$ order, while we calculate the next-to-leading-order terms only for Fig.~\ref{fig:feynman2}($h{\rm-}8$). The results are listed in Appendix~\ref{app:sumrule}. We shall numerically compare them with Eq.~(\ref{eq:piLL}) in the next subsection.

%
%=====================================================================================
%=====================================================================================
\subsection{Numerical Analyses}\label{sec:numerical}
%=====================================================================================
%=====================================================================================
%

In this subsection we perform numerical analyses using the sum rules obtained in the previous subsection. To do this we use the following values of current quark masses and various quark and gluon
condensates~\cite{Yang:1993bp,Herren:2017osy,Narison:2002pw,Gimenez:2005nt,Jamin:2002ev,Ioffe:2002be,Ovchinnikov:1988gk,colangelo}:
%%%%%%%%%%%%%%%%%%%%%%%%%%%%%%%%%%%%%%%%%%%%%%%%%%%%%%%%%%%%%%%%%%%%%%%%%%%%%%
\begin{eqnarray}
\nonumber \alpha_s(1~\mbox{GeV}) &=& 0.408 \pm 0.016 \, ,
\\ \nonumber m_u(1~\mbox{GeV}) &=& 3.0^{+0.7}_{-0.5} \mbox{ MeV} \, ,
\\ \nonumber m_d(1~\mbox{GeV}) &=& 6.3^{+0.7}_{-0.4} \mbox{ MeV} \, ,
\\ \langle g_c^2 GG \rangle &=& 0.48\pm 0.14 \mbox{ GeV}^4 \, ,
\label{eq:condensates}
\\ \nonumber \langle\bar qq \rangle &=& -(0.24 \pm 0.01)^3 \mbox{ GeV}^3 \, ,
\\ \nonumber \langle g_c \bar q \sigma G q\rangle &=& -M_0^2 \times\langle\bar qq\rangle \, ,
\\ \nonumber M_0^2 &=& (0.8\pm0.2) \mbox{ GeV}^2 \, .
\end{eqnarray}
%%%%%%%%%%%%%%%%%%%%%%%%%%%%%%%%%%%%%%%%%%%%%%%%%%%%%%%%%%%%%%%%%%%%%%%%%%%%%%
%

Using $\mu^2_{\rm \overline {MS}} = s_0 = M_B^2 = 1$~GeV$^2$ (standard QCD sum rule parameters for light scalar mesons~\cite{Chen:2006zh,Chen:2007xr,Groote:2014pva}), we numerically calculate Eq.~(\ref{eq:piLL}) to be:
\begin{eqnarray}
%%%%%%%%%%%%%%%%%%%%%%%%%%%%%%%%%%%%%%%%%%%%%%%%%%%%%%%%%%%%%%%%%%%%%%%%%%%%%%
&& \Pi_{LL}(s_0 = M_B^2 = 1~{\rm GeV}^2)
\label{eq:piLLnum}
\\ \nonumber &=& \Big( 1.1~[{\rm LO}] + 0.58~[{\rm NLO}] - 0.83~[{\rm D}^4] + 1.1~[{\rm D}^6] \Big)
\\ \nonumber && ~~~~~~~~~~~~~~~~~~~~~~~~~~~~~~~~~~~~~~~~~~~~~~ \cdot 10^{-9}~{\rm GeV}^{10} \, .
\end{eqnarray}
From the above expression, we find that the $\mathcal{O}(\alpha_s)$ corrections to the perturbative term ([NLO]) are similar to the leading-order perturbative term ([LO]). So do the two non-perturbative terms of $[{\rm D}={4}]$ and $[{\rm D}={6}]$. However, the summation of the non-perturbative terms is significantly smaller than the summation of the perturbative terms. Moreover, all the $[{\rm D} > 6]$ terms vanish, making this OPE converge very well.

Then we numerically calculate Eq.~(\ref{eq:piVV}), Eq.~(\ref{eq:piAA}), and Eq.~(\ref{eq:piVA}), which are obtained using $J_{VV}$, $J_{AA}$, and $J_{VA/AV}$, respectively:
\begin{eqnarray}
%%%%%%%%%%%%%%%%%%%%%%%%%%%%%%%%%%%%%%%%%%%%%%%%%%%%%%%%%%%%%%%%%%%%%%%%%%%%%%
&& {1\over4}\cdot\Pi_{VV}(s_0 = M_B^2 = 1~{\rm GeV}^2)
\label{eq:piVVnum}
\\ \nonumber &=& \Big( 1.1~[{\rm LO}] ~~~~~~~~~~~~~~ + 159~[{\rm D}^6] - 307~[{\rm D}^8]
\\ \nonumber && ~~~~~~~ + 28~[{\rm D}^{10}] + 6.1~[{\rm D}^{12}] + \cdots \Big)
\cdot 10^{-9}~{\rm GeV}^{10} \, ,
\\
&& {1\over4}\cdot\Pi_{AA}(s_0 = M_B^2 = 1~{\rm GeV}^2)
\label{eq:piAAnum}
\\ \nonumber &=& \Big( 1.1~[{\rm LO}] - 1.7~[{\rm D}^4] - 157~[{\rm D}^6] + 307~[{\rm D}^8]
\\ \nonumber && ~~~~~~~~ - 19~[{\rm D}^{10}] + 27~[{\rm D}^{12}] + \cdots \Big)
\cdot 10^{-9}~{\rm GeV}^{10} \, ,
\\
&& {1\over4}\cdot\Pi_{VA/AV}(s_0 = M_B^2 = 1~{\rm GeV}^2)
\label{eq:piAAnum}
\\ \nonumber &=& \Big( 1.1~[{\rm LO}] - 0.83~[{\rm D}^4] + 1.1~[{\rm D}^6]
\\ \nonumber && ~~~~~~~ - 4.1~[{\rm D}^{10}] - 16~[{\rm D}^{12}] + \cdots \Big)
\cdot 10^{-9}~{\rm GeV}^{10} \, .
\end{eqnarray}
We also numerically calculate Eq.~(\ref{eq:piLLP}) and Eq.~(\ref{eq:piLLN}), which are obtained using $J_{LL/LR;+}$ and $J_{LL/LR;-}$, respectively:
\begin{eqnarray}
%%%%%%%%%%%%%%%%%%%%%%%%%%%%%%%%%%%%%%%%%%%%%%%%%%%%%%%%%%%%%%%%%%%%%%%%%%%%%%
&& 2\cdot\Pi_{LL/LR;+}(s_0 = M_B^2 = 1~{\rm GeV}^2)
\label{eq:piLLPnum}
\\ \nonumber &=& \Big( 1.1~[{\rm LO}] - 0.83~[{\rm D}^4] + 1.1~[{\rm D}^6]
\\ \nonumber && ~~~~~~~ + 4.1~[{\rm D}^{10}] + 16~[{\rm D}^{12}] + \cdots \Big)
\cdot 10^{-9}~{\rm GeV}^{10} \, ,
\\
&& 2\cdot\Pi_{LL/LR;-}(s_0 = M_B^2 = 1~{\rm GeV}^2)
\label{eq:piLLNnum}
\\ \nonumber &=& \Big( 1.1~[{\rm LO}] - 0.83~[{\rm D}^4] + 1.1~[{\rm D}^6]
\\ \nonumber && ~~~~~~~ - 4.1~[{\rm D}^{10}] - 16~[{\rm D}^{12}] + \cdots \Big)
\cdot 10^{-9}~{\rm GeV}^{10} \, .
\end{eqnarray}
In the above expressions we have rescaled all the leading-order perturbative terms ([LO]) to be the same as Eq.~(\ref{eq:piLLnum}).

\begin{figure*}[hbtp]
\begin{center}
\includegraphics[width=0.384\textwidth]{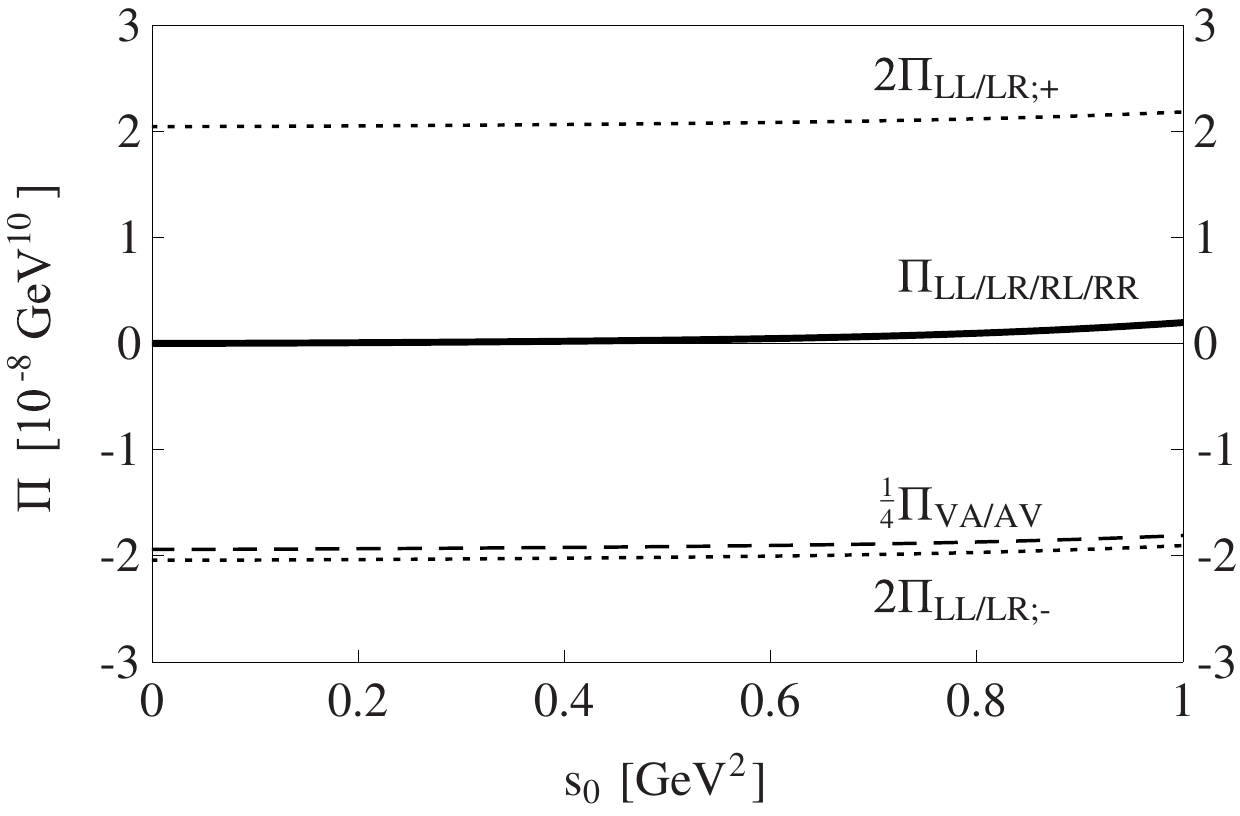}
~~~~~~~~
\includegraphics[width=0.40\textwidth]{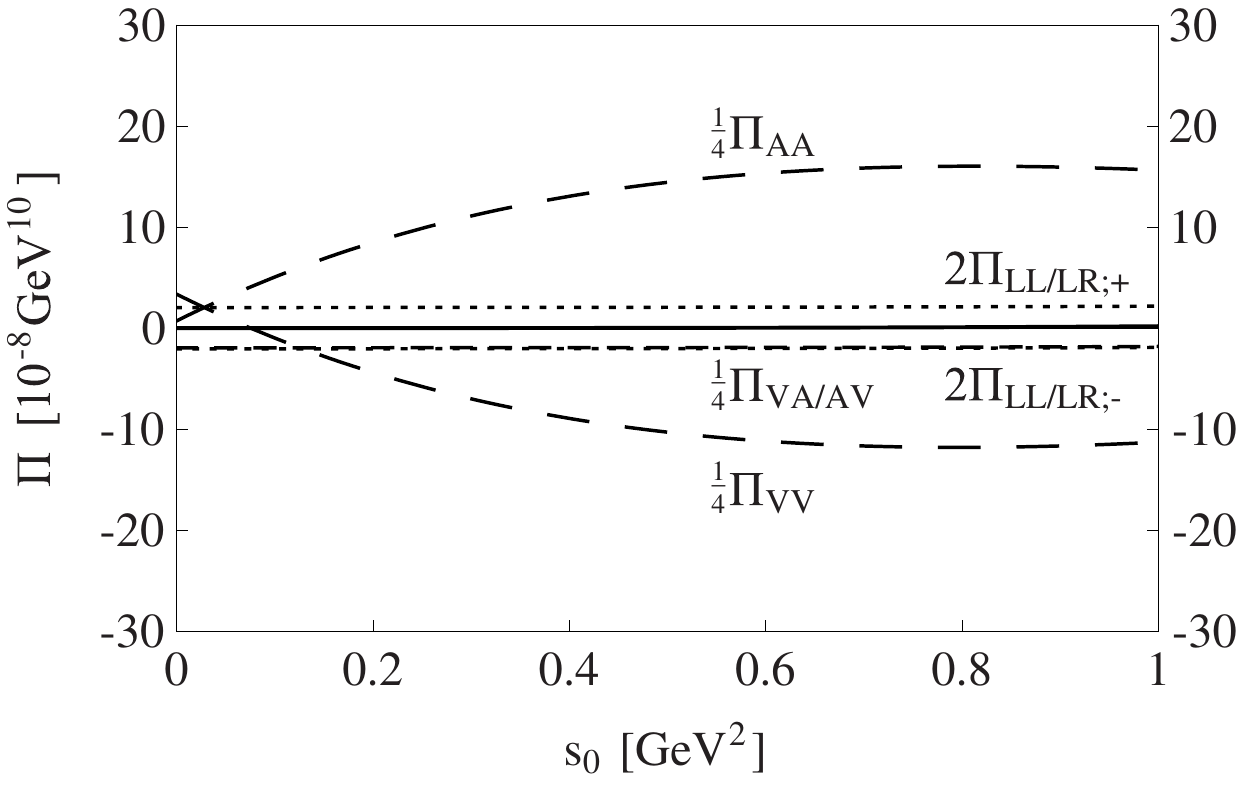}
\caption{The two-point correlation functions $\Pi_{LL/LR/RL/RR}(s_0, M_B^2)$ (solid), ${1\over4}\Pi_{VV}(s_0, M_B^2)$ (long-dashed), ${1\over4}\Pi_{AA}(s_0, M_B^2)$ (long-dashed), ${1\over4}\Pi_{VA/AV}(s_0, M_B^2)$ (short-dashed), $2\Pi_{LL/LR;+}(s_0, M_B^2)$ (dotted), and $2\Pi_{LL/LR;-}(s_0, M_B^2)$ (dotted), as functions of the threshold value $s_0$. These curves are obtained by setting $M_B^2 = 1$~GeV$^2$. ${1\over4}\Pi_{VV}$ and ${1\over4}\Pi_{AA}$ are only shown in the right panel, while the others are shown in both left and right panels.}
\label{fig:pi}
\end{center}
\end{figure*}

Compare Eq.~(\ref{eq:piLLnum}) and Eqs.~(\ref{eq:piVVnum}-\ref{eq:piLLNnum}), we find that the non-perturbative contributions to $J_{VV}$ and $J_{AA}$ are about one hundred times larger than the corresponding perturbative terms. Their contributions to $J_{VA/AV}$, $J_{LL/LR;+}$, and $J_{LL/LR;-}$ are also significantly larger than the corresponding perturbative terms; besides, these three OPEs do not (well) converge. We also show $\Pi_{LL}(s_0, M_B^2)$ in Fig.~\ref{fig:pi} as a function of the threshold value $s_0$, together with $\Pi_{VV}$, $\Pi_{AA}$, $\Pi_{VA/AV}$, $\Pi_{LL/LR;+}$, and $\Pi_{LL/LR;-}$. We find that $\Pi_{LL}(s_0, M_B^2)$ is significantly smaller than the others, especially in the low energy region.

%
%%%%%%%%%%%%%%%%%%%%%%%%%%%%%%%%%%%%%%%%%%%%%%%%%%%%%%%%%%%%%%%%%%%%%%%%%%%%%%
%%%%%%%%%%%%%%%%%%%%%%%%%%%%%%%%%%%%%%%%%%%%%%%%%%%%%%%%%%%%%%%%%%%%%%%%%%%%%%
%

Therefore, it is only $\Pi_{LL/LR/RL/RR}(s_0, M_B^2)$, the sum rules obtained using $J_{LL/LR/RL/RR}$, whose non-perturbative contributions are small and whose OPE well converges. This allows us to choose an abnormally small $s_0$ to perform further analyses in the next section.

%
%=====================================================================================
%=====================================================================================
\section{Phenomenological Analyses}\label{sec:phen}
%=====================================================================================
%=====================================================================================
%

In this section we study the two-point correlation function $\Pi_{LL}(s_0, M_B^2)$ at the hadron level. %It is extracted from the tetraquark current $J_{LL}$, while those extracted from $J_{LR/RL/RR}$ are the same.
To do this we assume that $J_{LL}$ couples to the tetraquark state $X_{LL}$ through
\begin{equation}
\langle 0| J_{LL} | X_{LL} \rangle = f_X \, ,
\end{equation}
and those coupled by $J_{LR/RL/RR}$ are $X_{LR/RL/RR}$, respectively. Then we express Eq.~(\ref{eq:pi}) in the form of the dispersion relation with a phenomenological spectral function:
%
%%%%%%%%%%%%%%%%%%%%%%%%%%%%%%%%%%%%%%%%%%%%%%%%%%%%%%%%%%%%%%%%%%%%%%%%%%%%%%
\begin{equation}
\Pi_{LL}(q^2)=\int^\infty_0\frac{\rho^{\rm phen}_{LL}(s)}{s-q^2-i\varepsilon}ds \, ,
\end{equation}
%%%%%%%%%%%%%%%%%%%%%%%%%%%%%%%%%%%%%%%%%%%%%%%%%%%%%%%%%%%%%%%%%%%%%%%%%%%%%%
%
where
%
%%%%%%%%%%%%%%%%%%%%%%%%%%%%%%%%%%%%%%%%%%%%%%%%%%%%%%%%%%%%%%%%%%%%%%%%%%%%%%
\begin{eqnarray}
\rho^{\rm phen}_{LL}(s) &=& \sum_n \delta(s - M^2_n) \langle0| J_{LL} |n\rangle \langle n| {J_{LL}^\dagger} |0\rangle
\\ \nonumber &=& f^2_X \delta(s-M^2_X) + \rm{higher~states} \, .
\label{eq:rhophen}
\end{eqnarray}
%%%%%%%%%%%%%%%%%%%%%%%%%%%%%%%%%%%%%%%%%%%%%%%%%%%%%%%%%%%%%%%%%%%%%%%%%%%%%%
%
For the second equation we have adopted a parametrization of one pole dominance for the ground state $X_{LL}$ together with a continuum contribution. After performing the Borel transformation on it, we obtain
\begin{equation}
%%%%%%%%%%%%%%%%%%%%%%%%%%%%%%%%%%%%%%%%%%%%%%%%%%%%%%%%%%%%%%%%%%%%%%%%%%%%%%
\mathcal{B}_{q^2 \rightarrow M_B^2}\Pi_{LL}(q^2) = f^2_X e^{-M_X^2/M_B^2} + \rm{higher~states} \, .
\label{eq:piphen}
\end{equation}

Comparing Eq.~(\ref{eq:piLL}) at the quark-gluon level and Eq.~(\ref{eq:piphen}) at the hadron level, we can use the quark-hadron duality to arrive at the sum rule equation
%
%%%%%%%%%%%%%%%%%%%%%%%%%%%%%%%%%%%%%%%%%%%%%%%%%%%%%%%%%%%%%%%%%%%%%%%%%%%%%%
\begin{equation}
f^2_Xe^{-M_X^2/M_B^2} = \Pi_{LL}(s_0, M_B^2) = \int^{s_0}_{0} \rho^{\rm OPE}_{LL}(s) e^{-s/M_B^2} ds \, ,
\label{eq:sumrule}
\end{equation}
%%%%%%%%%%%%%%%%%%%%%%%%%%%%%%%%%%%%%%%%%%%%%%%%%%%%%%%%%%%%%%%%%%%%%%%%%%%%%%
%
where we have approximated the contribution from continuum (higher) states by the spectral density of OPE above the threshold value $s_0$.

%
%=====================================================================================
%=====================================================================================
\subsection{Mass Estimation}\label{sec:mass}
%=====================================================================================
%=====================================================================================
%

In this subsection we estimate $M_X$, the mass of the state $X_{LL}$, using the sum rules given in Eq.~(\ref{eq:piLL}). It can be straightforwardly calculated by differentiating Eq.~(\ref{eq:sumrule}) with respect to ${1/M_B^2}$:
%
%%%%%%%%%%%%%%%%%%%%%%%%%%%%%%%%%%%%%%%%%%%%%%%%%%%%%%%%%%%%%%%%%%%%%%%%%%%%%%
\begin{equation}
M^2_X = \frac{\int^{s_0}_0  s \rho^{\rm OPE}_{LL}(s)~e^{-s/M_B^2}ds}{\int^{s_0}_0 \rho^{\rm OPE}_{LL}(s)~e^{-s/M_B^2} ds}\, .
\label{eq:LSR}
\end{equation}
%%%%%%%%%%%%%%%%%%%%%%%%%%%%%%%%%%%%%%%%%%%%%%%%%%%%%%%%%%%%%%%%%%%%%%%%%%%%%%
%
Before calculating $M_X$, we note that the method of QCD sum rules is actually a non-perturbative method. Hence, due to the very limited non-perturbative contributions, the present mass estimation should be treated with caution. Moreover, it is not well determined how to choose QCD sum rule parameters in the low energy region. In the present study we just use those values listed in Eqs.~(\ref{eq:condensates}).

%
%%%%%%%%%%%%%%%%%%%%%%%%%%%%%%%%%%%%%%%%%%%%%%%%%%%%%%%%%%%%%%%%%%%%%%%%%%%%%%
\begin{figure*}[hbtp]
\begin{center}
\includegraphics[width=0.3\textwidth]{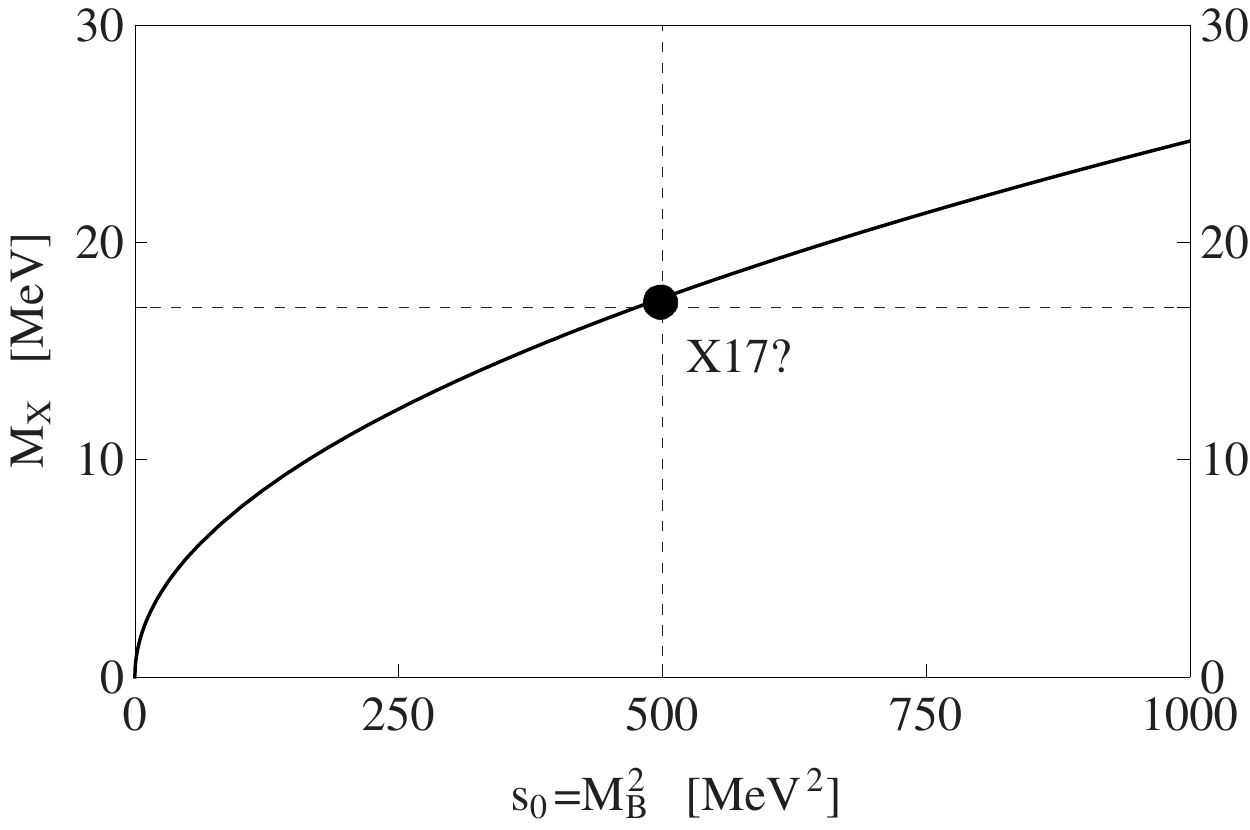}~~~
\includegraphics[width=0.3\textwidth]{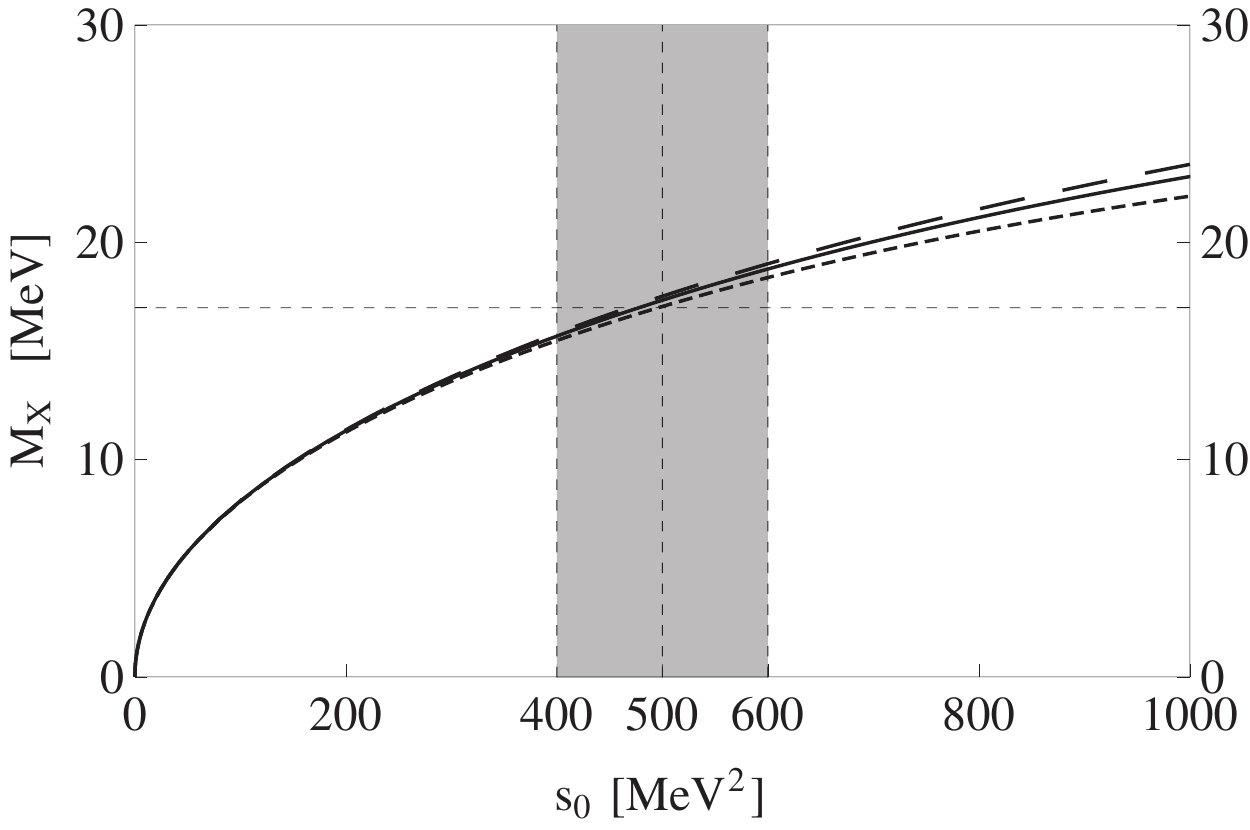}~~~
\includegraphics[width=0.3\textwidth]{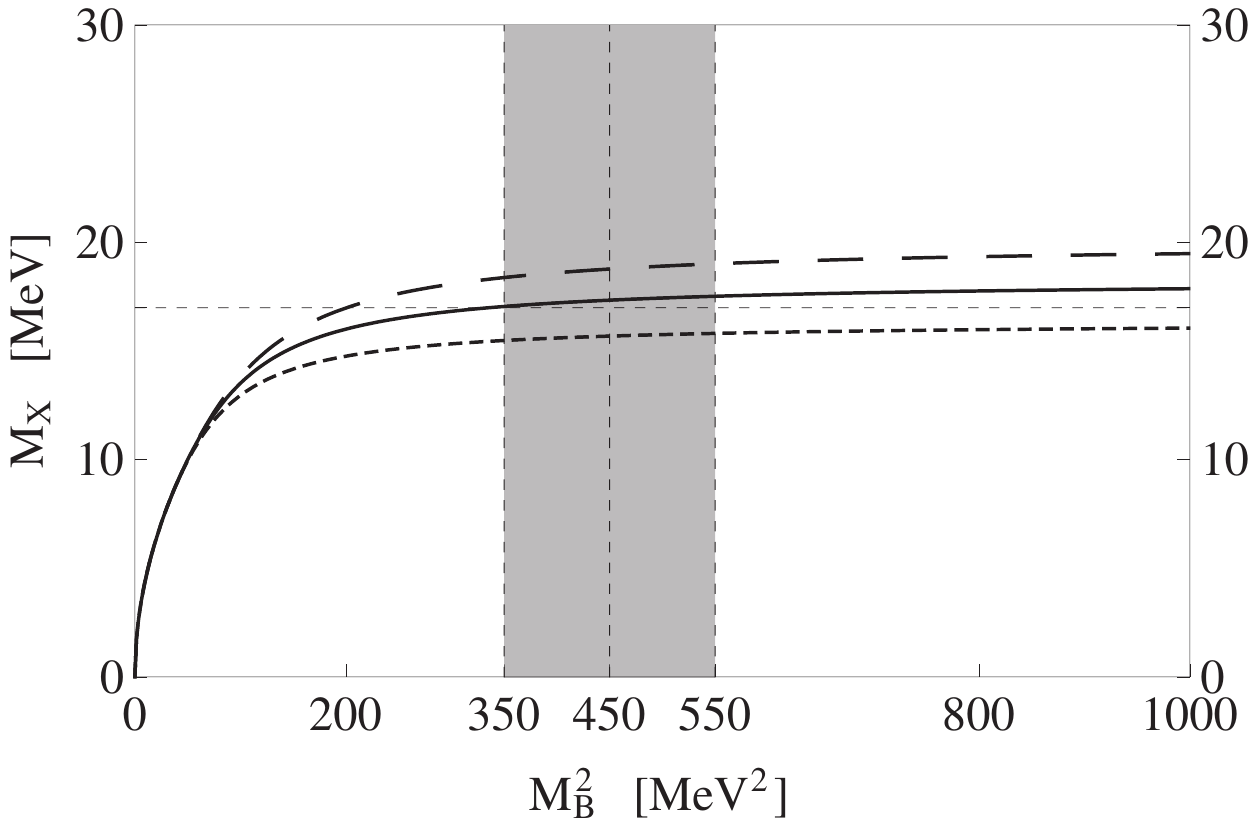}
\caption{The mass of the state $X_{LL}$, $M_X$, as a function of the threshold value $s_0$ (middle) and the Borel mass $M_B$ (right) individually, as well as $s_0 = M_B^2$ together (left). In the middle panel the short-dashed/solid/long-dashed curves are obtained by setting $M_B^2 = 350/450/550$~MeV$^2$, respectively. In the right panel the short-dashed/solid/long-dashed curves are obtained by setting $s_0 = 400/500/600$~MeV$^2$, respectively.}
\label{fig:JLLmass}
\end{center}
\end{figure*}
%%%%%%%%%%%%%%%%%%%%%%%%%%%%%%%%%%%%%%%%%%%%%%%%%%%%%%%%%%%%%%%%%%%%%%%%%%%%%%
%

To extract $M_X$ through Eq.~(\ref{eq:LSR}), we need to find proper working regions for the two free parameters of QCD sum rule method, the threshold value $s_0$ and the Borel mass $M_B$. The first criterion is usually to investigate the convergence of the OPE series, which is the cornerstone of a reliable QCD sum rule analysis. However, since all the $[{\rm D} > 6]$ terms vanish, the present OPE already converges very well.

Then we investigate the one-pole-dominance assumption by requiring the pole contribution (PC) to be larger than 30\%:
\begin{equation}
\mbox{PC} \equiv \left|\frac{ \Pi_{LL}(s_0, M_B^2) }{ \Pi_{LL}(\infty, M_B^2) }\right| \geq 30\% \, .
\label{eq:pole}
\end{equation}
This criterion can be more or less satisfied when choosing $M_B^2 \lesssim s_0$ in the low energy region. Accordingly, we show $M_X$ in the left panel of Fig.~\ref{fig:JLLmass} as a function of $s_0 = M_B^2$ together. The mass of the $X17$, $M_{X17} = 16.84 \pm 0.16 \pm 0.20$~MeV~\cite{Krasznahorkay:2015iga,Krasznahorkay:2019lyl}, can be arrived at by fine-tuning $s_0$ and $M_B^2$ to be both around $500$~MeV$^2$.

Finally, we choose $s_0 = 500$~MeV$^2$, and use Eq.~(\ref{eq:pole}) to determine $M_B^2 \leq 450$~MeV$^2$. Using these two values ($s_0 = 500$~MeV$^2$ and $M_B^2 = 450$~MeV$^2$), we obtain
\begin{equation}
M_{X} = 17.3^{+1.4}_{-1.7}~{\rm MeV} \, ,
\label{eq:mass}
\end{equation}
where the uncertainty is estimated by choosing $s_0 = 500\pm100$~MeV$^2$ and $M_B^2 = 450\pm100$~MeV$^2$ as well as using uncertainties of the QCD sum rule parameters listed in Eqs.~(\ref{eq:condensates}).

Therefore, our QCD sum rule results suggest it possible to interpret the $X17$ as a tetraquark state composed of four bare quarks. Since the four tetraquark currents $J_{LL/LR/RL/RR}$ lead to the same sum rules given in Eq.~(\ref{eq:piLL}), these four bare quarks can be $\bar{u}_L d_L \bar{d}_L u_L$, $\bar{u}_L d_L \bar{d}_R u_R$, $\bar{u}_R d_R \bar{d}_L u_L$, or $\bar{u}_R d_R \bar{d}_R u_R$. Neglecting the weak interaction, the Fierz transformations given in Eqs.~(\ref{fierzLL}-\ref{fierzRR}) tell that the first and fourth combinations are similar, the second and third combinations are similar, but the first and second combinations are different:
\begin{equation}
X_{LL} \sim X_{RR} \nsim X_{LR} \sim X_{RL} \, .
\end{equation}
Anyway, there can exist more than one states. Assuming their existence, we shall study their production and decay mechanisms in Sec.~\ref{sec:production} and Sec.~\ref{sec:decay}, respectively. We shall find the unique feature of this tetraquark assignment that we predict two almost degenerate states with significantly different widths.

For completeness, we also show $M_X$ in Fig.~\ref{fig:JLLmass} as a function of the threshold value $s_0$ (middle) and the Borel mass $M_B$ (right) individually. Its dependence on $M_B$ is weak when $M_B^2$ is around $450$~MeV$^2$. It does depend (moderately) on $s_0$, but this is reasonable because the two color-singlet $\bar u d$ and $\bar d u$ pairs can not be bound by the very limited non-perturbative QCD contributions so far. We shall further study this in the next subsection.

%
%=====================================================================================
%=====================================================================================
\subsection{Binding Mechanism}\label{sec:binding}
%=====================================================================================
%=====================================================================================
%

In the previous subsection we have found that non-perturbative QCD effects do not contribute much to the tetraquark states $X_{LL/LR/RL/RR}$ coupled by the currents $J_{LL/LR/RL/RR}$. Then a natural question arises: what force binds the two color-singlet $\bar u d$ and $\bar d u$ pairs?

To answer this, we use the ``chiral'' quark-antiquark currents $J_L = \bar{u}^a_L \gamma^\mu d^a_L$ and $J_R = \bar{u}^a_R \gamma^\mu d^a_R$ to perform QCD sum rule analyses. The sum rules obtained using the former are:
\begin{eqnarray}
%%%%%%%%%%%%%%%%%%%%%%%%%%%%%%%%%%%%%%%%%%%%%%%%%%%%%%%%%%%%%%%%%%%%%%%%%%%%%%
&& \Pi^{\mu\nu}_{L}(s_0, M_B^2) = \mathcal{B}_{q^2 \rightarrow M_B^2}\Pi^{\mu\nu}_{L}(q^2)
\label{eq:piL}
\\ \nonumber &\equiv& \mathcal{B}_{q^2 \rightarrow M_B^2} \left[ i\int d^4x~e^{iqx}~\langle 0 | \mathbb{T}\left[ J^\mu_{L}(x){J_{L}^{\nu,\dagger}}(0) \right] | 0 \rangle \right]
\\ \nonumber &\equiv& \left( g^{\mu\nu} - {q^\mu q^\nu \over q^2} \right) \cdot \Pi_{L}(s_0, M_B^2) + \cdots
\\ \nonumber &=& \left( g^{\mu\nu} - {q^\mu q^\nu \over q^2} \right) \times
\\ \nonumber && \Bigg(  \int^{s_0}_{0} {s \over 8 \pi^2} ~e^{-s/M_B^2}~ds - {1 \over 96 \pi^2} \langle g_c^2 GG \rangle
\\ \nonumber && ~~~ - {1 \over M_B^2} \left( {1\over12} m_u + {1\over12} m_d \right) \langle g_c\bar{q}\sigma Gq\rangle \Bigg) + \cdots
\, .
\end{eqnarray}
The sum rules obtained using the latter are the same:
\begin{equation}
%%%%%%%%%%%%%%%%%%%%%%%%%%%%%%%%%%%%%%%%%%%%%%%%%%%%%%%%%%%%%%%%%%%%%%%%%%%%%%
\Pi^{\mu\nu}_{L}(s_0, M_B^2) = \Pi^{\mu\nu}_{R}(s_0, M_B^2) \, .
\end{equation}
Very quickly, we find that the non-perturbative term related to the gluon condensate $\langle g_c^2 GG \rangle$ is non-zero.

\begin{figure}[hbtp]
\begin{center}
\subfigure[($k$)]{
\scalebox{0.15}{\includegraphics{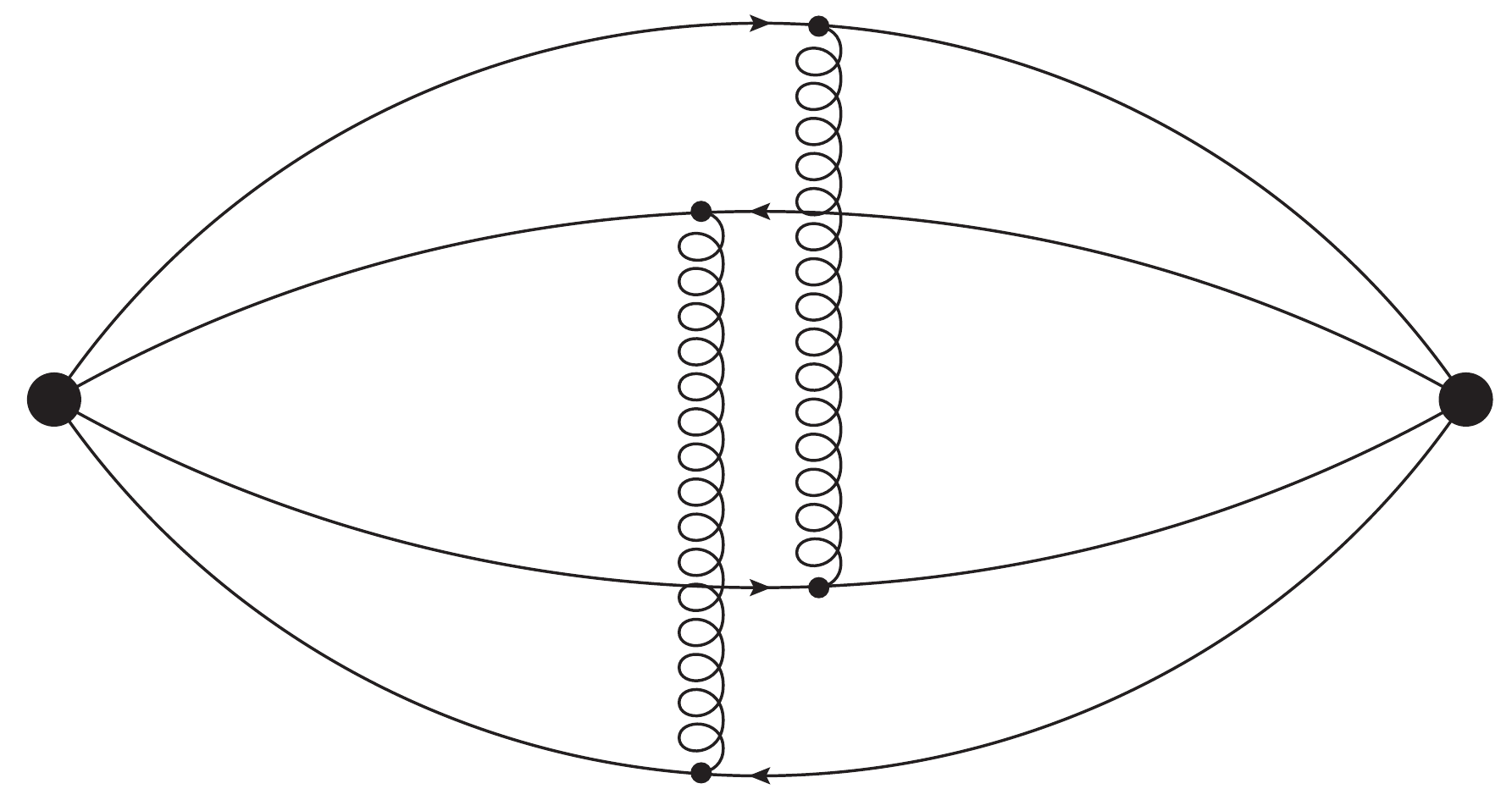}}}~~~~~
\subfigure[($l$)]{
\scalebox{0.15}{\includegraphics{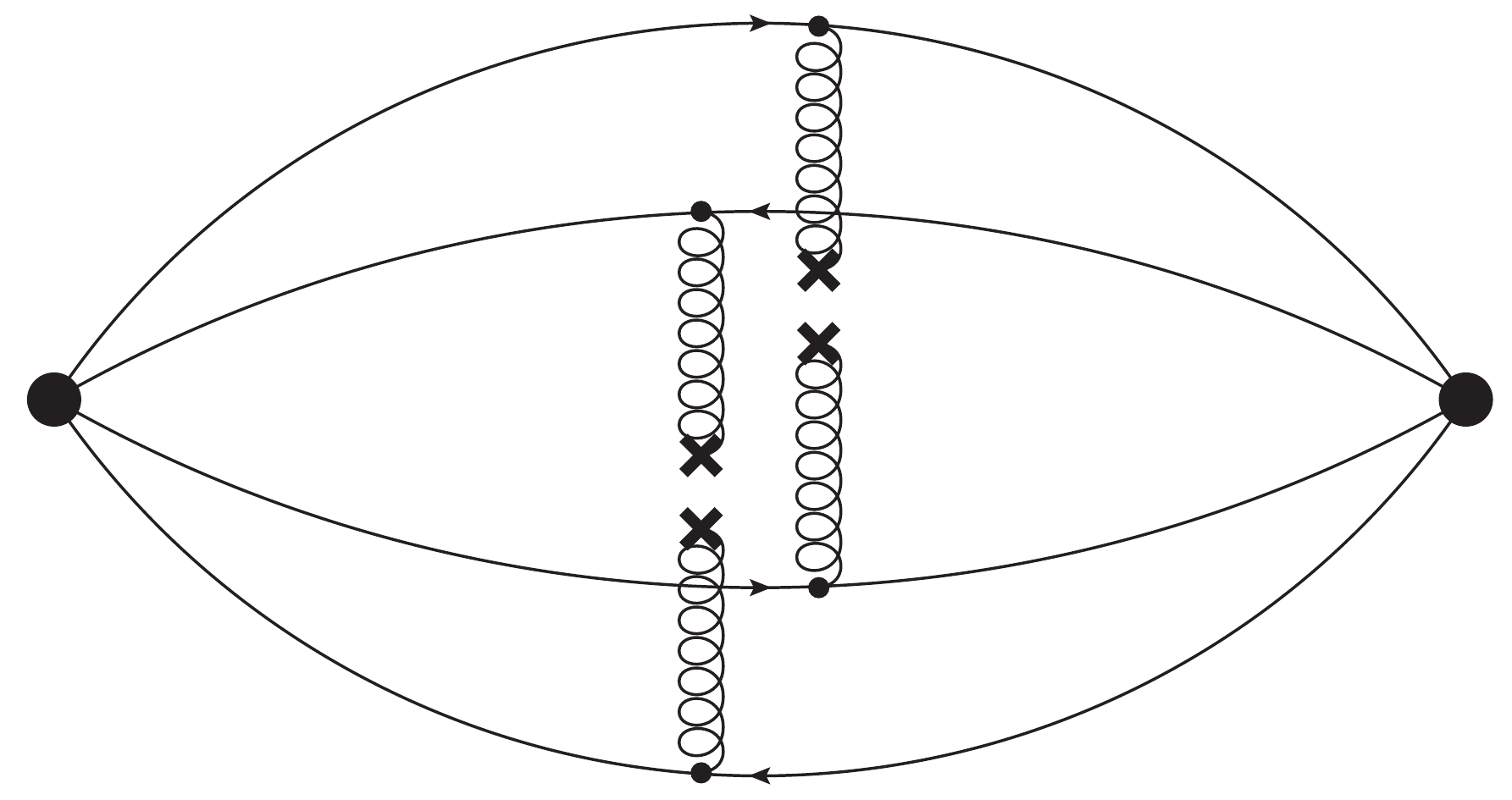}}}
\\[5mm]
$\Huge \bullet~\bullet~\bullet~~~\bullet~\bullet~\bullet$
\\[5mm]
\end{center}
\caption{Next-to-next-to-leading-order (NNLO) Feynman diagrams for tetraquark interpolating currents. They account for the residual strong interaction between the two color-singlet $\bar u d$ and $\bar d u$ pairs, but are not calculated in the present study.}
\label{fig:feynman3}
\end{figure}

Then it is interesting to find out why $\Pi_{LL}(s_0, M_B^2)$ given in Eq.~(\ref{eq:piLL}) does not contain such a term. To do this we write the expression for the two-point correlation function $\Pi^{\mu\nu}_{L}(x^2)$ before the Fourier-Borel transformations:
\begin{eqnarray}
%%%%%%%%%%%%%%%%%%%%%%%%%%%%%%%%%%%%%%%%%%%%%%%%%%%%%%%%%%%%%%%%%%%%%%%%%%%%%%
\Pi^{\mu\nu}_{L}(x^2) &\equiv& \langle 0 | \mathbb{T}\left[ J^\mu_{L}(x){J_{L}^{\nu,\dagger}}(0) \right] | 0 \rangle
\label{eq:piLmomentum}
\\ \nonumber &=& \left( - {g^{\mu\nu} \over 2 x^6} + {x^\mu x^\nu \over x^8} \right) \cdot {3 \over \pi^4}
\\ \nonumber && + \left( {g^{\mu\nu} \over 2 x^2} + {x^\mu x^\nu \over x^4} \right) \cdot { \langle g_c^2 GG \rangle \over 384 \pi^4}
+ \cdots
\, .
\end{eqnarray}
Hence, $\Pi^{\mu\nu}_{L}(x^2) \cdot \Pi^{\mu\nu}_{L}(x^2) \sim \Pi_{LL}(x^2)$ does not contain a non-zero $\langle g_c^2 GG \rangle$ term. Neither does its Fourier-Borel transformed version $\Pi_{LL}(s_0, M_B^2)$. Note that $\Pi_{LL}(x^2)$ does contain a non-zero $\langle g_c^2 GG \rangle^2$ term, but the contribution of this term is at the same level as Fig.~\ref{fig:feynman3}($l$), which we do not take into account in the present study (see discussions below).

Using $\mu^2_{\rm \overline {MS}} = s_0 = M_B^2 = 1$~GeV$^2$ and the QCD sum rule parameters listed in Eqs.~(\ref{eq:condensates}), we numerically calculate Eq.~(\ref{eq:piL}) to be:
\begin{eqnarray}
%%%%%%%%%%%%%%%%%%%%%%%%%%%%%%%%%%%%%%%%%%%%%%%%%%%%%%%%%%%%%%%%%%%%%%%%%%%%%%
&& \Pi_{L}(s_0 = M_B^2 = 1~{\rm GeV}^2)
\label{eq:piLnum}
\\ \nonumber &=& \Big( 3.3~[{\rm LO}] - 0.51~[{\rm D}^4] - 0.0086~[{\rm D}^6] \Big)
\cdot 10^{-3}~{\rm GeV}^{4} \, .
\end{eqnarray}
We find that the non-perturbative $\langle g_c^2 GG \rangle$ term is small compared to the leading-order perturbative term.

%
%%%%%%%%%%%%%%%%%%%%%%%%%%%%%%%%%%%%%%%%%%%%%%%%%%%%%%%%%%%%%%%%%%%%%%%%%%%%%%
\begin{figure}[hbtp]
\begin{center}
\includegraphics[width=0.48\textwidth]{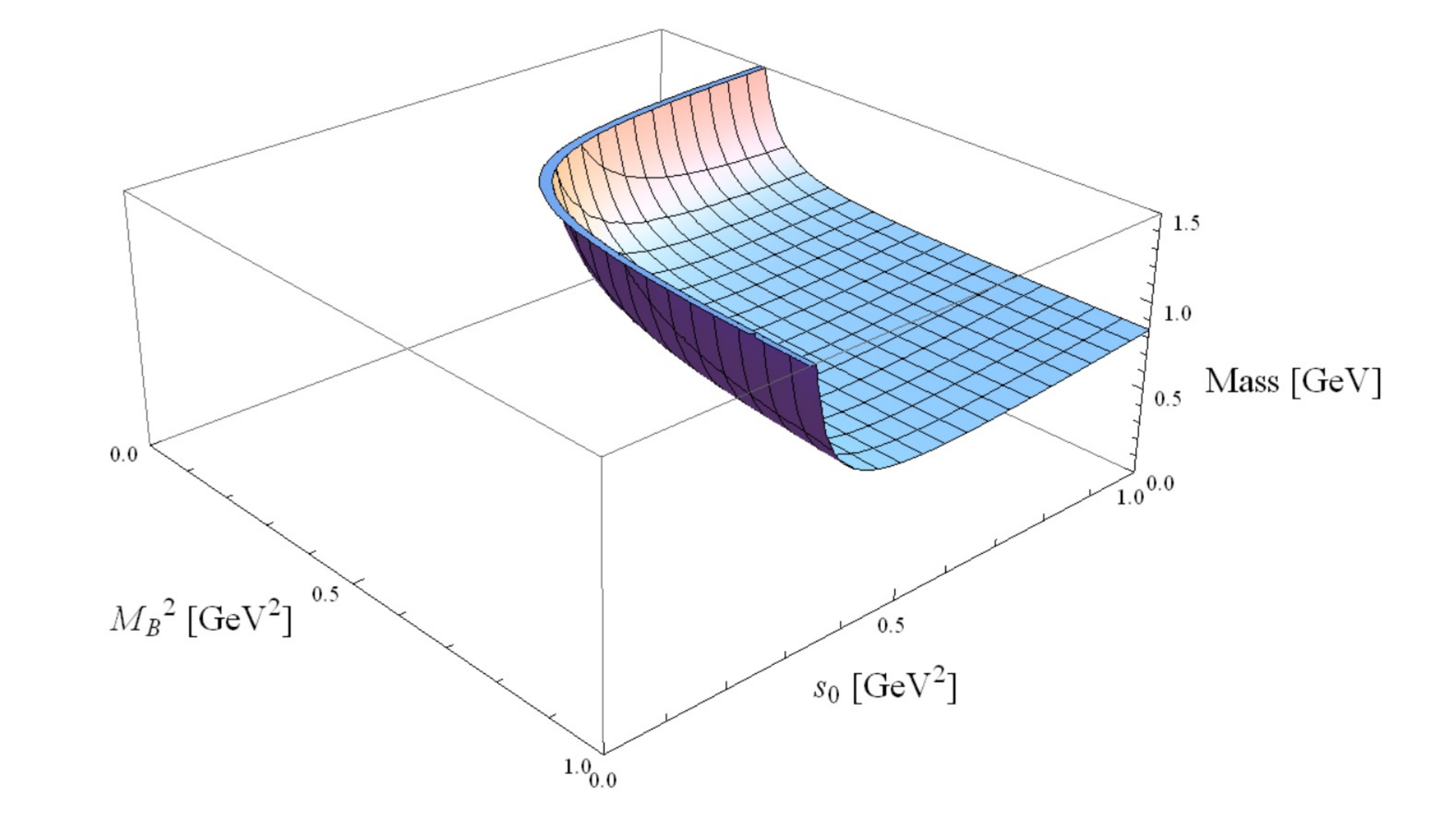}
\caption{The mass of the state $X_{L}$, coupled by the ``chiral'' quark-antiquark current $J_{L} = \bar{u}^a_L \gamma_\mu d^a_L$, as a function of the threshold value $s_0$ and the Borel mass $M_B$ simultaneously.}
\label{fig:JLmass}
\end{center}
\end{figure}
%%%%%%%%%%%%%%%%%%%%%%%%%%%%%%%%%%%%%%%%%%%%%%%%%%%%%%%%%%%%%%%%%%%%%%%%%%%%%%
%

However, let us assume that $J_{L}$ couples to some ``chiral'' state $X_L$ (different from $X_{LL}$ studied in the previous subsection), and follow the procedures used in the previous subsection to calculate its mass. As shown in Fig.~\ref{fig:JLmass} as a function of $s_0$ and $M_B$ simultaneously, this mass has a global minimum about 800~MeV. Therefore, although the non-perturbative $\langle g_c^2 GG \rangle$ term is small, its contribution is not small. This indicates that the tetraquark states $X_{LL/LR/RL/RR}$ can not simply fall apart into two quark-antiquark pairs, so it is still the confinement that binds the four quarks together.

Then we investigate the interactions between the two color-singlet $\bar u d$ and $\bar d u$ pairs, including the residual strong interaction and the direct electromagnetic interaction. In the present study we do not calculate the complicated next-to-next-to-leading-order (NNLO) Feynman diagrams shown in Fig.~\ref{fig:feynman3}, which account for the residual strong interaction between the $\bar u d$ and $\bar d u$ pairs. However, it is well known that the binding energy of the deuteron is $E_{strong} = -2.22$~MeV, which is also caused by the nuclear force {\it aka} residual strong interaction. Besides, we can use $r \approx 1$~fm to estimate the direct electromagnetic interaction between the $\bar u d$ and $\bar d u$ pairs to be $E_{electromagnetic} \approx -1.4$~MeV.

Therefore, the total binding energy between the two color-singlet $\bar u d$ and $\bar d u$ pairs inside the tetraquark states $X_{LL/LR/RL/RR}$ is at the MeV level (given that these two pairs are bound), that is at the same level as the non-perturbative QCD terms contained in Eq.~(\ref{eq:piLL}). Altogether, we can roughly estimate the binding energy among the four bare quarks to be also at the MeV level (given that these four bare quarks are bound). This makes it possible to produce an attractive potential, although much smaller than the sombrero-shape potential accounting for the spontaneous breaking of chiral symmetry, it can also provide a metastable equilibrium for the four bare quarks. This is the reason why we call $X_{LL/LR/RL/RR}$ the tetraquark states composed of ``bare quarks''. Assuming their existence, we shall study their production and decay mechanisms in the next three subsections.

%
%=====================================================================================
%=====================================================================================
\subsection{(Self-)Interactions}\label{sec:confine}
%=====================================================================================
%=====================================================================================
%

Assuming the existence of the tetraquark states $X_{LL/LR/RL/RR}$ composed of bare quarks, we study their interactions with the nucleus as well as their self-interactions in this subsection.

Firstly, we study the correlation functions between the tetraquark currents $J_{LL/LR/RL/RR}$ and the quark condensates. There are no interactions between $J_{LL}$ and the quark condensates:
\begin{eqnarray}
&& \Pi_{LL-\langle \bar{q}q\rangle}(q^2)
\label{eq:piLLQQ1}
\\ \nonumber &\equiv& i\int d^4x~e^{iqx}~\langle 0 | \mathbb{T}\left[ J_{LL}(x) \Big(\bar u(0) u(0)\Big) \right] | 0 \rangle
\\ \nonumber &=& 0 \, ,
\\
&& \Pi_{LL-\langle \bar{q}q\rangle^2}(q^2)
\label{eq:piLLQQ2}
\\ \nonumber &\equiv& i\int d^4x~e^{iqx}~\langle 0 | \mathbb{T}\left[ J_{LL}(x) \Big(\bar u(0) u(0) \bar d(0) d(0)\Big) \right] | 0 \rangle
\\ \nonumber &=& 0 \, ,
\end{eqnarray}
while $J_{LR}$ interacts closely with the quark condensates:
\begin{eqnarray}
&& \Pi_{LR-\langle \bar{q}q\rangle}(q^2)
\label{eq:piLRQQ1}
\\ \nonumber &\equiv& i\int d^4x~e^{iqx}~\langle 0 | \mathbb{T}\left[ J_{LR}(x) \Big(\bar u(0) u(0)\Big) \right] | 0 \rangle
\\ \nonumber &=& {3 \langle \bar{q}q\rangle \over 16 \pi^2} q^2 \ln({-q^2 \over \mu^2_{\rm \overline {MS}}}) + {\langle g_c^2 GG \rangle \langle\bar{q}q\rangle \over 64 \pi^2} {1 \over q^2}
+ {8 \pi \alpha_s \langle\bar{q}q\rangle^3 \over 9} {1 \over q^4} ,
\\
&& \Pi_{LR-\langle \bar{q}q\rangle^2}(q^2)
\label{eq:piLRQQ2}
\\ \nonumber &\equiv& i\int d^4x~e^{iqx}~\langle 0 | \mathbb{T}\left[ J_{LR}(x) \Big(\bar u(0) u(0) \bar d(0) d(0)\Big) \right] | 0 \rangle
\\ \nonumber &=& {1 \over 491520 \pi^6} q^8 \ln({-q^2 \over \mu^2_{\rm \overline {MS}}}) + {\langle g_c^2 GG \rangle \over 24576 \pi^6} q^4 \ln({-q^2 \over \mu^2_{\rm \overline {MS}}})
\\ \nonumber && - {\langle \bar{q}q\rangle^2 \over 96 \pi^2} q^2 \ln({-q^2 \over \mu^2_{\rm \overline {MS}}}) - { \langle\bar{q}q\rangle \langle g_c\bar{q}\sigma Gq\rangle \over 96 \pi^2} \ln({-q^2 \over \mu^2_{\rm \overline {MS}}})
\\ \nonumber && + ~{1 \over q^2}\left( - {\langle g_c\bar{q}\sigma Gq\rangle^2 \over 768 \pi^2} - {\langle g_c^2 GG \rangle \langle\bar{q}q\rangle^2 \over 1152 \pi^2} \right)
\\ \nonumber && + ~{1 \over q^4}\left( - {4 \pi \alpha_s \langle\bar{q}q\rangle^4 \over 81} - {\langle g_c^2 GG \rangle \langle\bar{q}q\rangle\langle g_c\bar{q}\sigma Gq\rangle \over 2304 \pi^2} \right) \, .
\end{eqnarray}
In the above expressions, we have used zero current quark masses ($m_{u/d} \rightarrow 0$). The results for $J_{RR}$ are the same as $J_{LL}$, and those for $J_{RL}$ are the same as $J_{LR}$. Eqs.~(\ref{eq:piLLQQ1}-\ref{eq:piLRQQ2}) indicate that the two states $X_{LL}$ and $X_{RR}$ are possibly produced more easily than the other two $X_{LR}$ and $X_{RL}$.

Secondly, because the diagonal correlation function $\Pi_{LL}$ given in Eq.~(\ref{eq:piLL}) only contains the very limited non-perturbative QCD contributions, it is interesting to compare it with the off-diagonal correlation functions:
\begin{equation}
{\bf \Pi} = \left(\begin{array}{cccc}
\Pi_{LL}    & \Pi_{LL-LR} & \Pi_{LL-RL} & \Pi_{LL-RR}
\\
\Pi_{LR-LL} & \Pi_{LR}    & \Pi_{LR-RL} & \Pi_{LR-RR}
\\
\Pi_{RL-LL} & \Pi_{RL-LR} & \Pi_{RL}    & \Pi_{RL-RR}
\\
\Pi_{RR-LL} & \Pi_{RR-LR} & \Pi_{RR-RL} & \Pi_{RR}
\end{array}\right) \, ,
\label{eq:matrix}
\end{equation}
where
\begin{equation}
\Pi_{i-j}(q^2) \equiv i\int d^4x~e^{iqx}~\langle 0 | \mathbb{T}\left[ J_{i}(x) J_{j}^\dagger(0) \right] | 0 \rangle \, .
\end{equation}
All the matrix elements have been given in Appendix~\ref{app:sumrule}, where the four diagonal elements are the same,
\begin{equation*}
\Pi_{LL} = \Pi_{LR} = \Pi_{RL} = \Pi_{RR} \, ,
\end{equation*}
and the six off-diagonal elements are divided into two groups,
\begin{eqnarray*}
%%%%%%%%%%%%%%%%%%%%%%%%%%%%%%%%%%%%%%%%%%%%%%%%%%%%%%%%%%%%%%%%%%%%%%%%%%%%%%
&& \Pi_{LL-LR} = \Pi_{LL-RL} = \Pi_{RR-LR} = \Pi_{RR-RL} \, ,
\\ && ~~~~~~~~~~~~~~~\, \Pi_{LL-RR} = \Pi_{LR-RL} \, .
\end{eqnarray*}

Using $\mu^2_{\rm \overline {MS}} = s_0 = M_B^2 = 1$~GeV$^2$ and the QCD sum rule parameters listed in Eqs.~(\ref{eq:condensates}), we numerically calculate Eq.~(\ref{eq:matrix}) to be:
\begin{equation}
{\bf \Pi} = \left(\begin{array}{cccc}
2.0    & -67    & -67    & 20
\\
-67    & 2.0    & 20     & -67
\\
-67    & 20     & 2.0    & -67
\\
20     & -67    & -67    & 2.0
\end{array}\right) \cdot 10^{-9}~{\rm GeV}^{10} \, .
\label{eq:matrix1}
\end{equation}
The off-diagonal correlation functions among $J_{LL/LR/RL/RR}$ are quite large, suggesting that the four states $X_{LL/LR/RL/RR}$ are highly mixed together.

However, if the chiral symmetry is restored so that the quark condensate vanishes ($\langle \bar q q \rangle \rightarrow 0$), Eq.~(\ref{eq:matrix}) becomes (the other parameters remain unchanged):
\begin{equation}
{\bf \Pi} = \left(\begin{array}{cccc}
1.7    & 0      & 0      & 0
\\
0      & 1.7    & 0      & 0
\\
0      & 0      & 1.7    & 0
\\
0      & 0      & 0      & 1.7
\end{array}\right) \cdot 10^{-9}~{\rm GeV}^{10} \, .
\label{eq:matrix2}
\end{equation}
The four diagonal elements are nearly the same, but the six off-diagonal elements all vanish. This suggests that $X_{LL/LR/RL/RR}$ are different states, and there can exist as many as four degenerate states.

From Eqs.~(\ref{eq:piLRQQ1}), (\ref{eq:piLRQQ2}), (\ref{eq:matrix1}), and (\ref{eq:matrix2}), we quickly realize that the production/existence of the tetraquark states $X_{LL/LR/RL/RR}$ is closely related to the chiral symmetry restoration, which will be further studied in the next subsection.

%
%=====================================================================================
%=====================================================================================
\subsection{Production Mechanism}\label{sec:production}
%=====================================================================================
%=====================================================================================
%

Assuming the existence of the tetraquark states $X_{LL/LR/RL/RR}$ composed of bare quarks, we study their production mechanism in this subsection. We only consider the state $X_{LL}$ coupled by the current $J_{LL}$, while the other three can be similarly investigated.

There are three well-known nuclear decay processes, the $\alpha$-, $\beta$-, and $\gamma$-decays. During these processes, the extra energy of the nucleus is (partly) taken away by an $\alpha$ particle, a $down$ quark changing to an $up$ quark and at the same time emitting an electron and a neutrino, and a photon, respectively. Besides them, there are a tremendous amount of sea quarks inside the proton and neutron, so it is interesting to investigate whether these sea quarks can also take away the extra energy, and induce a new type of nuclear decay process.

%
%%%%%%%%%%%%%%%%%%%%%%%%%%%%%%%%%%%%%%%%%%%%%%%%%%%%%%%%%%%%%%%%%%%%%%%%%%%%%%
\begin{figure}[hbtp]
\begin{center}
\includegraphics[width=0.45\textwidth]{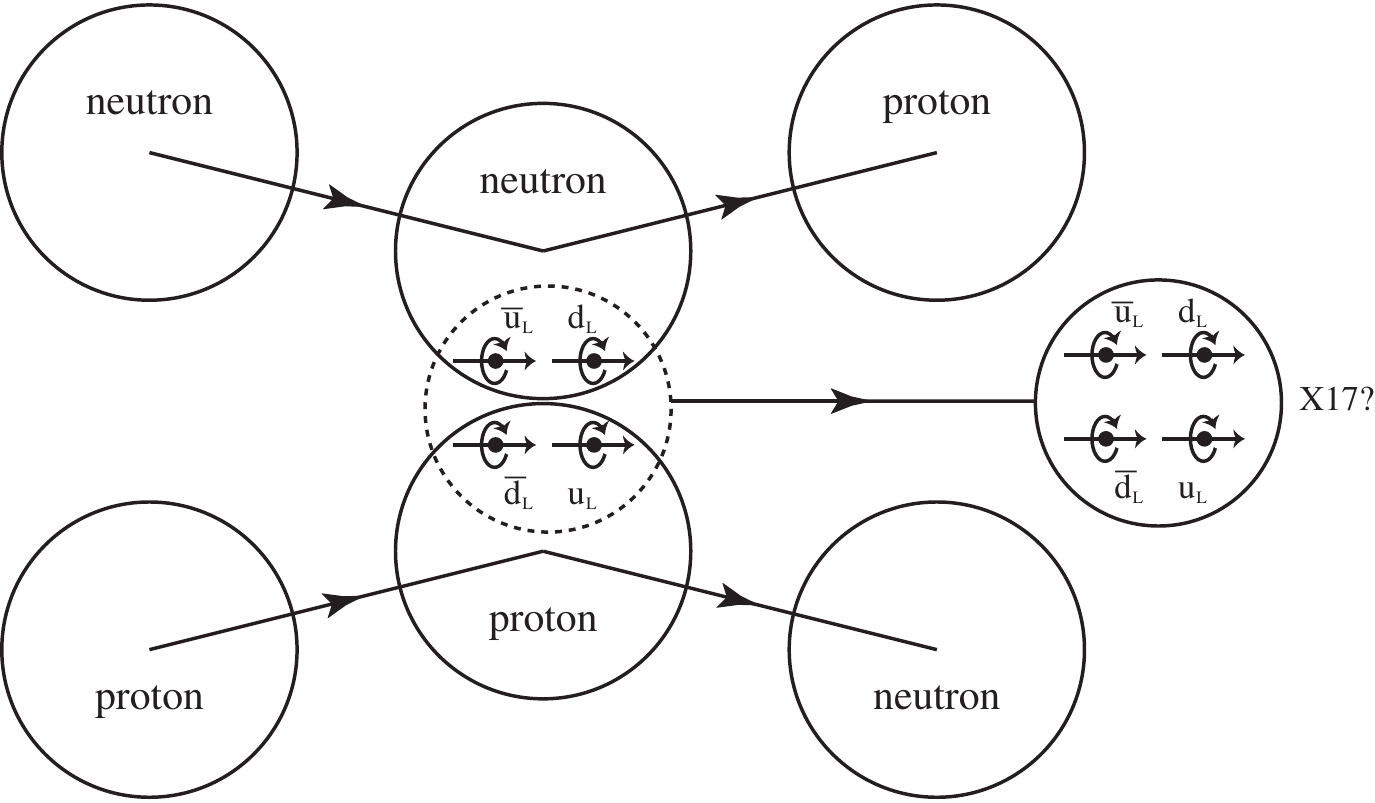}
\caption{Production mechanism of the tetraquark state $X_{LL}$ coupled by the tetraquark current $J_{LL} = \bar{u}^a_L \gamma_\mu d^a_L~\bar{d}_L^b \gamma^\mu u_L^b$.}
\label{fig:production}
\end{center}
\end{figure}
%%%%%%%%%%%%%%%%%%%%%%%%%%%%%%%%%%%%%%%%%%%%%%%%%%%%%%%%%%%%%%%%%%%%%%%%%%%%%%
%

To study this, we consider the hypothetical nuclear decay process depicted in Fig.~\ref{fig:production}, that is also the production process of the tetraquark state $X_{LL}$ coupled by $J_{LL}$. The conditions for this process to happen are:
\begin{itemize}

\item The two color-singlet $\bar u d$ and $\bar d u$ pairs must acquire enough energy, but not too much so that they are not bound any more;

\item Their spins and chiralities must satisfy certain conditions, so that non-perturbative QCD effects do not contribute much;

\item They must meet each other at boundaries of the proton and neutron.

\end{itemize}
Then we use the general quantum tunneling formula to estimate the probability for the $\bar u d$ and $\bar d u$ pairs to travel out of the neutron and proton together:
\begin{equation}
\left({\mathcal P}\right)^2 \approx \left( {16 E (V_0 - E) \over V_0^2} e^{ - {2a \over \hbar} \sqrt{2 \mu (V_0 - E)} } \right)^2 \approx 25\% \, ,
\end{equation}
where we have used $\mu \approx E \approx {M_X / 2} \approx 8.5$~MeV and $a \approx 1$~fm, and assumed the potential barrier to be $V_0 \approx T_c = 154 \pm 9$~MeV~\cite{Dick:2015twa,Bazavov:2017dus}.

Therefore, it is possible for the process depicted in Fig.~\ref{fig:production} to happen, although we are not able to estimate its production rate. We argue that the Atomki experiments~\cite{Krasznahorkay:2015iga,Krasznahorkay:2019lyl} might observe such a nuclear decay process. This process is similar to the evaporation or the sublimation process in some aspects: the color-singlet $\bar u d$ and $\bar d u$ (sea-quark) pairs are tightly confined inside the neutron and proton, while it is still possible for them to escape after absorbing enough energy, even when the critical temperature of chiral symmetry restoration is not reached.

%
%=====================================================================================
%=====================================================================================
\subsection{Decay Mechanism}\label{sec:decay}
%=====================================================================================
%=====================================================================================
%

Assuming the existence of the tetraquark states $X_{LL/LR/RL/RR}$ composed of bare quarks, we study their decay mechanism in this subsection. We only consider the two states $X_{LL}$ and $X_{LR}$ coupled by the currents $J_{LL}$ and $J_{LR}$, while the other two can be similarly investigated.

\begin{figure*}[hbtp]
\begin{center}
\includegraphics[width=0.40\textwidth]{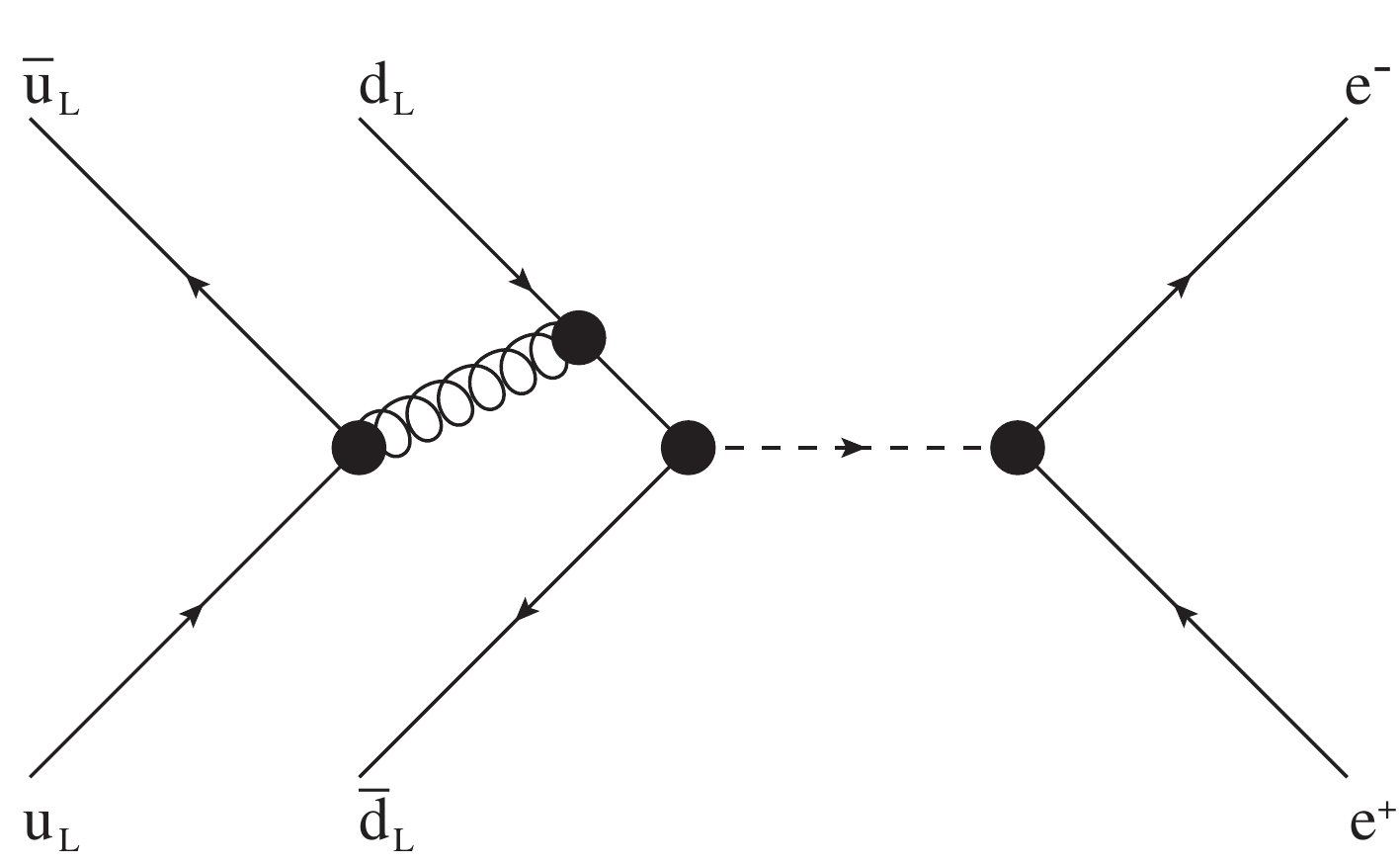}
~~~~~~~~~~
\includegraphics[width=0.40\textwidth]{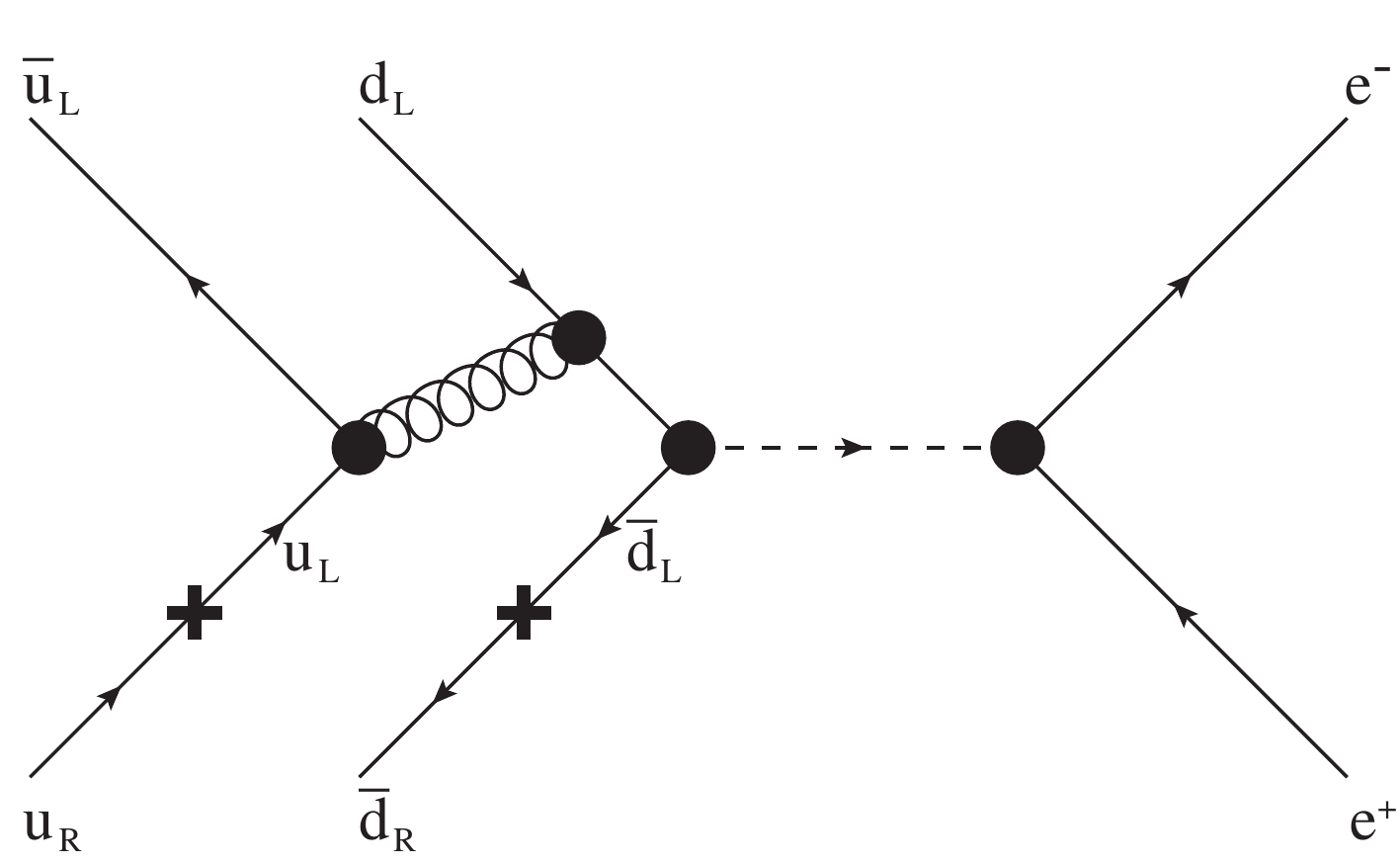}
\caption{Decay mechanism of the tetraquark states $X_{LL}$ (left) and $X_{LR}$ (right), coupled by the tetraquark currents $J_{LL} = \bar{u}^a_L \gamma_\mu d^a_L~\bar{d}_L^b \gamma^\mu u_L^b$ and $J_{LR} = \bar{u}^a_L \gamma_\mu d^a_L~\bar{d}_R^b \gamma^\mu u_R^b$, respectively.}
\label{fig:decay}
\end{center}
\end{figure*}

To do this we use Eqs.~(\ref{fierzLL}-\ref{fierzLR}) previously obtained using the Fierz transformation,
\begin{eqnarray*}
J_{LL} &=& \bar{u}^a_L \gamma_\mu d^a_L~\bar{d}_L^b \gamma^\mu u_L^b \rightarrow \bar{u}^a_L \gamma_\mu u_L^b~\bar{d}_L^b \gamma^\mu d^a_L \, ,
\\ J_{LR} &=& \bar{u}^a_L \gamma_\mu d^a_L~\bar{d}_R^b \gamma^\mu u_R^b \rightarrow -2~\bar{u}^a_L u_R^b~\bar{d}_R^b d^a_L \, ,
\end{eqnarray*}
to study the two hypothetical decay processes depicted in Fig.~\ref{fig:decay}:
\begin{itemize}

\item Left panel: when the state $X_{LL}$ decays, first the $u_L$ and $\bar u_L$ pair annihilates each other into a gluon, then the $d_L$ and $\bar d_L$ pair absorbs this gluon and annihilates each other into a photon, and finally this photon decays to an electron and a positron.

    There are both strong and electromagnetic processes in this decay. As we know, $\Gamma(\rho \to e^+ e^-) = 7.04 \pm 0.06$~keV and $\Gamma(\omega \to e^+ e^-) = 0.60 \pm 0.02$~keV~\cite{pdg}, etc. Accordingly, we roughly estimate the width of $X_{LL}$ to be at the eV level through, %using  through:
    \begin{eqnarray}
    && \Gamma(X_{LL} \to e^+ e^-)
    %\\ \nonumber &\sim& {\sigma(e^+ e^- \to \rho \rho) \over \sigma(e^+ e^- \to hadrons, \, s = m_\rho)} \times \Gamma(\rho \to e^+ e^-)
    \\ \nonumber &\sim& {\sigma(e^+ e^- \to \rho + others) \over \sigma(e^+ e^- \to hadrons; \, s = m_\rho^2)} \times \Gamma(\rho \to e^+ e^-) \, ,
    %\\ \nonumber &\sim& {0.35 nb \over 5 \times 10^{-4} mb} \times 7 keV
    %\\ \nonumber &\sim& {\rm eV}
    \end{eqnarray}
    and
    \begin{eqnarray}
    \nonumber &\sim& {\sigma(e^+ e^- \to \omega + others) \over \sigma(e^+ e^- \to hadrons; \, s = m_\omega^2)} \times \Gamma(\omega \to e^+ e^-) \, .
    \end{eqnarray}

\item Right panel: the decay process $X_{LR} \to e^+ e^-$ is a bit different from the above one. Here the annihilation of the $u_R$ and $\bar u_L$ pair is suppressed, and so does the annihilation of the $d_L$ and $\bar d_R$ pair. Therefore, the width of $X_{LR}$ is significantly smaller than $X_{LL}$, with the factor possibly being
    \begin{eqnarray}
    \left({m_u m_d \over s_0}\right)^2 \sim 10^{-3} \, ,
    \end{eqnarray}
    or even smaller. Here $s_0 = 500$~MeV$^2$ is the threshold value previously used in Sec.~\ref{sec:mass}.

\end{itemize}
Note that the above estimations are rather rough, and can differ much from the realistic values. Besides, in the present study we do not consider the weak interaction, which makes the two states $X_{LL}$ and $X_{RR}$ slightly different from each other.

%
%=====================================================================================
%=====================================================================================
\section{Summary and Conclusion}
\label{sec:summary}
%=====================================================================================
%=====================================================================================
%

In this paper we investigate the possible existence of tetraquark states composed of bare quarks, and study the possible assignment of the $X17$ as such a state. Based on our previous QCD sum rule studies~\cite{Chen:2006hy,Chen:2006zh,Chen:2007xr,Jiao:2009ra,Dong:2020okt}, we find two interpolating tetraquark currents
\begin{eqnarray*}
J_{LL} &=& \bar{u}^a_L \gamma_\mu d^a_L~\bar{d}_L^b \gamma^\mu u_L^b \, ,
\\ J_{LR} &=& \bar{u}^a_L \gamma_\mu d^a_L~\bar{d}_R^b \gamma^\mu u_R^b \, ,
\end{eqnarray*}
as well as their chiral partners
\begin{eqnarray*}
J_{RL} &=& \bar{u}^a_R \gamma_\mu d^a_R~\bar{d}_L^b \gamma^\mu u_L^b \, ,
\\ J_{RR} &=& \bar{u}^a_R \gamma_\mu d^a_R~\bar{d}_R^b \gamma^\mu u_R^b \, .
\end{eqnarray*}
We calculate their two-point correlation functions using the method of operator product expansion, and the results turn out to be the same, indicating there can exist as many as four degenerate states.

We find that all the non-perturbative QCD terms vanish in the limit of zero current quark masses, so non-perturbative QCD effects do not contribute much to them. As we know, masses of hadrons are always significantly larger than masses of their valence quarks inside, and the extra masses are mainly responsible by the non-perturbative nature of strong interaction in the low energy region. Considering this, our results indicate the possible existence of tetraquark states that have abnormal masses significantly smaller than normal hadron masses at the GeV level.

We assume that the tetraquark currents $J_{LL/LR/RL/RR}$ couple to the tetraquark states $X_{LL/LR/RL/RR}$, and use the method of QCD sum rules to estimate their masses to be
\begin{equation*}
M_{X} = 17.3^{+1.4}_{-1.7}~{\rm MeV} \, .
\end{equation*}
Note that this estimation should be treated with caution, because the method of QCD sum rules is actually a non-perturbative method, and in this case there are not so large non-perturbative contributions.

The above abnormally small mass value suggests it possible to interpret the $X17$ as a tetraquark state composed of four bare quarks. Since it can be separated into two color-singlet quark-antiquark pairs, we also study these two components using the same method of QCD sum rules, but find both of them to have a mass minimum about 800~MeV. Therefore, this tetraquark state can not simply fall apart into two quark-antiquark pairs, and it is still the confinement that binds the four quarks together. Besides, our results indicate that the production/existence of this tetraquark state is closely related to the chiral symmetry restoration.

Based on the tetraquark picture, we study the production process depicted in Fig.~\ref{fig:production}. We argue that the Atomki experiments~\cite{Krasznahorkay:2015iga,Krasznahorkay:2019lyl} might observe a new type of nuclear decay process, where sea quarks take away the extra energy of the nucleus. This process is similar to the evaporation or the sublimation process in some aspects: the color-singlet $\bar u d$ and $\bar d u$ (sea-quark) pairs are tightly confined inside the neutron and proton, while it is still possible for them to escape after absorbing enough energy, even when the critical temperature of chiral symmetry restoration is not reached.

Based on the tetraquark picture, we also study the decay processes depicted in Fig.~\ref{fig:decay}. The widths of $X_{LL}$ and $X_{RR}$ are roughly estimated to be at the eV level, while those of $X_{LR}$ and $X_{RL}$ to be significantly smaller. Therefore, we obtain a unique feature of our tetraquark assignment that we predict two almost degenerate states with significantly different widths. The Atomki experiment measured the width of $X17$ to be $\Gamma \approx 3.9 \times 10^{-5}$~eV~\cite{Krasznahorkay:2019lyl}, so it may be interpreted as $X_{LR/RL}$ within our framework, while $X_{LL/RR}$ still need to be observed. Accordingly, we propose to search for these two degenerate structures simultaneously in order to verify the present tetraquark assignment.

Before ending this paper, we would like to briefly discuss whether this tetraquark state is made by two quark-antiquark pairs ($[\bar q q][\bar q q]$) or one diquark-antidiquark pair ($[qq][\bar q \bar q]$). In the present study we have used the former, but it can be transformed to the latter through the Fierz transformation:
\begin{eqnarray}
J_{LL} &=& \bar{u}^a_L \gamma_\mu d^a_L~\bar{d}_L^b \gamma^\mu u_L^b \rightarrow 2~\bar{u}^a_L \mathbb{C} \bar d_L^{bT}~u_L^{bT} \mathbb{C} d^a_L \, ,
\\ \nonumber J_{LR} &=& \bar{u}^a_L \gamma_\mu d^a_L~\bar{d}_R^b \gamma^\mu u_R^b \rightarrow -\bar{u}^a_L \gamma_\mu \mathbb{C} \bar d_R^{bT}~u_R^{bT} \mathbb{C} \gamma_\mu d^a_L \, ,
\\ \nonumber J_{RL} &=& \bar{u}^a_R \gamma_\mu d^a_R~\bar{d}_L^b \gamma^\mu u_L^b \rightarrow -\bar{u}^a_R \gamma_\mu \mathbb{C} \bar d_L^{bT}~u_L^{bT} \mathbb{C} \gamma_\mu d^a_R \, ,
\\ \nonumber J_{RR} &=& \bar{u}^a_R \gamma_\mu d^a_R~\bar{d}_R^b \gamma^\mu u_R^b \rightarrow 2~\bar{u}^a_R \mathbb{C} \bar d_R^{bT}~u_R^{bT} \mathbb{C} d^a_R \, ,
\end{eqnarray}
where $\mathbb{C}$ is the charge-conjugation operator. Hence, the latter diquark-antidiquark combination is also possible within our framework, while some of the discussions in Sec.~\ref{sec:binding} need to be accordingly modified.

Finally, we give a note here. In this paper we study the possible assignment of the $X17$ as a tetraquark state composed of bare quarks. There have been many other possible explanations~\cite{Feng:2016jff,Feng:2016ysn,Gu:2016ege,Neves:2016nek,Kahn:2016vjr,Fayet:2016nyc,Dror:2017nsg,Das:2019ycx,Jentschura:2020zlr,Alexander:2016aln,Battaglieri:2017aum,Liang:2016ffe,Jia:2016uxs,Kitahara:2016zyb,Ellwanger:2016wfe,Chen:2016tdz,Kozaczuk:2016nma,Chiang:2016cyf,Dror:2017ehi,Kozaczuk:2017per,Jia:2018mkc,DelleRose:2018pgm,DelleRose:2018eic,Hati:2020fzp,Bauer:2017ris,Alves:2017avw,Kirpichnikov:2020tcf,Dusaev:2020gxi,Zhang:2017zap,DelleRose:2017xil,Fornal:2017msy,Jiang:2018uhs,Chen:2019ivz,Nam:2019osu,Pulice:2019xel,Krasnikov:2019dgh,Koch:2020ouk,Wong:2020hjc,Veselsky:2020ewb}, and the present one provides an alternative assignment. In this assignment the $X17$ behaves as the ``singular point'' of the QCD sum rule method. However, it could be the case that: a) the realistic ``singular point'' of the strong interaction does not exist, and b) the realistic one does exist, but different from the present QCD sum rule ``singular point''. To verify this, further experimental studies and other theoretical studies are definitely needed.

\appendix

\begin{widetext}
\section{QCD sum rule expressions}
\label{app:sumrule}

In this appendix we list the QCD sum rule expressions calculated in this paper. Our calculations are partly performed using the software $Mathematica$ with the package $FeynCalc$~\cite{Shtabovenko:2020gxv}.

First we give a more detailed expression for the sum rules obtained using the tetraquark current $J_{LL}$:
\begin{eqnarray}
%%%%%%%%%%%%%%%%%%%%%%%%%%%%%%%%%%%%%%%%%%%%%%%%%%%%%%%%%%%%%%%%%%%%%%%%%%%%%%
&& \Pi_{LL}(s_0, M_B^2) = \mathcal{B}_{q^2 \rightarrow M_B^2}\Pi_{LL}(q^2) \equiv \mathcal{B}_{q^2 \rightarrow M_B^2} \left[ i\int d^4x~e^{iqx}~\langle 0 | \mathbb{T}\left[ J(x)_{LL}{J_{LL}^\dagger}(0) \right] | 0 \rangle \right]
\\ \nonumber &=& \int^{s_0}_{0} \Bigg( {s^4 \over 81920 \pi^6}  \left( 1 + {4\alpha_s \over \pi} \right)
+ {s^3 \over 2048 \pi^6} \Big( -m_u^2 - m_d^2 \Big)
\\ \nonumber && ~~~~~~~~ + {s^2 \over 256 \pi^4} \Bigg( \Big( m_u + m_d \Big) \langle\bar{q}q\rangle + {3\over4\pi^2} \Big( m_u^4 + 4 m_u^2 m_d^2 + m_d^4 \Big) \Bigg)
\\ \nonumber && ~~~~~~~~ + {s \over 256 \pi^4}   \Bigg( \Big( m_u + m_d \Big) \langle g_c\bar{q}\sigma Gq\rangle - \Big( 6 m_u^3 + 12 m_u^2 m_d + 12 m_u m_d^2 + 6 m_d^3 \Big) \langle\bar{q}q\rangle - {9\over\pi^2} \Big( m_u^4 m_d^2 + m_u^2 m_d^4 \Big) \Bigg)
\\ \nonumber && ~~~~~~~~ + {1 \over 128 \pi^4} \Bigg( \Big( 6 m_u^4 m_d + 12 m_u^3 m_d^2 + 12 m_u^2 m_d^3 + 6 m_u m_d^4 \Big) \langle\bar{q}q\rangle + \Big( 2 m_u^2 + 8 m_u m_d + 2 m_d^2 \Big) \pi^2 \langle\bar{q}q\rangle^2
\\ \nonumber && ~~~~~~~~~~~~~~~~~~~~~~~ - \Big( m_u^3 +2m_u^2m_d +2m_um_d^2 + m_d^3 \Big) \langle g_c\bar{q}\sigma Gq\rangle + {9\over2\pi^2}m_u^4m_d^4 \Bigg) \Bigg)~e^{-s/M_B^2}~ds
\\ \nonumber &&  + {1 \over 128 \pi^4} \Bigg( - \Big( 6 m_u^4 m_d^3 + 6 m_u^3 m_d^4 \Big) \langle\bar{q}q\rangle - \Big( 8 m_u^3 m_d + 8 m_u^2 m_d^2 + 8 m_u m_d^3 \Big) \pi^2 \langle\bar{q}q\rangle^2
\\ \nonumber && ~~~~~~~~~~~~~ + \Big( m_u^4m_d + 2m_u^3m_d^2 + 2m_u^2m_d^3 + m_um_d^4 \Big) \langle g_c\bar{q}\sigma Gq\rangle + \Big( {2\over3}m_u^2 + {8\over3}m_um_d + {2\over3}m_d^2 \Big) \pi^2 \langle\bar{q}q\rangle\langle g_c\bar{q}\sigma Gq\rangle \Bigg)
\\ \nonumber &&  - {1 \over 128 \pi^4 M_B^2} \Bigg( \Big( 2 m_u^4 m_d^2 +8 m_u^3 m_d^3 + 2 m_u^2 m_d^4 \Big) \pi^2 \langle\bar{q}q\rangle^2 + \Big( {8\over3} m_u^2 m_d + {8\over3} m_u m_d^2 \Big) \pi^4 \langle\bar{q}q\rangle^3
\\ \nonumber && ~~~~~~~~~~~~~ - \Big( m_u^4m_d^3 + m_u^3m_d^4 \Big) \langle g_c\bar{q}\sigma Gq\rangle + \Big( {1\over18}m_u^2 + {2\over9}m_um_d + {1\over18}m_d^2 \Big) \pi^2 \langle g_c\bar{q}\sigma Gq\rangle^2
\\ \nonumber && ~~~~~~~~~~~~~ - \Big( {8\over3} m_u^3 m_d + {8\over3} m_u^2 m_d^2 + {8\over3} m_u m_d^3 \Big) \pi^2 \langle\bar{q}q\rangle\langle g_c\bar{q}\sigma Gq\rangle \Bigg) + \cdots
\, .
\end{eqnarray}
The sum rules obtained using the other three tetraquark currents $J_{LR/RL/RR}$ are the same:
\begin{equation}
%%%%%%%%%%%%%%%%%%%%%%%%%%%%%%%%%%%%%%%%%%%%%%%%%%%%%%%%%%%%%%%%%%%%%%%%%%%%%%
\Pi_{LL}(s_0, M_B^2) = \Pi_{LR}(s_0, M_B^2) = \Pi_{RL}(s_0, M_B^2) = \Pi_{RR}(s_0, M_B^2) \, .
\end{equation}
Because we do not keep all the $\mathcal{O}(m_q^2)$ terms in the quark propagator given in Eq.~(\ref{eq:quark}), we do not use this long expression in the present study.

We also give the sum rules for the off-diagonal correlation functions among the four tetraquark currents $J_{LL}$, $J_{LR}$, $J_{RL}$, and $J_{RR}$, as well as the sum rules obtained using the eight tetraquark currents $J_{VV}$, $J_{AA}$, $J_{VA}$, $J_{AV}$, $J_{LL;+}$, $J_{LL;-}$, $J_{LR;+}$, and $J_{LR;-}$. For these sum rules, we keep the terms up to the $\mathcal{O}(m_q)$ order, and omit all the disconnected diagrams and most of the next-to-leading-order terms except Fig.~\ref{fig:feynman2}($h{\rm-}8$).

The sum rules for the off-diagonal correlation function between the tetraquark currents $J_{LL}$ and $J_{LR}$ are:
\begin{eqnarray}
%%%%%%%%%%%%%%%%%%%%%%%%%%%%%%%%%%%%%%%%%%%%%%%%%%%%%%%%%%%%%%%%%%%%%%%%%%%%%%
&& \Pi_{LL-LR}(s_0, M_B^2) = \mathcal{B}_{q^2 \rightarrow M_B^2}\Pi_{LL-LR}(q^2) \equiv \mathcal{B}_{q^2 \rightarrow M_B^2} \left[ i\int d^4x~e^{iqx}~\langle 0 | \mathbb{T}\left[ J_{LL}(x){J_{LR}^\dagger}(0) \right] | 0 \rangle \right]
\label{eq:piLLLR}
\\ \nonumber &=& \int^{s_0}_{0} \Bigg(
 {s^2 \over 512 \pi^4} \Big( - m_u - m_d \Big) \langle \bar{q}q\rangle
 + {s \over 1024 \pi^4} \Bigg(  16 \pi^2 \langle\bar{q}q\rangle^2 - \Big( 3 m_u + 3 m_d \Big) \langle g_c\bar{q}\sigma Gq\rangle \Bigg)
\\ \nonumber && ~~~~~~~~ + {1 \over 1024 \pi^4} \Bigg( 16 \pi^2 \langle\bar{q}q\rangle \langle g_c\bar{q}\sigma Gq\rangle + \Big( m_u + m_d \Big) \langle g_c^2 GG \rangle \langle\bar{q}q\rangle \Bigg) \Bigg)~e^{-s/M_B^2}~ds
\\ \nonumber && + {1 \over 4096 \pi^4} \Bigg( 8 \pi^2 \langle g_c\bar{q}\sigma Gq\rangle^2 - {16 \pi^2 \over 3} \langle g_c^2 GG \rangle \langle\bar{q}q\rangle^2 + \Big( {256\over3} m_u + {256\over3} m_d \Big) \pi^4 \langle\bar{q}q\rangle^3
 + \Big( m_u + m_d \Big) \langle g_c^2 GG \rangle \langle g_c\bar{q}\sigma Gq\rangle \Bigg)
\\ \nonumber && - {1 \over M_B^2} \Bigg( -{1 \over 1536 \pi^2 } \langle g_c^2 GG \rangle \langle\bar{q}q\rangle \langle g_c\bar{q}\sigma Gq\rangle + \Big( {1\over72} m_u + {1\over72} m_d \Big) \langle\bar{q}q\rangle^2 \langle g_c\bar{q}\sigma Gq\rangle \Bigg) + \cdots \, .
\end{eqnarray}
We obtain identical sum rules for the following off-diagonal correlation functions:
\begin{equation}
%%%%%%%%%%%%%%%%%%%%%%%%%%%%%%%%%%%%%%%%%%%%%%%%%%%%%%%%%%%%%%%%%%%%%%%%%%%%%%
\Pi_{LL-LR}(s_0, M_B^2) = \Pi_{LL-RL}(s_0, M_B^2) = \Pi_{RR-LR}(s_0, M_B^2) = \Pi_{RR-RL}(s_0, M_B^2) \, .
\end{equation}

The sum rules for the off-diagonal correlation function between the tetraquark currents $J_{LL}$ and $J_{RR}$ are:
\begin{eqnarray}
%%%%%%%%%%%%%%%%%%%%%%%%%%%%%%%%%%%%%%%%%%%%%%%%%%%%%%%%%%%%%%%%%%%%%%%%%%%%%%
&& \Pi_{LL-RR}(s_0, M_B^2) = \mathcal{B}_{q^2 \rightarrow M_B^2}\Pi_{LL-RR}(q^2) \equiv \mathcal{B}_{q^2 \rightarrow M_B^2} \left[ i\int d^4x~e^{iqx}~\langle 0 | \mathbb{T}\left[ J_{LL}(x){J_{RR}^\dagger}(0) \right] | 0 \rangle \right]
\label{eq:piLLRR}
\\ \nonumber &=& \Big( -{1\over6} m_u -{1\over6} m_d \Big) \langle\bar{q}q\rangle^3
 - {1 \over M_B^2} \Bigg(  - {8\pi \alpha_s \over 27} \langle\bar{q}q\rangle^4 - \Big( {1\over8} m_u + {1\over8} m_d \Big) \langle\bar{q}q\rangle^2 \langle g_c\bar{q}\sigma Gq\rangle \Bigg) + \cdots
\, .
\end{eqnarray}
We obtain identical sum rules for the off-diagonal correlation function between $J_{LR}$ and $J_{RL}$:
\begin{equation}
%%%%%%%%%%%%%%%%%%%%%%%%%%%%%%%%%%%%%%%%%%%%%%%%%%%%%%%%%%%%%%%%%%%%%%%%%%%%%%
\Pi_{LL-RR}(s_0, M_B^2) = \Pi_{LR-RL}(s_0, M_B^2) \, .
\end{equation}

The sum rules obtained using the tetraquark current $J_{VV}$ are:
\begin{eqnarray}
%%%%%%%%%%%%%%%%%%%%%%%%%%%%%%%%%%%%%%%%%%%%%%%%%%%%%%%%%%%%%%%%%%%%%%%%%%%%%%
&& \Pi_{VV}(s_0, M_B^2) = \mathcal{B}_{q^2 \rightarrow M_B^2}\Pi_{VV}(q^2)
\label{eq:piVV}
\\ \nonumber &=& \int^{s_0}_{0} \Bigg( {s^4 \over 20480 \pi^6}
 + {s \over 128 \pi^4} \Bigg( 16 \pi^2 \langle\bar{q}q\rangle^2 - \Big( m_u + m_d \Big) \langle g_c\bar{q}\sigma Gq\rangle \Bigg)
\\ \nonumber && ~~~~~~~~ + {1 \over 128 \pi^4} \Bigg( 16 \pi^2 \langle\bar{q}q\rangle \langle g_c\bar{q}\sigma Gq\rangle + \Big( m_u + m_d \Big) \langle g_c^2 GG \rangle \langle\bar{q}q\rangle \Bigg) \Bigg)~e^{-s/M_B^2}~ds
\\ \nonumber && + {1 \over 512 \pi^4} \Bigg( 8 \pi^2 \langle g_c\bar{q}\sigma Gq\rangle^2 - {16 \pi^2 \over 3} \langle g_c^2 GG \rangle \langle\bar{q}q\rangle^2 - \Big( 256 m_u + 256 m_d \Big) \pi^4 \langle\bar{q}q\rangle^3
 + \Big( m_u + m_d \Big) \langle g_c^2 GG \rangle \langle g_c\bar{q}\sigma Gq\rangle \Bigg)
\\ \nonumber && - {1 \over M_B^2} \Bigg( -{1 \over 192 \pi^2 } \langle g_c^2 GG \rangle \langle\bar{q}q\rangle \langle g_c\bar{q}\sigma Gq\rangle - {32\pi \alpha_s \over 27} \langle\bar{q}q\rangle^4 - \Big( {7\over18} m_u + {7\over18} m_d \Big) \langle\bar{q}q\rangle^2 \langle g_c\bar{q}\sigma Gq\rangle \Bigg) + \cdots
\, .
\end{eqnarray}

The sum rules obtained using the tetraquark current $J_{AA}$ are:
\begin{eqnarray}
%%%%%%%%%%%%%%%%%%%%%%%%%%%%%%%%%%%%%%%%%%%%%%%%%%%%%%%%%%%%%%%%%%%%%%%%%%%%%%
&& \Pi_{AA}(s_0, M_B^2) = \mathcal{B}_{q^2 \rightarrow M_B^2}\Pi_{AA}(q^2)
\label{eq:piAA}
\\ \nonumber &=& \int^{s_0}_{0} \Bigg( {s^4 \over 20480 \pi^6}+ {s^2 \over 32 \pi^4} \Big( m_u + m_d \Big) \langle\bar{q}q\rangle
 + {s \over 128 \pi^4} \Bigg( - 16 \pi^2 \langle\bar{q}q\rangle^2 + \Big( 5 m_u + 5 m_d \Big) \langle g_c\bar{q}\sigma Gq\rangle \Bigg)
\\ \nonumber && ~~~~~~~~ + {1 \over 128 \pi^4} \Bigg( - 16 \pi^2 \langle\bar{q}q\rangle \langle g_c\bar{q}\sigma Gq\rangle - \Big( m_u + m_d \Big) \langle g_c^2 GG \rangle \langle\bar{q}q\rangle \Bigg) \Bigg)~e^{-s/M_B^2}~ds
\\ \nonumber && + {1 \over 512 \pi^4} \Bigg( - 8 \pi^2 \langle g_c\bar{q}\sigma Gq\rangle^2 + {16 \pi^2 \over 3} \langle g_c^2 GG \rangle \langle\bar{q}q\rangle^2 - \Big( m_u +  m_d \Big) {1280\pi^4\over3} \langle\bar{q}q\rangle^3
 - \Big( m_u + m_d \Big) \langle g_c^2 GG \rangle \langle g_c\bar{q}\sigma Gq\rangle \Bigg)
\\ \nonumber && - {1 \over M_B^2} \Bigg( {1 \over 192 \pi^2 } \langle g_c^2 GG \rangle \langle\bar{q}q\rangle \langle g_c\bar{q}\sigma Gq\rangle - {32\pi \alpha_s \over 27} \langle\bar{q}q\rangle^4 - \Big( {11\over18} m_u + {11\over18} m_d \Big) \langle\bar{q}q\rangle^2 \langle g_c\bar{q}\sigma Gq\rangle \Bigg) + \cdots
\, .
\end{eqnarray}

The sum rules obtained using the tetraquark currents $J_{VA}$ and $J_{AV}$ are the same:
\begin{eqnarray}
%%%%%%%%%%%%%%%%%%%%%%%%%%%%%%%%%%%%%%%%%%%%%%%%%%%%%%%%%%%%%%%%%%%%%%%%%%%%%%
&& \Pi_{VA}(s_0, M_B^2) = \Pi_{AV}(s_0, M_B^2)
\label{eq:piVA}
\\ \nonumber &=& \int^{s_0}_{0} \Bigg( {s^4 \over 20480 \pi^6}+ {s^2 \over 64 \pi^4} \Big( m_u + m_d \Big) \langle\bar{q}q\rangle
 + {s \over 64 \pi^4} \Big( m_u + m_d \Big) \langle g_c\bar{q}\sigma Gq\rangle \Bigg)~e^{-s/M_B^2}~ds
\\ \nonumber && + \Big( {2\over3} m_u + {2\over3} m_d \Big) \langle\bar{q}q\rangle^3
- {1 \over M_B^2} \Bigg( {32\pi \alpha_s \over 27} \langle\bar{q}q\rangle^4 + \Big( {1\over2} m_u + {1\over2} m_d \Big) \langle\bar{q}q\rangle^2 \langle g_c\bar{q}\sigma Gq\rangle \Bigg) + \cdots
\, .
\end{eqnarray}

The sum rules obtained using the tetraquark currents $J_{LL;+}$ and $J_{LR;+}$ are the same:
\begin{eqnarray}
%%%%%%%%%%%%%%%%%%%%%%%%%%%%%%%%%%%%%%%%%%%%%%%%%%%%%%%%%%%%%%%%%%%%%%%%%%%%%%
&& \Pi_{LL;+}(s_0, M_B^2) = \Pi_{LR;+}(s_0, M_B^2)
\label{eq:piLLP}
\\ \nonumber &=& \int^{s_0}_{0} \Bigg( {s^4 \over 163840 \pi^6}+ {s^2 \over 512 \pi^4} \Big( m_u + m_d \Big) \langle\bar{q}q\rangle
 + {s \over 512 \pi^4} \Big( m_u + m_d \Big) \langle g_c\bar{q}\sigma Gq\rangle \Bigg)~e^{-s/M_B^2}~ds
\\ \nonumber && + \Big( - {1\over12} m_u - {1\over12} m_d \Big) \langle\bar{q}q\rangle^3
- {1 \over M_B^2} \Bigg( - {4\pi \alpha_s \over 27} \langle\bar{q}q\rangle^4 - \Big( {1\over16} m_u + {1\over16} m_d \Big) \langle\bar{q}q\rangle^2 \langle g_c\bar{q}\sigma Gq\rangle \Bigg) + \cdots
\, .
\end{eqnarray}

The sum rules obtained using the tetraquark currents $J_{LL;-}$ and $J_{LR;-}$ are the same:
\begin{eqnarray}
%%%%%%%%%%%%%%%%%%%%%%%%%%%%%%%%%%%%%%%%%%%%%%%%%%%%%%%%%%%%%%%%%%%%%%%%%%%%%%
&& \Pi_{LL;-}(s_0, M_B^2) = \Pi_{LR;-}(s_0, M_B^2)
\label{eq:piLLN}
\\ \nonumber &=& \int^{s_0}_{0} \Bigg( {s^4 \over 163840 \pi^6}+ {s^2 \over 512 \pi^4} \Big( m_u + m_d \Big) \langle\bar{q}q\rangle
 + {s \over 512 \pi^4} \Big( m_u + m_d \Big) \langle g_c\bar{q}\sigma Gq\rangle \Bigg)~e^{-s/M_B^2}~ds
\\ \nonumber && + \Big( {1\over12} m_u + {1\over12} m_d \Big) \langle\bar{q}q\rangle^3
- {1 \over M_B^2} \Bigg( {4\pi \alpha_s \over 27} \langle\bar{q}q\rangle^4 + \Big( {1\over16} m_u + {1\over16} m_d \Big) \langle\bar{q}q\rangle^2 \langle g_c\bar{q}\sigma Gq\rangle \Bigg) + \cdots
\, .
\end{eqnarray}
\end{widetext}

\end{document}